\documentclass[12pt]{article}

\usepackage{color}
\usepackage{array}
\usepackage{epsfig}
\usepackage{amssymb}
\usepackage{graphics,graphpap}

\usepackage{amsmath}

\renewcommand{\thefootnote}{\fnsymbol{footnote}}
\newcommand{\bal}{\begin{align}}

\def \thl {{\theta_l}}
\def \thK {{\theta_{K^*}}}
\def \azeL{{A_0^L}}
\def \azeR{{A_0^R}}
\def \apaL{{A_\parallel^L}}
\def \apaR{{A_\parallel^R}}
\def \apeL{{A_\perp^L}}
\def \apeR{{A_\perp^R}}
\def \re{\text{Re}}
\def \im{\text{Im}}
\def \kstar{{K^*}}
\def \eff{{\text{eff}}}
\def \braket#1#2#3{\langle #1|#2| #3\rangle}

\allowdisplaybreaks
\begin{document}

%%%%%%%%%% Title page
\begin{titlepage}
\begin{flushright}
\begin{tabular}{l}
IPPP/08/58\\
DCPT/08/116\\
TUM--HEP--696/08
\end{tabular}
\end{flushright}
\vskip1.5cm
\begin{center}
{\Large \bf \boldmath
Symmetries and Asymmetries of $B\to K^*\mu^+\mu^-$ Decays\\[5pt] in
the Standard Model and Beyond}
\vskip1.3cm 
{\sc
Wolfgang Altmannshofer\footnote{wolfgang.altmannshofer@ph.tum.de}$^{,1}$,
Patricia Ball\footnote{Patricia.Ball@durham.ac.uk}$^{,1,2}$,
Aoife Bharucha\footnote{a.k.m.bharucha@durham.ac.uk}$^{,2}$,
Andrzej~J.~Buras\footnote{andrzej.buras@ph.tum.de}$^{,1,3}$,
David~M.~Straub\footnote{david.straub@ph.tum.de}$^{,1}$
and 
Michael~Wick\footnote{michael.wick@ph.tum.de}$^{,1}$
} \vskip0.5cm
$^1$ {\em Physik-Department, Technische Universit\"at M\"unchen, 85748 
Garching, Germany}\\
\vskip0.4cm
        $^2$ {\em IPPP, Department of Physics,
University of Durham, Durham DH1 3LE, UK}\\
\vskip0.4cm
$^3$ {\em TUM Institute for Advanced Study, Technische Universit\"at
  M\"unchen, 80333 M\"unchen, Germany}

\vskip1.5cm

\vskip1.5cm

{\large\bf Abstract\\[10pt]} \parbox[t]{\textwidth}{
The rare decay $B\to K^*(\to K\pi)\mu^+\mu^-$ is regarded as one of
the crucial channels for $B$ physics as the polarization of the $K^*$ 
allows a precise angular reconstruction resulting in many observables
that offer new important tests of the Standard Model and its
extensions. These angular observables can be expressed in terms of 
CP-conserving and CP-violating quantities which we study in terms of
the full form factors calculated from QCD sum rules on the light-cone, 
including QCD factorization corrections. We investigate all 
observables in the context of the Standard Model and various New
Physics models, in particular the Littlest Higgs model with T-parity
and various MSSM scenarios, identifying those observables with  
small to moderate 
dependence on hadronic quantities and large impact of New Physics. 
One important result of our studies is that new CP-violating phases
will produce clean signals in CP-violating asymmetries. We also identify a 
number of correlations between various observables which will allow a 
clear distinction between different New Physics scenarios.
}

\end{center}
\end{titlepage}

\setcounter{footnote}{0}
\renewcommand{\thefootnote}{\arabic{footnote}}
\renewcommand{\theequation}{\arabic{section}.\arabic{equation}}

\newpage

%%%%%%%%%%%%%%%%% Sec1: Introduction

\section{Introduction}\label{sec:1}
\setcounter{equation}{0}

The penguin-induced 
flavour-changing neutral current (FCNC) transitions $b\to s$ and $b\to d$ 
are among the most valuable probes of flavour physics. They are
characterized by their high sensitivity to New Physics (NP)
contributions and the particularly large impact of short-distance QCD
corrections to the relevant observables like branching ratios and more
local quantities, see Ref.~\cite{review} for a review. The decay
$b\to s\gamma$ has probably been the most popular FCNC transition 
ever since its first experimental observation as $B\to K^*\gamma$ at
CLEO in 1993 \cite{cleo}. Despite its considerable success as
benchmark probe, in connection with electroweak precision observables
\cite{gw}, its usefulness is limited by the number of observables it
gives access to -- the branching ratio and CP asymmetries
like the time-dependent CP asymmetry in $B\to K^*\gamma$ \cite{Ball:2006cva}.

Much more versatile in this respect
is the decay $b\to s\ell^+\ell^-$ with the possibility to measure,
for instance, the differential decay rate in the leptons' invariant mass.
One can also
construct asymmetries, like the well-known forward-backward
asymmetry ($A_{\rm FB}$), with differing sensitivity to NP
effects. A full angular analysis of $B\to K^*(\to K\pi)\ell^+\ell^-$
would give access to a multitude of observables \cite{Kruger:1999xa}.
The downside of such measurements -- low statistics -- has
started to be overcome at the $B$ factories BaBar and Belle, with recent
measurements of the forward-backward asymmetry in several bins in the
lepton invariant mass and the $K^*$'s polarization 
\cite{BaBar_angular,BelleICHEP}. Current experimental results are compiled in
Tab.~\ref{tab:exp}. 
\begin{table}[b]
\centering
\renewcommand{\arraystretch}{1.4}
\begin{tabular}{|c||c|c|c|} 
\hline
Experiment & BaBar \cite{BRBaBar} & Belle \cite{BelleICHEP} & CDF \cite{BRCDF}
\\\hline\hline
$\text{BR}(B \to K^* \mu^+  \mu^- ) \times 10^7$
&  $11.1\pm 1.9\pm 0.7$ & $10.8^{+1.0}_{-1.0}\pm 0.9$ & $8.1\pm 3.0\pm 1.0$
\\\hline
Number of $B\bar B$ events
& $384\times 10^6$ & $657\times 10^6$ & --\\
\hline
\end{tabular}
\renewcommand{\arraystretch}{1}
\caption[]{\small Experimental results for the branching ratio of $B \to K^*
\mu^ + \mu ^ - $; the region around the charm resonances with $B\to
K^* \psi (\to \mu^+\mu^-)$ is excluded. The first error is
statistics, the second systematics.}\label{tab:exp}
\end{table}
The absolute number of events observed is still rather small (230 at
Belle \cite{BelleICHEP}),  making $B\to K^*\ell^+\ell^-$ one of the rarest $B$
decays ever observed -- at least if the resonance-dominated region
around the charmonium resonances with $B\to K^* \psi (\to
\ell^+\ell^-)$ is excluded.
This situation will improve once the LHC experiments 
have started taking data, allowing one
to probe the short-distance physics governing $b\to
s\ell^+\ell^-$ at an unprecedented level of the angular
spectrum: a recent study by the LHCb collaboration
\cite{LHCbstudy} predicts
7200 signal events (an improvement by an order of magnitude from
the present situation) with a data set of $2\,\text{fb}^{-1}$, which
corresponds to one nominal year of running. 

One difference between the experimental reach of $B$
factories and LHC, though, is the preference of the latter for {\em
  exclusive} channels, mainly realized as $B\to
K^*\ell^+\ell^-$.
This implies that the analysis of this channel requires control not only
over short-distance perturbative effects, described by
Wilson coefficients in the relevant effective Hamiltonian, 
but also long-distance non-perturbative effects, described largely,
but not completely,
by form factors. It is the objective of our paper to provide
such an analysis, based on QCD factorization \cite{BF00,BFS01,BFS04}, including
a full set of form factors calculated from QCD sum rules on the
light-cone \cite{LCSR}, and the dominant effects suppressed for large $b$ quark
mass. As the LHC has increased sensitivity to charged particles
  in the final state, we focus on the decays of neutral $B$'s, $\bar
  B^0\to \bar K^{*0}(\to K^-\pi^+)\mu^+\mu^-$ and its CP-conjugate 
$ B^0\to K^{*0}(\to K^+\pi^-)\mu^+\mu^-$, which have the additional
  advantage that the flavour of the decaying $B$ meson ($B^0$ or 
$\bar B^0$) is unambiguously tagged by the final state. We also focus on
  $\ell=\mu$ which can be cleanly measured at the LHC; see
  Ref.~\cite{hillerkruger} for a discussion of $\mu\leftrightarrow e$ effects.

$B\to K^*\mu^+\mu^-$ decays have been investigated by many
authors of whom we can cite only a few. 
In 1999, Ali et al.\ calculated the dilepton mass spectrum
and $A_{\rm FB}$ in the SM and various SUSY scenarios using na\"\i\/ve
factorization and QCD sum rules on the light cone \cite{Ali99}. Later
it was shown by Beneke et al.\ \cite{BFS01,BFS04} 
that $B\to K^*\mu^+\mu^-$ admits a systematic theoretical
description using QCD factorization in the heavy quark limit
$m_b\to\infty$. This limit is relevant for small invariant lepton masses and 
reduces the number of independent
form factors from 7 to 2. Spectator effects, neglected in na\"\i\/ve
factorization, also become calculable. The drawback is that
corrections to that limit are only partially known: for instance, in
Ref.~\cite{FM02} power-suppressed effects relevant for
isospin asymmetries were calculated.\footnote{Very recently, BaBar has
reported a positive result \cite{BaBarisospin}
for a deviation of the isospin asymmetry
from the SM prediction, which so far, however,
has not been confirmed by Belle \cite{BelleICHEP}.} In Ref.~\cite{Ali06}, a
calculation of $B\to K^*\mu^+\mu^-$ using soft-collinear theory
(SCET) was presented. More recently, two analyses appeared which also
use QCD factorization and focus on possible NP effects in CP
asymmetries \cite{Hiller08} and on observables available from 
angular distributions \cite{Hurth08}, respectively.  There is also
vast literature on NP analyses, with varying degrees of reliability
of theoretical input for long-distance QCD effects and scope of observables
considered. One intrinsic NP contribution, for instance, comes
from an extended scalar sector. Most studies
available so far, with the notable exception of Ref.~\cite{kim},
focus on the effects of such contributions on $A_{\rm FB}$, which
turn out to be small, while we shall argue that the effect can best
be seen in one particular angular observable not considered before, 
see Sec.~\ref{sec:3}.

In the present paper, we aim to improve on previous studies in the
following way:
\begin{itemize}
\item we include the full set of 7 form factors, rather than the 2
  form factors in the heavy quark limit, calculated from QCD sum
  rules on the light-cone; we show that our set of form factors fulfills all
  correlations required in the heavy quark limit, which has never been
  demonstrated before for any form factor calculation;
\item we give an up-to-date prediction of the $B\to
  K^*(\to K\pi)\mu^+\mu^-$ observables in the SM and
  shall argue that the bulk of power-suppressed 
  corrections is due to the difference between the full QCD form factors
  and their heavy quark limit;
\item we study all angular observables in the decay $B\to
  K^*(\to K\pi)\mu^+\mu^-$ and identify those with small sensitivity to
  hadronic and large sensitivity to NP effects;
\item we include the effects of scalar and pseudoscalar operators,
  which are extremely suppressed in the SM, on all angular observables; 
\item we study the effects of various NP models,
  including several manifestations of the 
  MSSM and the Littlest Higgs model with T-parity. 
\end{itemize}
Our main results are collected in Sec.~\ref{sec:7}.

Our paper is organized as follows: in Sec.~\ref{sec:2} we 
review  the theoretical framework, based on the trinity of effective
Hamiltonian, form factors and QCD factorization. 
In Sec.~\ref{sec:3} we discuss the (rather 
involved) kinematics of the decay and define the basic observables in
the process. 
Sec.~\ref{sec:4} gives a short overview over the NP models whose
effects we study.  In Sec.~\ref{sec:5} we define
observables satisfying the requirements of theoretical cleanliness and 
high sensitivity to NP
effects. Section~\ref{sec:6}, the centre part of our paper, contains the
phenomenological analysis of those observables in the SM, in a model-independent way and in several selected
NP scenarios.
We conclude in Sec.~\ref{sec:7}. In the appendices we
review the kinematics of four-body decays and show that for large $b$
quark mass the form factors calculated, in Sec.~\ref{sec:2}, from QCD
sum rules on the light-cone fulfill the relations imposed by
heavy-quark symmetry.

%%%%%%%%%%%%%%%%% Sec2: Theoretical Framework

\section{Theoretical Framework}\label{sec:2}

The theoretical framework which allows one to calculate the decay
amplitude of $B\to K^* \mu^+\mu^-$ 
is quite involved and 
 requires three different steps which are described in this section:
\begin{itemize}
\item the separation of short-distance (QCD, weak interaction and new
  physics) effects from long-distance QCD in an effective Hamiltonian
  ${\cal H}_{\eff}$;
\item the calculation of matrix elements of local quark bilinear
  operators $J$ of type $\langle K^* | J | B\rangle$ (form factors);
\item the calculation of effects of 4-quark operators in
  ${\cal H}_{\eff}$ which give rise to so-called non-factorizable
  corrections and can be calculated using QCD factorization (QCDF).
\end{itemize}
QCDF is only valid for small invariant dilepton mass $q^2\sim
O(1\,{\rm GeV}^2)$, or, equivalently, large $K^*$ energy $E\sim O(m_B/2)$,
which implies certain cuts on $q^2$ or $E$. In this paper, we restrict
ourselves to $1\,{\rm GeV}^2< q^2 < 6\,{\rm GeV}^2$. The reasons will
be discussed in Sec.~\ref{2.4}. 
Obviously, {\em all} the above steps need to be under good control for a
reliable prediction of the decay. We will discuss them in turn and
also explain our strategy for calculating the $B\to K^* \mu^+\mu^-$
amplitude.

%%%%%%%%%%%%%%%%% Sec2.1: Effective Hamiltonian

\subsection{Effective Hamiltonian}

The effective Hamiltonian for $b\to s \mu^+\mu^-$ transitions is
given by \cite{bobeth,bobeth02}
\begin{equation} \label{eq:Heff}
    {\cal H}_{\eff} = - \frac{4\,G_F}{\sqrt{2}}\left(
\lambda_t {\cal H}_{\eff}^{(t)} + \lambda_u {\cal
  H}_{\eff}^{(u)}\right)
\end{equation}
with the CKM combination $\lambda_i=V_{ib}V_{is}^*$ and
\begin{eqnarray*}
{\cal H}_{\eff}^{(t)} 
& = & 
C_1 \mathcal O_1^c + C_2 \mathcal O_2^c + \sum_{i=3}^{6} C_i 
\mathcal O_i + \sum_{i=7,8,9,10,P,S} (C_i \mathcal O_i + C'_i \mathcal
O'_i)\,,
\\
{\cal H}_{\eff}^{(u)} 
& = & 
C_1 (\mathcal O_1^c-\mathcal O_1^u)  + C_2(\mathcal O_2^c-\mathcal
O_2^u)\,.
\end{eqnarray*}
Although the contribution of ${\cal H}_{\eff}^{(u)}$ is doubly
Cabibbo-suppressed with respect to that of ${\cal H}_{\eff}^{(t)}$ and
hence often dropped, it proves relevant for certain observables
sensitive to complex phases of decay amplitudes, so we keep it. 
The operators $\mathcal O_{i\leq 6}$ are identical to the $P_i$ given
in Ref.~\cite{bobeth}, while the remaining ones are given by
\begin{align}
\label{eq:O7}
{\mathcal{O}}_{7} &= \frac{e}{g^2} m_b
(\bar{s} \sigma_{\mu \nu} P_R b) F^{\mu \nu} ,&
{\mathcal{O}}_{7}^\prime &= \frac{e}{g^2} m_b
(\bar{s} \sigma_{\mu \nu} P_L b) F^{\mu \nu} ,\\
\label{eq:O8}
{\mathcal{O}}_{8} &= \frac{1}{g} m_b
(\bar{s} \sigma_{\mu \nu} T^a P_R b) G^{\mu \nu \, a} ,&
{\mathcal{O}}_{8}^\prime &= \frac{1}{g} m_b
(\bar{s} \sigma_{\mu \nu} T^a P_L b) G^{\mu \nu \, a} ,\\
\label{eq:O9}
{\mathcal{O}}_{9} &= \frac{e^2}{g^2} 
(\bar{s} \gamma_{\mu} P_L b)(\bar{\mu} \gamma^\mu \mu) ,&
{\mathcal{O}}_{9}^\prime &= \frac{e^2}{g^2} 
(\bar{s} \gamma_{\mu} P_R b)(\bar{\mu} \gamma^\mu \mu) ,\\
\label{eq:O10}
{\mathcal{O}}_{10} &=\frac{e^2}{g^2}
(\bar{s}  \gamma_{\mu} P_L b)(  \bar{\mu} \gamma^\mu \gamma_5 \mu) ,&
{\mathcal{O}}_{10}^\prime &=\frac{e^2}{g^2}
(\bar{s}  \gamma_{\mu} P_R b)(  \bar{\mu} \gamma^\mu \gamma_5 \mu) ,\\
\label{eq:OS}
{\mathcal{O}}_{S} &=\frac{e^2}{16\pi^2}
m_b (\bar{s} P_R b)(  \bar{\mu} \mu) ,&
 {\mathcal{O}}_{S}^\prime &=\frac{e^2}{16\pi^2}
m_b (\bar{s} P_L b)(  \bar{\mu} \mu) ,\\
\label{eq:OP}
{\mathcal{O}}_{P} &=\frac{e^2}{16\pi^2}
m_b (\bar{s} P_R b)(  \bar{\mu} \gamma_5 \mu) ,&
 {\mathcal{O}}_{P}^\prime &=\frac{e^2}{16\pi^2}
m_b (\bar{s} P_L b)(  \bar{\mu} \gamma_5 \mu),
\end{align}
where $g$ is the strong coupling constant and $P_{L,R}=(1\mp
\gamma_5)/2$. $m_b$ denotes the running $b$ quark
mass in the $\overline{\rm MS}$ scheme. 
The primed operators with opposite chirality to the unprimed ones 
vanish or are highly suppressed in the SM, as are $\mathcal O_{S,P}$. 
We neglect the contributions of $\mathcal O_i^\prime$ for $1\leq i\leq 6$. 
These operators are generated in some NP scenarios,
for instance in left-right symmetric models or through gluino 
contributions in a general MSSM, but their
impact is either heavily constrained or turns out to be very small 
generically.

The Wilson coefficients $C_i$ in (\ref{eq:Heff}) encode
  short-distance physics and possible NP effects. They are calculated at the
matching scale $\mu=m_W$, in a perturbative expansion in powers of
$\alpha_s(m_W)$, and are then evolved down to scales $\mu \sim
m_b$ according to the solution of the renormalization group
equations. Any NP contributions enter through $C_i(m_W)$, while
  the evolution to lower scales is determined by the SM.
The inclusion of the factors 
$16\pi^2/g^2=4\pi/\alpha_s$ in 
the definition of the operators $\mathcal{O}_{i\geq 7}$ 
and the corresponding primed operators serves to allow a more
transparent organization of the expansion of their Wilson 
coefficients in perturbation theory: 
all $C_i$ are expanded as
\begin{equation}\label{2.8}
C_i = C_i^{(0)} + \frac{\alpha_s}{4\pi}\, C_i^{(1)} + 
\left(\frac{\alpha_s}{4\pi}\right)^2 C_i^{(2)} + O(\alpha_s^3)\,,
\end{equation}
where $C_i^{(0)}$ is the tree-level contribution, which vanishes 
 for all operators but $\mathcal O_2$. In our normalization of operators 
also $C_9^{(0)}$ is non-zero. $C_i^{(n)}$ denotes an $n$-loop
contribution.
In our paper we aim at next-to-next-to-leading logarithmic (NNLL)
accuracy, which requires the calculation of the matching conditions
at $\mu=m_W$ to two-loop accuracy. This has been done in
Ref.~\cite{bobeth}. NP contributions, on the other hand, will be included
to one-loop accuracy only.\footnote{An explicit calculation of 
two-loop  corrections  in the MSSM \cite{bobeth02} shows that they are small.}

Two-loop accuracy in the matching requires the inclusion of
anomalous dimensions in the renormalization-group equations
to three-loop accuracy. The corresponding $O(\alpha_s^3)$ entries in
the $10\times 10$ SM anomalous dimension matrix have been calculated
in Refs.~\cite{gorbahn,gorbahn2}. On the other hand, the 
operators $\mathcal O_{S,P}^{(')}$ are given in terms
of conserved currents, i.e.\ they carry no scale-dependence, they do
not mix with other operators and their
Wilson coefficients are given by the coefficients at the matching
scale. $\mathcal O_9$ is also given by conserved currents, but mixes with
$\mathcal O_{1,\dots,6}$, via diagrams with a virtual photon decaying into
$\mu^+\mu^-$. Additional scale dependence in $C_9$ comes from 
the factor $1/g^2$. The latter dependence is also present in $C_{10}$, 
which otherwise would be scale independent.

In Tab.~\ref{tab:wilson} we give all the SM values of the Wilson
coefficients to NNLL accuracy.
As we shall see below, in Eq.~(\ref{eq:matrixelement}), 
$C_{7,9}$ always appear in a particular combination with other $C_i$
in matrix elements. It hence proves convenient to define effective coefficients
$C_{7,9}^{(\prime)\rm eff}$, and also $C_{8,10}^{(\prime)\rm eff}$, 
which are given by
\cite{ceff}
\begin{eqnarray}
C_7^{\rm eff} & = & \frac{4\pi}{\alpha_s}\, C_7 -\frac{1}{3}\, C_3 -
\frac{4}{9}\, C_4 - \frac{20}{3}\, C_5\, -\frac{80}{9}\,C_6\,,
\nonumber\\
C_8^{\rm eff} & = & \frac{4\pi}{\alpha_s}\, C_8 + C_3 -
\frac{1}{6}\, C_4 + 20 C_5\, -\frac{10}{3}\,C_6\,,
\nonumber\\
C_9^{\rm eff} & = & \frac{4\pi}{\alpha_s}\,C_9 + Y(q^2)\,,
\nonumber\\
C_{10}^{\rm eff} & = & \frac{4\pi}{\alpha_s}\,C_{10}\,,\qquad
C_{7,8,9,10}^{\prime,\rm eff} = \frac{4\pi}{\alpha_s}\,C'_{7,8,9,10}\,,
\\
{\rm with}\quad
Y(q^2) & = & h(q^2,m_c) \left( \frac{4}{3}\, C_1 + C_2 + 6 C_3 + 60 C_5\right)
\nonumber\\
& & {}-\frac{1}{2}\,h(q^2,m_b) \left( 7 C_3 + \frac{4}{3}\,C_4 + 76 C_5
  + \frac{64}{3}\, C_6\right)
\nonumber\\
& & {}-\frac{1}{2}\,h(q^2,0) \left( C_3 + \frac{4}{3}\,C_4 + 16 C_5
  + \frac{64}{3}\, C_6\right)
\nonumber\\
& & {} + \frac{4}{3}\, C_3 + \frac{64}{9}\, C_5 + \frac{64}{27}\,
C_6\,.
\end{eqnarray} 
The function
\begin{equation}
h(q^2,m_q) = -\frac{4}{9}\, \left( \ln\,\frac{m_q^2}{\mu^2} - \frac{2}{3}
- z \right) - \frac{4}{9}\, (2+z) \sqrt{|z-1|} \times 
\left\{
\begin{array}{l@{\quad}l}
\displaystyle\arctan\, \frac{1}{\sqrt{z-1}} & z>1\\[10pt]
\displaystyle\ln\,\frac{1+\sqrt{1-z}}{\sqrt{z}} - \frac{i\pi}{2} & z \leq 1
\end{array}
\right.
\end{equation}
with $z=4 m_q^2/q^2$, is related to the basic fermion loop.

\begin{table}
\renewcommand{\arraystretch}{1.4}
\addtolength{\arraycolsep}{1pt}
$$
\begin{array}{|c|c|c|c|c|c|c|c|c|c|}
\hline
 C_1(\mu) &   C_2(\mu) &  C_3(\mu) &  C_4(\mu) & C_5(\mu) &  C_6(\mu)
& C_7^{\rm eff}(\mu) & C_8^{\rm eff}(\mu) & C_9^{\rm eff}(\mu)-Y(q^2) & 
C_{10}^{\rm eff}(\mu)\\\hline
-0.257 & 1.009 & -0.005 & -0.078  & 0.000 & 0.001 & -0.304 & -0.167 &
4.211 & -4.103  \\
\hline\hline
\bar C_1(\mu) &  \bar C_2(\mu) & \bar C_3(\mu) & \bar C_4(\mu) & 
\bar C_5(\mu) & \bar C_6(\mu) & \multicolumn{1}{c|}{C_7'{}^{\rm
    eff}(\mu)} & \multicolumn{1}{c|}{C_8'{}^{\rm
    eff}(\mu)}\\\cline{1-8}
-0.128 & 1.052 & 0.011 & -0.032 & 0.009 & -0.037 & \multicolumn{1}{c|}{-0.006}
& \multicolumn{1}{c|}{-0.003}\\
\cline{1-8}
\end{array}
$$
\renewcommand{\arraystretch}{1}
\addtolength{\arraycolsep}{-1pt}
\caption[]{\small 
SM Wilson coefficients at the scale $\mu=m_b=4.8$\,GeV,
to NNLL accuracy. All other Wilson coefficients are heavily
suppressed in the SM.
The ``barred'' $\bar C_i$ are related to $C_i$ as defined in Ref.~\cite{BFS01}.
Input: $\alpha_s(m_W)=0.120$, $\alpha_s(m_b)=0.214$, 
obtained from $\alpha_s(m_Z)=0.1176$ \cite{pdg}, 
using three-loop evolution. We also use $m_t(m_t)=162.3\,$GeV \cite{EWWG},
$m_W = 80.4\,$GeV and $\sin^2\theta_W = 0.23$.
}\label{tab:wilson}
\end{table}

We shall see below that $B\to K^*(\to K\pi)\mu^+\mu^-$does not allow
access to all the above coefficients separately: for instance, only
the combinations $C_S-C_S'$ and $C_P-C_P'$ enter the decay
amplitude.

%%%%%%%%%%%%%% Sec2.2: Form Factors

\subsection{Form Factors}\label{sec:ff}

The $B\to K^*$ matrix elements of the operators $\mathcal
O_{7,9,10,S,P}^{(\prime)}$ can be expressed in
terms of seven form factors which depend on the momentum transfer $q^2$ between
the $B$ and the $K^*$ ($q^\mu = p^\mu - k^\mu$):
\begin{eqnarray}
\lefteqn{
\langle \bar K^*(k) | \bar s\gamma_\mu(1-\gamma_5) b | \bar B(p)\rangle  =  
-i \epsilon^*_\mu (m_B+m_{K^*})
A_1(q^2) + i (2p-q)_\mu (\epsilon^* \cdot q)\,
\frac{A_2(q^2)}{m_B+m_{K^*}}}\hspace*{2.8cm}\nonumber\\
&& {}+  i
q_\mu (\epsilon^* \cdot q) \,\frac{2m_{K^*}}{q^2}\,
\left[A_3(q^2)-A_0(q^2)\right] +
\epsilon_{\mu\nu\rho\sigma}\epsilon^{*\nu} p^\rho k^\sigma\,
\frac{2V(q^2)}{m_B+m_{K^*}},\hspace*{0.5cm}\label{eq:SLFF}\\
{\rm with\ }A_3(q^2) & = & \frac{m_B+m_{K^*}}{2m_{K^*}}\, A_1(q^2) -
\frac{m_B-m_{K^*}}{2m_{K^*}}\, A_2(q^2)\mbox{~~and~~} 
A_0(0) =  A_3(0);\label{eq:A30}\\[-1cm]\nonumber
\end{eqnarray}
\begin{eqnarray}
\lefteqn{\langle \bar K^*(k) | \bar s \sigma_{\mu\nu} q^\nu (1+\gamma_5) b |
\bar B(p)\rangle = i\epsilon_{\mu\nu\rho\sigma} \epsilon^{*\nu}
p^\rho k^\sigma \, 2 T_1(q^2)}\nonumber\\
& & {} + T_2(q^2) \left[ \epsilon^*_\mu
  (m_B^2-m_{K^*}^2) - (\epsilon^* \cdot q) \,(2p-q)_\mu \right] + T_3(q^2) 
(\epsilon^* \cdot q) \left[ q_\mu - \frac{q^2}{m_B^2-m_{K^*}^2}\, (2p-q)_\mu
\right],\nonumber\\[-10pt]\label{eq:pengFF}
\end{eqnarray}
with $T_1(0) = T_2(0)$. $\epsilon_\mu$ is the polarization vector of
the $K^*$. The form factors $A_i$ and $V$ are observables, i.e.\
scale independent, while the $T_i$ depend on the renormalization scale
$\mu$. 

$A_0$ is also the form factor of the pseudoscalar current:
\begin{equation}\label{eq:A0}
\langle \bar K^* |\partial_\mu A^\mu | \bar B\rangle = (m_b+m_s)
\langle \bar K^* |\bar s i\gamma_5 b | \bar B\rangle = 2 m_{K^*}
(\epsilon^* \cdot q) A_0(q^2).
\end{equation}

The form factors are hadronic quantities and call for a
non-perturbative calculation. No lattice calculation of a full set of
form factors is available yet. As a recent result we quote a (quenched) 
value for $T_1(0)$ relevant for $B\to K^*\gamma$:
$T_1(0) = 0.24\pm 0.03^{+0.04}_{-0.01}$ \cite{T1latt}. Preliminary results from
an alternative lattice calculation of $T_1(0)$ have been reported in
Ref.~\cite{wingate}. At present, a more promising method for
calculating form factors at large energies of the final-state meson
(i.e.\ at small $q^2$)
is offered by QCD sum rules on the light-cone (LCSRs) (s.\
Ref.~\cite{LCSR} for reviews).   
This method combines standard QCD sum rule techniques with
the information on light-cone hadron distribution amplitudes (DAs)
familiar from the theory of exclusive processes \cite{exclusive}. It
has been applied to $B\to K^*$ form factors in, for instance,
Refs.~\cite{BB98,BZ04}. 
The key idea is to
consider a correlation function of the $b\to s$ current and a current with
the quantum numbers of the $B$ meson, sandwiched between the vacuum and
the $K^*$. For large (negative) virtualities of these currents, the
correlation function is, in coordinate-space, dominated by light-like distances
 and can be expanded around the
light-cone. In contrast to the short-distance expansion
employed in conventional QCD sum rules \`a la
Shifman/Vainshtein/Zakharov \cite{SVZ}, where
non-perturbative effects are encoded in vacuum expectation values
of local operators with
vacuum quantum numbers, the condensates, LCSRs
rely on the factorization of the underlying correlation function into
genuinely non-perturbative and universal hadron DAs
$\phi$. The DAs are convoluted with process-dependent amplitudes $T_H$,
which similarly to Wilson coefficients 
can be
calculated in perturbation theory, schematically
\begin{equation}\label{eq:3}
\mbox{correlation function~}\sim \sum_n T_H^{(n)}\otimes \phi^{(n)}.
\end{equation}
The sum runs over contributions with increasing twist, labelled by
$n$, and $\otimes$ means integration over the longitudinal momenta of
the partons described by $\phi^{(n)}$.
We shall see below that contributions of non-leading twist are
suppressed by increasing powers of $m_{K^*}/m_b$.
The same correlation function can, on the other hand, be written as a
dispersion-relation, in the virtuality of the current coupling to the
$B$ meson. Equating dispersion-representation and
light-cone expansion, and separating the $B$ meson contribution from
that of higher one- and multi-particle states, one obtains a relation
(QCD sum rule) for the form factor.

For $B\to K^*$ form factors the relevant correlation function is
\begin{eqnarray}\label{eq5}
i\int d^4y e^{-ipy} \langle \bar K^*(p)|TJ_\mu(0) j_B^\dagger(y)|0\rangle
\propto \Pi(q^2)\end{eqnarray}
 with $j_B = \bar d i\gamma_5 b$ and $J_\mu = \bar s\gamma_\mu
 (1-\gamma_5)b$, $\bar s\sigma_{\mu\nu} q^\nu (1+\gamma_5)b$ or 
$\bar s i\gamma_5 b$. The factor of proportionality contains
 four-vectors with open indices and/or mass factors like $(m_B+m_{K^*})$
 etc., which are irrelevant for  dynamics.
LCSRs for all 7 form factors except for $A_0$ are available at $O(\alpha_s)$ 
accuracy for  twist-2 and-3 and tree-level accuracy for twist-4
 contributions \cite{BZ04}. For this paper, we have also calculated
 the LCSR for $A_0$ to the same accuracy. 
The
correlation function $\Pi(q^2)$, calculated for unphysical
$p^2$, can be written as dispersion-relation over its physical cut. Singling
out the contribution of the $B$ meson, one has, for the pseudoscalar current
 $J_\mu = \bar s i\gamma_5 b$,
\begin{equation}\label{eq:corr}
\Pi(q^2) =  A_0(q^2) \, \frac{m_B^2f_B}{m_b}\,\frac{1}{m_B^2-p^2}
+ \mbox{\rm higher poles and cuts},
\end{equation}
where $f_B$ is the leptonic decay constant of the $B$ meson,
\begin{equation}\label{fB}
f_Bm_B^2=m_b\langle B| \bar b i\gamma_5 d|0\rangle\,.
\end{equation}
In the framework of LCSRs one does not use (\ref{eq:corr}) as it stands,
but performs a  Borel-trans\-for\-ma\-tion,
\begin{equation}\label{eq:9}
\hat{B}\,\frac{1}{t-p^2} = \frac{1}{M^2} \exp(-t/M^2),
\end{equation}
with the Borel-parameter $M^2$; this transformation enhances the
ground-state $B$ meson contribution to the dispersion-representation of $\Pi$.
The next step is to invoke quark-hadron
duality to approximate the contributions of hadrons other than the
ground-state $B$ meson by the imaginary part of the light-cone
expansion of $\Pi$, so that
\begin{eqnarray}
\hat{B}{\Pi^{\rm LC}} & = &
\frac{1}{M^2}\, \frac{m_B^2f_B}{m_b} \,A_0(q^2)\,e^{-m_B^2/M^2} +
\frac{1}{M^2}\, \frac{1}{\pi}\int_{s_0}^\infty \!\! dt \, {\rm
Im}{\Pi^{\rm LC}}(t) \, \exp(-t/M^2)\,.\hspace*{0.5cm}\label{eq:SR}
\end{eqnarray}
Subtracting the integral from both sides, Eq.~(\ref{eq:SR}) becomes 
the LCSR for $A_0$. $s_0$ is the so-called continuum
threshold, which separates the ground-state from the continuum
contribution. 
As with standard QCD sum rules, the use of quark-hadron 
duality above $s_0$ and the
choice of $s_0$ itself introduce a certain model-dependence (or
systematic error) in the final result for the form factor.

As an explicit example for a LCSR, we quote the tree-level result for
$T_1(0)$ as given in Ref.~\cite{BZ06}:
\begin{eqnarray}
\lefteqn{\frac{m_B^2 f_B}{m_b}\, T_1(0) e^{-m_B^2/M^2} = 
f_{K^*}^\perp m_b \int_{u_0}^1 du \, e^{-m_b^2/(uM^2)}
\,\frac{\phi_\perp(u)}{2u}} \nonumber\\
&&{}+ f_{K^*}^\parallel m_{K^*} \int_{u_0}^1 du \, e^{-m_b^2/(uM^2)}\left[
  \frac{\Phi(u)}{2u} + \frac{1}{2}\,g_\perp^{(v)}(u) +
  \frac{1}{8u}\left( 1 - u \frac{d}{du}\right)
  g_\perp^{(a)}(u)\right.\nonumber\\
&&\hspace*{1.3cm}{}
\left. -\frac{1}{u}\frac{d}{du} \int_0^u d\alpha_1 \int_0^{\bar
    u}d\alpha_2 \,\frac{u-\alpha_1}{2\alpha_3^2}\,
\left({\cal A}(\underline{\alpha})\vphantom{\frac{1}{2}} +
  {\cal V}(\underline{\alpha})\right)\right]\nonumber\\
&&{}+ f_{K^*}^\perp m_b \frac{m_{K^*}^2}{m_b^2}  \int_{u_0}^1
du \, e^{-m_b^2/(uM^2)}\left[\frac{1}{2}\frac{d}{du}\left\{
  \vphantom{\frac{1}{2}} u\bar u
  \phi_\perp(u) + 2 I_L(u) + u H_3(u)
  \right.\right.\nonumber\\
&&\hspace*{1.3cm}\left.{}-\int_0^u d\alpha_1 \int_0^{\bar u}
  d\alpha_2 \,\frac{1}{\alpha_3} \left(
  S(\underline{\alpha}) - \tilde S(\underline{\alpha}) +
  T_1^{(4)}(\underline{\alpha}) - T_2^{(4)}(\underline{\alpha}) + 
T_3^{(4)}(\underline{\alpha}) - 
T_4^{(4)}(\underline{\alpha})\right)\right\}\nonumber\\
&&{}\hspace*{1.3cm}\left. -
  \frac{1}{8}\, u \frac{d^2}{du^2}\,{\mathbb A}_\perp(u)\right],\\
&\equiv & m_b \int_{u_0}^1 du \, e^{-m_b^2/(uM^2)}\left[ f_{K^*}^\perp  R_1(u)
  + f_{K^*}^\parallel\,\frac{m_{K^*}}{m_b} \,R_2(u) + f_{K^*}^\perp
  \left(\frac{m_{K^*}}{m_b}\right)^2 R_3(u)\right],\label{SR}
\end{eqnarray}
where $u_0$ is given by $m_b^2/s_0$. $f_{K^*}^\parallel$ 
and $f_{K^*}^\perp$ are the decay constants of, respectively,
longitudinally and transversely polarized $K^*$ mesons.
$\phi_\perp$, $\Phi$, $g_\perp^{(v,a)}$,
$I_L$ and $H_3$ are DAs and integrals
thereof, as defined in Ref.~\cite{BZ04}. 
$\cal A$, $\cal {V}$, $S$, $\tilde S$ and $T_i^{(4)}$
are three-particle DAs. The precise definitions of all these DAs as
well as explicit parameterizations can
be found in Refs.~\cite{DAs}. In a slight abuse of language, we shall call
$R_1$ the twist-2 contribution to the sum rule, $R_2$ twist-3 and
$R_3$ twist-4. $u$ is the longitudinal momentum fraction of
the quark in a two-particle Fock state of the final-state vector
meson, whereas $\alpha_{1,2,3}$, with $\sum \alpha_i=1$, are the
momentum fractions of the partons in a three-particle state. The
light-cone expansion is accurate up to terms of order
$(m_{K^*}/m_b)^3$. Up-to-date results for these DAs can be found in
Ref.~\cite{DAs}. Although
we only write down the tree-level expression for the form factor, radiative
corrections are known for $R_1$ \cite{BB98} and the two-particle
contributions to $R_2$ \cite{BZ04}, and will be included in the
numerical analysis. All scale-dependent quantities are calculated at
the (infra-red) 
factorization scale $\mu^2_F = m_B^2-m_b^2$. The form factor itself
carries an ultra-violet scale dependence. As a default, we choose
$\mu=m_b$ for that ultra-violet scale.

It is clearly visible from the above formula that the respective
weight of various contributions is controlled by the parameter
$m_{K^*}/m_b$; the next term in the light-cone expansion contains 
twist-3, -4 and -5 DAs and is 
of order $(m_{K^*}/m_b)^3$. 
Numerically, the expansion works very well, with the $O(m_{K^*}^2/m_b^2)$ terms contributing less than 5\% to the
LCSRs. Note that the expansion is in $m_{K^*}/m_b$ only for
$q^2=0$. For $q^2>0$, the expansion parameter is, see App.~\ref{app:B}, $m_b
m_{K^*}/(m_b^2-q^2)\approx m_{K^*}/(2E)$, $E$ being the energy of the
$K^*$. Obviously, the smaller $E$ (and the
larger $q^2$), the more relevant the higher-twist terms. For $E\to
m_{K^*}$, the light-cone expansion breaks down.

The LCSR method sketched above does not rely on $m_b$ being a large
(or hard) scale -- LCSRs also work very well for $D$ meson decays, see
Ref.~\cite{Ddecays}. In order to calculate the $B\to K^*\mu^+\mu^-$
decay amplitude, however, knowing the form factors is not enough:
there are additional terms which can be calculated using QCD
factorization, see the next subsection. 

As for numerics, we collect the most important input parameters in
Tab.~\ref{tab:numinput}. We evaluate the sum rules at $M^2=8\,$GeV$^2$
and choose $s_0$ such that the minimum in $M^2$ is at
$8\,$GeV$^2$. The resulting $s_0$ lie all between $33\,{\rm GeV}^2$ and
$37\,{\rm GeV}^2$. Note that the LCSRs return results for $f_B$ times
the form factor, rather than the form factor itself. Hence the LCSR
must be divided by $f_B$ which, as one can see from
Tab.~\ref{tab:numinput}, comes with a rather large error. It is well
known that the resulting prediction for, for instance, $T_1(0)$ is on
the high side compared with the experimental result that can be
extracted from the branching ratio of $B\to K^*\gamma$, assuming the
absence of New Physics \cite{BFS01}. For this reason we {\em fix} the
value of $f_B$ to reproduce the
experimental value $T_1^{\rm exp}(0) = 0.268$ \cite{Ball:2006eu}. This
corresponds to setting $f_B = 0.186\,$GeV -- well within the allowed
range quoted in Tab.~\ref{tab:numinput}. As the  observables
we calculate  are ratios, the normalization and the precise value of $f_B$
cancel in the end -- which is why we neglect the residual experimental
and theoretical error of $T_1^{\rm exp}(0)$ and  $f_B$ of about
$7\%$ \cite{Ball:2006eu}.  
The resulting values of the
form factors at $q^2=0$ are given in Tab.~\ref{tab:FFs}. 
With $f_B$ fixed, the errors
of the form factors become rather small and are below 20\%.

\begin{figure}
\centering
\includegraphics[width=0.45\textwidth]{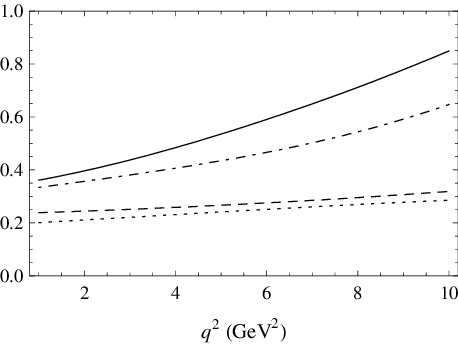}
\qquad
\includegraphics[width=0.45\textwidth]{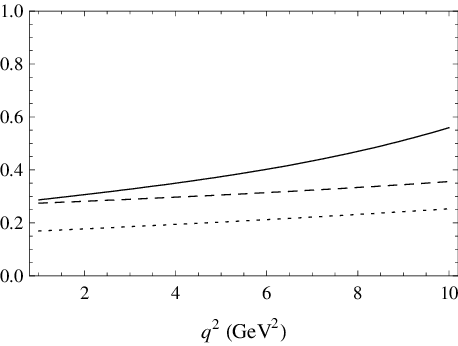}
\caption[]{\small Form factors from LCSRs for central values of input
  parameters. Left: Solid curve: $A_0$, long dashes: $A_1$, short dashes: $A_2$, dot-dashed curve: $V$. Right: Solid curve: $T_1$, long dashes: $T_2$, short dashes: $T_3$.}\label{fig:1}
\end{figure}

As for the $q^2$-dependence, it follows directly from the sum
rules. In Fig.~\ref{fig:1} we plot the central values of all form
factors as functions of $q^2$.

\begin{table}[tb]
\addtolength{\arraycolsep}{3pt}
\renewcommand{\arraystretch}{1.3}
$$
\begin{array}{|c|c|c|c|}\hline
\multicolumn{4}{|c|}{\mbox{$B$ parameters}}\\\hline
f_B\mbox{~\cite{Onogi}} & \lambda_{B}(\mu_h) 
\mbox{~\cite{BZ06}} & \multicolumn{2}{l|}{\mu_h} \\\hline
200(25)\,{\rm MeV} & 0.51(12)\,{\rm GeV} 
& \multicolumn{2}{l|}{2.2\,{\rm GeV}}\\\hline\hline
\multicolumn{4}{|c|}{\mbox{$K^*$ parameters}}\\\hline
f_{K^*}^\parallel & f_{K^*}^\perp(2{\rm GeV}) & a_1^{\perp,\parallel}
(2{\rm GeV}) & 
a_2^{\perp,\parallel}(2{\rm GeV})\\\hline
220(5)\,{\rm MeV} & 163(8)\,{\rm MeV} & 0.03(3) & 0.08(6)\\\hline\hline
\multicolumn{4}{|c|}{\mbox{quark masses}}\\\hline
m_b(m_b)\mbox{~\cite{steinhauser}} & 
m_c(m_c)\mbox{~\cite{steinhauser}} & \multicolumn{2}{l|}{
m_t(m_t)\mbox{~\cite{EWWG}}}\\\hline
4.20(4)\,{\rm GeV} & 1.30(2)\,{\rm GeV} & 
\multicolumn{2}{l|}{162.3(1.1)\,{\rm GeV}} \\\hline
\end{array}
$$
%\vspace*{-20pt}
\caption[]{\small Numerical values of hadronic input 
parameters. $a_i^{\perp,\parallel}$ are parameters of the twist-2
$K^*$ DAs and are taken from Ref.~\cite{DAs}, from where we  also
take all higher-twist parameters not included in the table. 
}\label{tab:numinput}
\end{table}
\begin{table}[tb]
$$
\addtolength{\arraycolsep}{3pt}
\renewcommand{\arraystretch}{1.3}
\begin{array}{|c|c|c|c|}
\hline
A_0(0) & A_1(0) & A_2(0) & V(0)\\\hline
0.333\pm 0.033 & 0.233 \pm 0.038 & 0.190\pm 0.039 & 0.311\pm 0.037\\
\hline\hline
T_1(0) & T_3(0) & \xi_\parallel(0) & \xi_\perp(0)\\\hline
0.268\pm 0.045 & 0.162\pm 0.023 & 0.118\pm 0.008 & 0.266\pm 0.032\\
\hline
\end{array}
$$
\caption[]{\small LCSR results for $q^2=0$. $T_2(0)=T_1(0)$.
  The scale-dependent form factors
  $T_i$ and $\xi_{\parallel,\perp}$ are evaluated at
  $\mu=4.8\,$GeV. The soft form factors $\xi_{\perp,\parallel}$ are
  introduced in Sec.~\ref{sec:2.3}. 
  The error is calculated from varying $s_0$ by $\pm
  2\,$GeV$^2$, $M^2$ by $\pm 2\,$GeV$^2$ and all hadronic input
  parameters according to their uncertainties given in
  Tab.~\ref{tab:numinput}, except for $f_B$, see text.}\label{tab:FFs}
\end{table}

%%%%%%%%%%%%%% Sec2.3: QCD Factorization

\subsection{QCD Factorization}\label{sec:2.3}

In addition to terms proportional to the form factors, the $B\to
K^*\mu^+\mu^-$ amplitude also contains certain ``non-factorizable''
effects that do not correspond to form factors. They are related to
matrix elements of the purely hadronic operators $\mathcal O_1$ to
$\mathcal O_6$ and the chromomagnetic-dipole operator $\mathcal O_8$ 
with additional (virtual) photon emission. These effects can, in the
combined heavy quark and large energy limit, be calculated using 
QCD factorization (QCDF) methods \cite{BF00,BFS01,BFS04}. Here large energy
 means large energy of the $K^*$, $E\sim O(m_B/2)$. $E$ is related
to $q^2$, the dilepton mass, by
\begin{equation}
2 m_B E = m_B^2 + m_{K^*}^2 - q^2\,.
\end{equation}
For the phenomenological analysis in later sections, we require $E> 2.1\,$GeV, 
which corresponds to $q^2<6\,{\rm GeV}^2$, well below the charm
threshold. We would like to stress here that QCDF {\em does not work}
for large $q^2$ above the charm resonances -- here the only
theoretical prediction we have are the contributions to the $B\to
K^*\mu^+ \mu^-$ matrix element given in terms of the form
factors, which is probably a reasonable approximation at the 10 to 20\%
level.

In the heavy quark and large energy limit, the number of 
independent form factors reduces from 7 to 2 which correspond to the
polarization of the $K^*$ (transversal or longitudinal) 
and are usually denoted by $\xi_{\perp}$ and $\xi_{\parallel}$.
Neglecting for the moment $O(\alpha_s)$ corrections, one can define
the $\xi$'s as \cite{BFS04}
\begin{eqnarray}
\xi_\perp(q^2) &=& \frac{m_B}{m_B+m_{K^*}}\, V(q^2)\,,\label{3.17}\\
\xi_\parallel(q^2) &=& \frac{m_B+m_{K^*}}{2E}\,A_1(q^2) -
\frac{m_B-m_{K^*}}{m_B}\, A_2(q^2)\,.\label{3.18}
\end{eqnarray}

At this point we would like to recall that the large energy limit is
also familiar from the description of hard perturbative QCD processes
\`a la Brodsky-Lepage \cite{BroLe}. For instance, the electromagnetic
 form factor of the $\pi$ can be
factorized into a convolution of the DAs of the initial and final
state pion and a perturbative hard scattering amplitude, schematically
\begin{equation}
F_\pi(Q^2) \sim \phi_\pi\otimes T_H \otimes \phi_\pi\,.
\end{equation}
As in the previous subsection, the $\otimes$ stands for integration of
the longitudinal momenta of all hard partons in the process.
Note that $T_H$ is due to hard-gluon exchange between the $\pi$ constituents
and hence is $O(\alpha_s)$. In this ``hard mechanism'' only valence-quark
configurations contribute and all quarks have large longitudinal momentum.
The subtlety with heavy meson decays, however, is that the
power-counting of various contributions in $1/m_b$ differs from
that of the $\pi$ EM form factor and similar processes, and that
a second mechanism contributes at the same (leading) order in $1/m_b$.
This is the so-called soft or Feynman mechanism, where not all partons in the
meson participate in the hard subprocess; it involves highly asymmetric
configurations where the spectator quark stays soft, but the light
quark produced in the weak decay has large energy. This mechanism
is described by the $\xi$ form factors, which, as a
consequence, are also called soft form factors, and the factorization
formula for $B\to K^*$ decay form factors reads, schematically:
\begin{equation}\label{eqF}
F(q^2) = D \xi(E) + \phi_B \otimes T_H \otimes \phi_{K^*} + O(1/m_b)\,,
\end{equation}
where $D=1+O(\alpha_s)$ includes hard corrections to the weak vertex and $E$
is the energy of the $K^*$. As made explicit by the
last term on the right-hand side, the above formula is not exact, but
will receive corrections (both soft and hard) which are suppressed by
powers of $m_b$. These corrections are unknown to date.
An additional complication is that the separation between the
``soft'' contributions included in the form factor $\xi$ and the
``hard'' contributions in the convolution is not clear-cut and
requires the definition of a factorization scheme \cite{BF00}.
In Ref.~\cite{BFS04}, the factorization scheme is defined by
(\ref{3.17}) and (\ref{3.18}), absorbing the hard and hard-spectator
corrections into the definition of the 
$\xi$'s. Eqs.~(\ref{3.17}) and (\ref{3.18}) are then valid to all orders in
$\alpha_s$.
The redundancy of 5 form factors in the heavy quark limit induces a
number of relations between them. We discuss these relations, and test
their validity at finite $b$ quark mass, in App.~\ref{app:B}.

Coming back to the non-factorizable corrections to $B\to
K^*\mu^+\mu^-$ mentioned before, it turns out that they can be
included in a factorization formula very similar to that for form factors:
apart from overall factors and the Lorentz structure, the relevant
terms in the decay amplitude can be written as \cite{BFS04}
\begin{eqnarray}
\mathcal T_a^{(i)} & = & \xi_a C_a^{(i)} + \phi_B \otimes T_a^{(i)} \otimes \phi_{a,K^*} + O(1/m_b),
\end{eqnarray}
with $a=\perp,\parallel$ and $i=u,t$.
Note that the $C_a^{(i)}$ in the above formula are {\em not} Wilson
coefficients.
They do, however, contain both factorizable corrections related to the
rewriting of the full QCD form factors by the $\xi$'s using (\ref{eqF})
and non-factorizable corrections related to the matrix elements of
hadronic operators with virtual photon emission. Also note that
$\mathcal O_{7,9,10}$ only contain a 2-quark operator and hence
do not induce non-factorizable corrections.

%%%%%%%%%%%%%% Sec2.4: Differential Decay Distribution and ...

\subsection{Our Strategy}\label{2.4}

Based on the above discussion, our strategy for calculating 
$B\to K^*(\to K\pi)\mu^+\mu^-$ decays is the following:
\begin{itemize}
\item we predict observables in the dilepton mass range $1\,{\rm GeV}^2 < q^2 
< 6\,{\rm GeV}^2$;
\item we include the main source of power-suppressed corrections by using the
full QCD form factors in the na\"i\/vely factorized amplitude, and the $\xi$ 
form factors in the QCDF corrections;
\item we concentrate on the prediction of observables which are independent of
the absolute values of form factors, and only depend on their ratios;
\item for the error analysis, we employ the correlated errors between form 
factors, which follow from the light-cone sum rules;
\item we include new-physics effects in the Wilson coefficients 
$C_{7,9,10,S,P}$,
and their primed counterparts, but not in the other $C_i$.
\end{itemize}

A few comments are in order. Obviously QCDF breaks down close to the charm 
resonances. Technically, this shows up as a threshold at $q^2=4m_c^2$. In 
order to stay sufficiently below the threshold, we set $q^2_{\rm max}=
6\,{\rm GeV}^2$. On the other hand, for small $q^2$ close to the kinematical 
minimum, the decay amplitude is dominated by the photon pole and by just one 
Wilson coefficient, $C_7^\text{eff}$. Hence, 
by probing the region of very small $q^2$, one does not get any new information
as compared to the well-studied radiative decay $b\to s\gamma$. In addition,
as the photon virtuality is small, there could be (unknown) 
resonance contributions from $\rho$ or other mesons. In order to avoid this 
region, we set $q^2_{\rm min}=
1\,{\rm GeV}^2$.

As for power-suppressed corrections, we view QCDF as an expansion in two small 
parameters, $1/m_b$ and $\alpha_s$ and restrict ourselves to the first order
in these parameters. While $O(\alpha_s)$ corrections are completely covered by 
QCDF, those in $1/m_b$ are not included. One source of such corrections are 
obviously the differences between the 7 full QCD form factors and the 2 soft
form factors $\xi$. Since the light-cone sum rules allow a calculation of all
7 form factors, all these corrections are included in our calculation. An 
obvious question, though, is whether these are {\em all} $\alpha_s^0/m_b$ 
corrections or not. As we shall see in Sec.~\ref{sec:NLO}, 
one QCDF correction to na\"i\/ve factorization
is weak annihilation, where the quarks in the $B$ meson annihilate. This 
contribution is of leading order in $1/m_b$, with unknown power-suppressed
corrections. For the decay processes we consider, this contribution comes
with small Wilson coefficients, so the impact of $1/m_b$ corrections is 
negligible. Another potential source of $1/m_b$ corrections are such with
formal $\alpha_s/m_b$ counting, but an end-point divergence in the convolution
integral. Such divergent integrals were found in power-suppressed corrections
relevant for isospin violation \cite{FM02}. They signal the breakdown of QCDF
and call for soft contributions to mend the divergence. Such soft contributions
involve soft gluons and hence $\alpha_s$ is not to be evaluated at a hard 
scale, but becomes non-perturbative, thus rendering this contribution 
$O(1/m_b)$ in our counting. While it is not known how to calculate such 
contributions in the context of QCDF, similar contributions do occur 
in light-cone sum rules and are described by three-particle distribution 
amplitudes of type $\langle 0 | \bar q G s | \bar K^*\rangle$, with $G$ the 
gluonic field-strength tensor. Although these contributions could, 
{\em a priori}, be large due to soft-gluon effects, it turns out that
such three-particle couplings are numerically small \cite{DAs} and that, as
mentioned in Sec.~\ref{sec:ff}, their contribution to form factors is 
negligible. Based on this, we do not expect any sizeable effects of such terms
at $O(1/m_b)$ and conclude that the main source of power-suppressed 
corrections are those from form factors.

%%%%%%%%%%%%%% Sec3: Differential Decay Distribution and ...

\section{Differential Decay Distribution and Spin Amplitudes}\label{sec:3}
\setcounter{equation}{0}

In this section we discuss the kinematics of the 4-body decay $B\to
K^*(\to K\pi)\mu^+\mu^-$, define the angular
observables in the spectrum and derive
explicit formulas in terms of form factors and Wilson coefficients.

%%%%%%%%%%%%%% Sec3.1: Differential Decay Distribution

\subsection{Differential Decay Distribution}

The actual decay being observed in experiment is not $B\to
K^*\mu^+\mu^-$, but $B\to K^*(\to K\pi)\mu^+\mu^-$. 
As discussed in Ref.~\cite{Kruger:1999xa}, the
additional information provided by the angle between $K$ and $\pi$
is sensitive to the polarization of the $K^*$ and thus provides an
additional probe of the effective Hamiltonian. 

The matrix element of the effective Hamiltonian (\ref{eq:Heff})
for the decay $B\to K^*(\to
K\pi)\mu^+\mu^-$ can be written, in na\"\i\/ve factorization, as
\begin{equation}\label{eq:matrixelement}
\begin{split}
{\mathcal M}\ =& \frac{G_F\alpha}{\sqrt{2}\pi}V_{tb}^{}V_{ts}^*\bigg\{
\bigg[ \braket{K \pi}{\bar{s}\gamma^{\mu}({C_9^\text{eff}P_L+
C_9^{\prime\text{eff}} P_R})b}{\bar B} \\
&-\frac{2m_b}{q^2}
\braket{K\pi}{\bar{s}i\sigma^{\mu\nu}q_{\nu}(C_7^\text{eff} P_R+
C_7^{\prime\text{eff}}  P_L)b}{\bar B}\bigg]
(\bar{\mu}\gamma_{\mu}\mu)\\
&+ \braket{K\pi}{\bar{s}\gamma^{\mu}({C_{10}^\text{eff} P_L+
C_{10}^{\prime\text{eff}} P_R})b}{\bar B}(\bar{\mu}\gamma_{\mu}\gamma_5 \mu) \\
& {+\braket{K \pi}{\bar{s} ({C_S P_R+C_S^\prime P_L}) b}{\bar B}  
(\bar{\mu}\mu)} 
 {+\braket{K \pi}{\bar{s} ({C_P P_R+C_P^\prime P_L}) b}{\bar B}  
(\bar{\mu}\gamma_5\mu)} \bigg \}.
\end{split}
\end{equation}
To express the $B\to K\pi$ matrix elements in terms of the $B\to K^*$ 
form factors discussed in Sec.~\ref{sec:ff}, one assumes
that the $K^*$ decays resonantly\footnote{%
For a study of off-resonance effects, see Ref.~\cite{Grinstein:2005ud}.
}. Then, one can use a narrow-width 
approximation by making the following replacement in the squared $K^*$ 
propagator:
\begin{equation}
 \frac{1}{(k^2-m_{K^*}^2)^2+(m_{K^*}\Gamma_{K^*})^2} \xrightarrow{\Gamma_{K^*} \ll m_{K^*}} \frac{\pi}{m_{K^*}\Gamma_{K^*}} \delta(k^2-m_{K^*}^2).
\end{equation}
In this way, the form factors are independent of the $K^*K\pi$ coupling $g_{K^*K\pi}$ \cite{Kruger:1999xa,Kim:2000dq}, because it cancels between the vertex factor and the width
\begin{equation}
 \Gamma_{K^*} = \frac{g_{K^*K\pi}^2}{48 \pi} m_{K^*} \beta^3,
\end{equation}
where
\begin{equation}
 \beta = \frac{1}{m_{K^*}^2} \left[ m_{K^*}^4  + m_K^4 + m_\pi^4 - 2 (m_{K^*}^2 m_K^2+ m_K^2 m_\pi^2  + m_{K^*}^2 m_\pi^2) \right]^{1/2}.
\end{equation}

Writing the matrix elements in Sec.~\ref{sec:ff} as
\begin{equation}
\langle \bar K^*(k) | J_\mu | \bar B(p)\rangle  =  \epsilon^{*\nu} A_{\nu\mu},
\end{equation}
where $A_{\nu\mu}$ contains the $B\to  K^*$ form factors, the
corresponding $B\to K\pi$ matrix element can then be expressed as
\begin{equation}
\langle \bar K(k_1)\pi(k_2) | J_\mu | \bar B(p)\rangle  =  - D_{K^*}(k^2) \,  W^\nu A_{\nu\mu},
\end{equation}
where \cite{Kruger:1999xa}
\begin{gather}
| D_{K^*}(k^2) |^2 =  g_{K^*K\pi}^2 \frac{\pi}{m_{K^*}\Gamma_{K^*}} \delta(k^2-m_{K^*}^2) = \frac{48\pi^2}{\beta^3 m_{K^*}^2} \delta(k^2-m_{K^*}^2),
\\
 W^\mu = K^\mu-\frac{m_K^2-m_\pi^2}{k^2} \, k^\mu, \qquad k^\mu = k_1^\mu+k_2^\mu, \qquad K^\mu = k_1^\mu-k_2^\mu.
\end{gather}

With an on-shell $K^*$, the decay is completely described by four independent kinematical variables: the dilepton invariant mass squared $q^2$ and the three angles $\thK$, $\thl$ and $\phi$ as defined in App.~\ref{sec:kinematics}.
Squaring the matrix element, summing over spins of the final state particles and making use of the kinematical identities sketched in App.~\ref{sec:kinematics}, one obtains the full angular decay distribution of $\bar B^0\to \bar K^{*0}(\to K^-\pi^+)\mu^+\mu^-$:
\begin{equation}\label{eq:d4Gamma}
  \frac{d^4\Gamma}{dq^2\, d\cos\theta_l\, d\cos\theta_{K^*}\, d\phi} =
   \frac{9}{32\pi} I(q^2, \theta_l, \theta_{K^*}, \phi)\,,
\end{equation}
where
\begin{align} \label{eq:angulardist}
  I(q^2, \thl, \thK, \phi)& = 
      I_1^s \sin^2\thK + I_1^c \cos^2\thK
      + (I_2^s \sin^2\thK + I_2^c \cos^2\thK) \cos 2\thl
\nonumber \\       
    & + I_3 \sin^2\thK \sin^2\thl \cos 2\phi 
      + I_4 \sin 2\thK \sin 2\thl \cos\phi 
\nonumber \\       
    & + I_5 \sin 2\thK \sin\thl \cos\phi
\nonumber \\      
    & + (I_6^s \sin^2\thK +
      {I_6^c \cos^2\thK})  \cos\thl 
      + I_7 \sin 2\thK \sin\thl \sin\phi
\nonumber \\ 
    & + I_8 \sin 2\thK \sin 2\thl \sin\phi
      + I_9 \sin^2\thK \sin^2\thl \sin 2\phi\,.
\end{align}
The corresponding expression for the CP-conjugated mode 
$B^0\to K^{*0}(\to K^+\pi^-)\mu^+\mu^-$ is
\begin{equation}\label{eq:d4Gammabar}
  \frac{d^4 \bar{\Gamma}}{dq^2\, d\cos\theta_l\, d\cos\theta_{K^*}\, d\phi} =
   \frac{9}{32\pi} \bar{I}(q^2, \theta_l, \theta_{K^*}, \phi)\,.
\end{equation}
The function $\bar{I}(q^2, \theta_l, \theta_{K^*}, \phi)$ is obtained from (\ref{eq:angulardist}) by the replacements \cite{Kruger:1999xa}
\begin{equation}\label{eq:IIbar}
 I_{1,2,3,4,7}^{(a)} \longrightarrow \bar{I}_{1,2,3,4,7}^{(a)}\,, 
\qquad I_{5,6,8,9}^{(a)} \longrightarrow - \bar{I}_{5,6,8,9}^{(a)}\,,
\end{equation}
where $\bar{I}^{(a)}_i$ equals $I^{(a)}_i$ with all weak phases conjugated.
The minus sign in (\ref{eq:IIbar}) is a result of our convention that,
while $\theta_{K^*}$ is the angle between the $\bar K^{*0}$ and the
$K^-$ flight direction or between the $K^{*0}$ and the $K^+$,
respectively, the angle $\theta_l$ is measured between the $\bar K^{*0}$
($K^{*0}$) and the lepton $\mu^-$ in {\em both} modes. Thus, a CP
transformation interchanging lepton and antilepton leads to the
transformations $\theta_l\to\theta_l-\pi$ and $\phi\to-\phi$, as can
be seen from Eqs.~(\ref{eq:thetaYZ}) and (\ref{eq:thetaX}).
This convention agrees with
Refs.~\cite{Kruger:1999xa,Hiller08,Kruger:2005ep}, but is different
from the convention used in some experimental publications 
\cite{LHCbstudy}, where $\theta_l$ is defined as the angle between 
$K^{*0}$ and $\mu^+$ in the $B^0$ decay, but between $\bar K^{*0}$ and 
$\mu^-$ in the $\bar B^0$ decay.

The angular coefficients $I^{(a)}_i$, which are functions of $q^2$
only, are usually expressed in terms of $\bar K^*$ transversity
amplitudes. Since we want to explicitly keep lepton-mass effects and 
include also contributions from scalar and pseudoscalar operators,
this step deserves a closer look.

%%%%%%%%%%%%% Sec3.2: Transversity Amplitudes

\subsection{Transversity Amplitudes}\label{sec:3.2}

To introduce the transversity amplitudes, consider for the moment the decay $B\to K^* V^*$, with the $B$ meson decaying to an on-shell $K^*$ and a virtual photon or $Z$ boson (which can later decay into a lepton-antilepton pair). The amplitude for this process can be written as
\begin{equation} \label{eq:BVV}
\mathcal M_{(m,n)}(B\to K^* V^*) = \epsilon_{K^*}^{*\mu}(m) \, M_{\mu\nu} \,\epsilon_{V^*}^{*\nu}(n)
\end{equation}
where $\epsilon_{V^*}^\mu(n)$ is the polarization vector of the virtual gauge boson, which can be transverse ($n=\pm$), longitudinal ($n=0$) or timelike ($n=t$). In the $B$ meson rest frame, the four basis vectors can be written as \cite{Kim:2000dq,Faessler:2002ut}
\begin{align}
 \epsilon_{V^*}^\mu(\pm) & = (0,1,\mp i,0) / \sqrt{2}, \\
 \epsilon_{V^*}^\mu(0) & = (-q_z,0,0,-q_0) / \sqrt{q^2}, \\
\label{eq:epst}
 \epsilon_{V^*}^\mu(t) & = (q_0,0,0,q_z) / \sqrt{q^2},
\end{align}
where $q^\mu=(q_0,0,0,q_z)$ is the four-momentum vector of the gauge boson. They satisfy the orthonormality and completeness relations
\begin{align}
 \epsilon_{V^*}^{*\mu}(n)  \epsilon_{V^* \, \mu}(n') = g_{nn'}, \\
 \sum_{n,n'} \epsilon_{V^*}^{*\mu}(n)  \epsilon_{V^*}^\nu(n') g_{nn'} = g^{\mu\nu},
\end{align}
where $n,n'=t,\pm,0$ and $g_{nn'}=\text{diag}(+,-,-,-)$.

The $K^*$, on the other hand, is on shell and thus has only three polarization states, $\epsilon_{K^*}^\mu(m)$ with $m=\pm,0$, which read in the $B$ rest frame
\begin{align}
 \epsilon_{K^*}^\mu(\pm) & = (0,1,\pm i,0) / \sqrt{2}, \\
 \epsilon_{K^*}^\mu(0) & = (k_z,0,0,k_0) / m_{K^*},
\end{align}
where $k^\mu=(k_0,0,0,k_z)$ is the four-momentum vector of the $K^*$ (note that $k_z=-q_z$). They satisfy the relations
\begin{align}
 \epsilon_{K^*}^{*\mu}(m)  \epsilon_{K^* \, \mu}(m') = - \delta_{mm'}, \\
 \sum_{m,m'} \epsilon_{K^*}^{*\mu}(m)  \epsilon_{K^*}^\nu(m')\, \delta_{mm'} = - g^{\mu\nu} + \frac{k^\mu k^\nu}{m_{K^*}^2}.
\end{align}
The helicity amplitudes $H_0$, $H_{+}$ and $H_{-}$ can now be
projected out from $M_{\mu\nu}$ by contracting  with the explicit 
polarization vectors in (\ref{eq:BVV}),
\begin{equation}
\label{eq:H+-0}
 H_m = \mathcal M_{(m,m)}(B\to K^* V^*), \qquad m=0,+,-.
\end{equation}
 Alternatively, one can work with the transversity amplitudes defined as \cite{Kruger:2005ep}
\begin{equation}
\label{eq:Aperppara0}
 A_{\perp,\parallel} = (H_{+1} \mp H_{-1})/\sqrt{2}, ~~ A_0 \equiv H_0.
\end{equation}

In contrast to the decay of $B$ to two (on-shell) vector mesons, to which this formalism can also be applied, there is an additional transversity amplitude in the case of $B\to K^* V^*$ because the gauge boson is virtual, namely
\begin{equation}
 A_t = \mathcal M_{(0,t)}(B\to K^* V^*),
\end{equation}
which corresponds to a $K^*$ polarization vector which is longitudinal in the $K^*$ rest frame and a $V^*$ polarization vector which is timelike in the $V^*$ rest frame.\footnote{%
Unlike sometimes stated in the literature, $A_t$ does not correspond to a timelike polarization of the $K^*$ meson. As mentioned above, the $K^*$ decays on the mass shell and thus has only three polarization states.}

If we now consider the subsequent decay of the gauge boson into a lepton-antilepton pair, the amplitude becomes
\begin{equation} \label{eq:BVVll}
 \mathcal M(B\to K^* V^* (\to\mu^+\mu^-))(m) \propto \epsilon_{K^*}^{*\mu}(m) \, M_{\mu\nu} \, \sum_{n,n'} \epsilon_{V^*}^{*\nu}(n) \epsilon_{V^*}^{\rho}(n') \, g_{nn'} \,
 (\bar\mu \gamma_\rho P_{L,R} \mu).
\end{equation}
This amplitude can now be expressed in terms of six transversity amplitudes $A_{\perp,\parallel,0}^L$ and $A_{\perp,\parallel,0}^R$, where $L$ and $R$ refer to the chirality of the leptonic current,
as well as the seventh transversity amplitude $A_t$. 
The reason that for $A_t$ no separate left-handed and right-handed parts have to be considered can be seen as follows.
Noticing that the timelike polarization vector in (\ref{eq:epst}) is simply given by $\epsilon^\mu_{V^*}(t)=q^\mu/\sqrt{q^2}$, one can see from current conservation,
\begin{equation}
\label{eq:currcons}
q^\mu (\bar{\mu} \gamma^\mu \mu) = 0, \qquad
q^\mu (\bar{\mu} \gamma^\mu \gamma_5 \mu) = 2 i m_\mu (\bar{\mu} \gamma_5 \mu),
\end{equation}
that the timelike component of the $V^*$ can only couple to an axial-vector current. In addition, this shows that $A_t$ vanishes in the limit of massless leptons.

Now, having shown that the amplitude of the sequential decay 
$B\to K^* V^* (\to\mu^+\mu^-)$ can be expressed in terms of seven
transversity amplitudes, it is clear that this is true for all 
contributions of the operators $\mathcal O_7^{(\prime)}$,  $\mathcal
O_9^{(\prime)}$ and  $\mathcal O_{10}^{(\prime)}$ to the decay of
interest, $B\to K^{*}(\to K\pi)\mu^+\mu^-$, regardless of whether they
originate from virtual gauge boson exchange (i.e.\ photon or $Z$ penguin diagrams) or from box diagrams.

Does this also apply to decays mediated not by a vector, but a 
scalar and pseudoscalar operator?
Inspecting Eqs.~(\ref{eq:OS}), (\ref{eq:OP}) and (\ref{eq:currcons}),
one can see that the combination $(\mathcal O_{P}-\mathcal O_{P}')$
can be absorbed into the transversity amplitude $A_t$, because it
couples to axial-vector currents, just like the timelike component of
a virtual gauge boson. However, this is not possible for the scalar
operators $\mathcal O_{S}^{(\prime)}$. Therefore, the inclusion of
scalar operators in the decay $B\to {K}^{*}(\to K\pi)\mu^+\mu^-$ requires the introduction of a an additional, ``scalar'' transversity amplitude, which we denote $A_S$.

To summarize, the treatment of the decay ${B}\to{K}^{*}(\to K\pi)\mu^+\mu^-$ by decomposition of the amplitude into seven transversity amplitudes $A_{\perp,\parallel,0}^{L,R}$ and $A_t$ is sufficient as long as the operators $\mathcal O_{7,9,10}^{(\prime)}$ and $\mathcal O_{P}^{(\prime)}$ are considered, but has to be supplemented by an additional, eighth transversity amplitude $A_S$ once contributions from scalar operators are taken into account.

Finally, we give the explicit form of the eight transversity amplitudes (up to corrections of $O(\alpha_s)$, whose discussion we postpone until Sec.~\ref{sec:NLO}):
\begin{align}
A_{\perp L,R}  &=  N \sqrt{2} \lambda^{1/2} \bigg[ 
\left[ (C_9^\eff + C_9^{\eff\prime}) \mp (C_{10}^\eff + C_{10}^{\eff\prime}) \right] \frac{ V(q^2) }{ m_B + m_\kstar} 
 + \frac{2m_b}{q^2} (C_7^\eff + C_7^{\eff\prime}) T_1(q^2)
\bigg], \label{3.46}\\
A_{\parallel L,R}  & = - N \sqrt{2}(m_B^2 - m_\kstar^2) \bigg[ \left[ (C_9^\eff - C_9^{\eff\prime}) \mp (C_{10}^\eff - C_{10}^{\eff\prime}) \right] 
\frac{A_1(q^2)}{m_B-m_\kstar}
\nonumber\\
& \qquad +\frac{2 m_b}{q^2} (C_7^\eff - C_7^{\eff\prime}) T_2(q^2)
\bigg],\\
A_{0L,R}  &=  - \frac{N}{2 m_\kstar \sqrt{q^2}}  \bigg\{ 
 \left[ (C_9^\eff - C_9^{\eff\prime}) \mp (C_{10}^\eff - C_{10}^{\eff\prime}) \right]
\nonumber\\
 & \qquad \times 
\bigg[ (m_B^2 - m_\kstar^2 - q^2) ( m_B + m_\kstar) A_1(q^2) 
 -\lambda \frac{A_2(q^2)}{m_B + m_\kstar}
\bigg] 
\nonumber\\
& \qquad + {2 m_b}(C_7^\eff - C_7^{\eff\prime}) \bigg[
 (m_B^2 + 3 m_\kstar^2 - q^2) T_2(q^2)
-\frac{\lambda}{m_B^2 - m_\kstar^2} T_3(q^2) \bigg]
\bigg\},\label{3.48} \\
 A_t  &= \frac{N}{\sqrt{q^2}}\lambda^{1/2} \left[ 2 (C_{10}^\eff - C_{10}^{\eff\prime}) + \frac{q^2}{m_\mu} (C_{P} - C_{P}^\prime)  \right] A_0(q^2) ,\\
\label{3.50}
 A_S  &= - 2N \lambda^{1/2} (C_{S} - C_{S}^\prime)  A_0(q^2) ,
\end{align}
where
\begin{equation}
N= V_{tb}^{\vphantom{*}}V_{ts}^* \left[\frac{G_F^2 \alpha^2}{3\cdot 2^{10}\pi^5 m_B^3}
 q^2 \lambda^{1/2}
\beta_\mu \right]^{1/2},
\end{equation}
with $\lambda= m_B^4  + m_{K^*}^4 + q^4 - 2 (m_B^2 m_{K^*}^2+ m_{K^*}^2 q^2  + m_B^2 q^2)$ and $\beta_\mu=\sqrt{1-4m_\mu^2/q^2}$.

%%%%%%%%%%%%%%%%%%%%% Sec3.3: Angular Coefficients

\subsection{Angular Coefficients}\label{sec:AC}

With the eight transversity amplitudes defined in the preceding subsection, the angular coefficients $I_i$ in (\ref{eq:angulardist}) can be written as
\begin{align}
\label{eq:AC-first}
  I_1^s & = \frac{(2+\beta_\mu^2)}{4} \left[|\apeL|^2 + |\apaL|^2 + (L\to R) \right] 
            + \frac{4 m_\mu^2}{q^2} \re\left(\apeL^{}\apeR^* + \apaL^{}\apaR^*\right), 
\\
  I_1^c & =  |\azeL|^2 +|\azeR|^2  + \frac{4m_\mu^2}{q^2} 
               \left[|A_t|^2 + 2\re(\azeL^{}\azeR^*) \right] + \beta_\mu^2 |A_S|^2 ,
\\
  I_2^s & = \frac{ \beta_\mu^2}{4}\left[ |\apeL|^2+ |\apaL|^2 + (L\to R)\right],
\\
  I_2^c & = - \beta_\mu^2\left[|\azeL|^2 + (L\to R)\right],
\\
  I_3 & = \frac{1}{2}\beta_\mu^2\left[ |\apeL|^2 - |\apaL|^2  + (L\to R)\right],
\\
  I_4 & = \frac{1}{\sqrt{2}}\beta_\mu^2\left[\re (\azeL^{}\apaL^*) + (L\to R)\right],
\\
  I_5 & = \sqrt{2}\beta_\mu\left[\re(\azeL^{}\apeL^*) - (L\to R) 
- \frac{m_\mu}{\sqrt{q^2}}\, \re(\apaL {A_S^*}+\apaR {A_S^*})
\right],
\\
\label{eq:I6c}
  I_6^s  & = 2\beta_\mu\left[\re (\apaL^{}\apeL^*) - (L\to R) \right],
\\
   I_6^c  &  =
 4 \beta_\mu  \frac{m_\mu}{\sqrt{q^2}}\, \re \left[ \azeL {A_S^*} + (L\to R) \right],
\\
  I_7 & = \sqrt{2} \beta_\mu \left[\im (\azeL^{}\apaL^*) - (L\to R) 
+ \frac{m_\mu}{\sqrt{q^2}}\, {\im}(\apeL {A_S^*}+\apeR {A_S^*})
\right],
\\
  I_8 & = \frac{1}{\sqrt{2}}\beta_\mu^2\left[\im(\azeL^{}\apeL^*) + (L\to R)\right],
\\
\label{eq:AC-last}
  I_9 & = \beta_\mu^2\left[\im (\apaL^{*}\apeL) + (L\to R)\right].
\end{align}

A few comments are in order:
\begin{itemize}
\item In contrast to the transversity amplitudes themselves, the angular coefficients $I_i$ are all physical observables. In fact, they contain the complete information that can be extracted from the measurement of the decay $\bar{B}^0\to\bar{K}^{*0}(\to K^-\pi^+)\mu^+\mu^-$. We will discuss in Sec.~\ref{sec:5} which combinations of the angular coefficients constitute \textit{theoretically clean} observables.
\item In the limit of massless leptons, the well-known relations $I_1^s = 3 I_2^s$ and $I_1^c = -I_2^c$ hold.
\item The coefficient $I_6^c$ vanishes unless contributions from
  scalar operators \textit{and} lepton mass effects are taken into
  account. Therefore, to our knowledge, it has never been considered
  in the literature before. However, it is a potentially good observable for scalar currents. We will come back to this point in Sec.~\ref{sec:scalar}.
\end{itemize}

%%%%%%%%%%%%%%%%%%%% Sec3.4: Inclusion of NLO corrections

\subsection{Additional Corrections to Transversity 
Amplitudes}\label{sec:NLO}

As mentioned in Sec.~\ref{sec:3.2}, the transversity amplitudes
(\ref{3.46}) to (\ref{3.50}) do not include effects from spectator
interactions, which do induce, on the one hand, $O(\alpha_s)$ 
corrections and, on the other hand, corrections from weak
annihilation (WA). These corrections have been calculated within the
QCD factorization (QCDF) 
framework in Refs.~\cite{BFS01} and \cite{BFS04} in terms of the soft 
form factors $\xi_{\perp}$ and $\xi_{\parallel}$ discussed in 
Sec.~\ref{sec:2.3}. 

In Ref.~\cite{BFS01}, there are two types of $O(\alpha_s)$ 
corrections, factorizable and non-factorizable. The factorizable
corrections arise when expressing the full form factors in terms of 
$\xi_{\parallel}$ and $\xi_{\perp}$, are given by the radiative
corrections in Eqs.~(\ref{B.1})--(\ref{B.4}) and therefore are
redundant in our set-up. The only exception arises upon expressing the
running $b$ quark mass in the operators $\mathcal O_{7,8}$,
Eqs.~(\ref{eq:O7}) and (\ref{eq:O8}), by a mass parameter in a
different renormalization scheme. In the numerical analysis, however, we 
use the running $b$ quark mass in the $\overline{\rm MS}$ scheme,
so all factorizable $O(\alpha_s)$ 
corrections calculated in Refs.~\cite{BFS01,BFS04} have to be dropped.

The second QCDF correction to the transversity amplitude in
Sec.~\ref{sec:3.2} is given by the WA contribution,
$T^{(0)}_{\parallel,-}(u,\omega)$ in the notation of Ref.~\cite{BFS01}. It is
induced by the penguin operators $\mathcal O_3$ and $\mathcal O_4$ and
hence is numerically small, see Tab.~\ref{tab:wilson}. This is a term
which is leading in $1/m_b$ and $O(\alpha_s)$, so in
principle one should also include power-suppressed and radiative
corrections.  However, in view of its small size, we feel
justified in neglecting them. As discussed in Ref.~\cite{BFS04}, there
are further WA corrections which are suppressed by one
power of $m_b$ with respect to the leading terms. For the leading CKM
amplitude in $\lambda_t$, see Eq.~(\ref{eq:Heff}), these are again
due to penguin-annihilation diagrams and hence can be neglected due to
the smallness of the Wilson coefficients. The WA contribution to the
$\lambda_u$ amplitude vanishes for $B^0\to K^{*0}\mu^+\mu^-$, but
contains, for $B^+\to K^{*+}\mu^+\mu^-$, the factor
$C_2\approx 1$ and hence should be included for this process. 
As, in this work, however, 
we focus on neutral $B$ meson decays, we can neglect all WA
contributions except for $T^{(0)}_{\parallel,-}(u,\omega)$.

On introducing the chirality-flipped operators, the $\mathcal{T}_{\perp,\parallel}^{(t,u)}$ introduced in
Sec.~\ref{sec:2.3} are promoted to $\mathcal{T}_{\perp,\parallel}^{\pm(t,u)}$ corresponding to the notations of
Ref.~\cite{LunghiMatias}. In terms of these quantities, we can define the additional corrections to the transversity amplitudes\footnote{%
 It should be noted that the functions $F_{1,2,u}^{(7,9)}$ entering the non-factorizable corrections are defined with a different overall sign in Refs.~\cite{BFS04} and \cite{Seidel04}.
}:
\begin{align}
\Delta A^{\rm QCDF}_{\perp L,R} & = \sqrt{2} N\,
\frac{2 m_b}{q^2}\, (m_B^2-q^2) (\mathcal T_\perp^{+(t),{\rm WA+nf}}+\hat\lambda_u\mathcal T_\perp^{+(u)})\,,
\nonumber\\
\Delta A^{\rm QCDF}_{\parallel L,R} & = -\sqrt{2}N\,\frac{2m_b}{q^2}\,
(m_B^2-q^2) (\mathcal T_\perp^{-(t),{\rm WA+nf}}+\hat\lambda_u\mathcal T_\perp^{-(u)})\,,
\nonumber \\
\Delta A^{\rm QCDF}_{0 L,R}  &=  
\frac{N (m_B^2-q^2)^2}{m_{K^*}m_B^2\sqrt{q^2}}\, m_b 
(\mathcal T_\parallel^{-(t),{\rm WA+nf}}+\hat\lambda_u\mathcal T_\parallel^{-(u)})\,.
\end{align}
The superscript, WA+nf, on $\mathcal T_\perp^{\pm(t)}$ indicates that only contributions from WA and non-factorizable $O(\alpha_s)$ corrections are to be included. In accordance with Ref.~\cite{Hiller08}, we define  $\hat\lambda_u=\lambda_u/\lambda_t$.
The total transversity amplitudes are given by the expressions in
(\ref{3.46})--(\ref{3.48}) plus the above terms $\Delta A^{\rm QCDF}$.
Note there are no corrections to $A_t$ or $A_S$.

%%%%%%%%%%%%%%%%% Sec4: Testing the SM and its Extensions

\section{Testing the SM and its Extensions}\label{sec:4}
\setcounter{equation}{0}

\subsection{Preliminaries}
The multitude of observables accesssible in $B\to K^*\mu^+\mu^-$ decays
allows one to test the SM and its extensions more locally than is possible
 through global quantities like branching ratios and the dimuon mass spectrum.
The goal of this section is to describe very briefly the extensions of the SM
that we will analyse numerically in Sec.~\ref{sec:6}. To this end we
distinguish different classes of models using two properties:
\begin{itemize}
\item
the presence or absence of additional operators in the effective weak
 Hamiltonian relative to the SM ones,
\item
the presence or absence of new sources of flavour and CP violation beyond the
CKM matrix.
\end{itemize}
  When appropriate, we will also comment on the correlation between
  observables in $B\to K^*\mu^+\mu^-$ and other important observables
  in $B$ physics, such as the mass difference in the neutral $B_{d,s}$
  meson systems, $\Delta M_{d,s}$, and the time-dependent CP
  asymmetries $S$ in various decay channels. These are: $S_{\psi
    K_S}$, measured in $B_d\to J/\psi K_S$, which in the SM equals
  $\sin 2\beta$, $\beta$ being one of the angles of the unitarity
  triangle; $S_{\phi K_S}$, originating from the $b\to s\bar s s$
  penguin-decay $B_d\to \phi K_S$, which in the SM also equals
  $\sin 2\beta$, but is sensitive to new CP-violating phases in $b\to
  s$ transitions; $S_{\psi\phi}$, measured in $B_s\to J/\psi \phi$,
  given by the $B_s$ mixing phase, which in the SM is close to
  zero.\footnote{There are, however, hints from the Tevatron that this
    phase might actually be large \cite{Bsmix}.}

First, however, we would like to stress the importance of the observables in
$B\to K^* (\to K\pi) \mu^+\mu^-$ for tests of the SM.

\subsection{Standard Model}
The importance of the observables discussed in the present paper for 
tests of the SM originates from the following facts:
\begin{itemize}
\item
Several of the observables we consider are predicted to be strongly suppressed in
the SM or even vanish so that New Physics (NP) effects can be seen more easily than in 
the branching ratio of $B\to K^*\mu^+\mu^-$  which is measured to be 
 consistent with the SM expectations.
\item
The relatively small number of relevant SM parameters which are already
well constrained by a number of processes allows rather definitive 
predictions for many observables subject mainly to the theoretical uncertainty
of form factors.
\item
In certain cases the sign of a given observable
has a unique prediction in the SM, which can be tested more easily than the magnitude itself. 
\end{itemize}
However, to use these facts in a meaningful way, it is essential to have reliable
calculations of the relevant form factors.
In fact the use of the improved form factors presented in our paper and the consideration of correlations between the uncertainties of the different form factors allows one to obtain rather reliable predictions for angular coefficients in the SM.

\subsection{Models with Minimal Flavour Violation (MFV)}
\label{sec:MFV}

The simplest class of extensions of the SM are models with constrained MFV (CMFV)
in which the operators are SM-like, all  flavour violating transitions
are governed by the CKM matrix and also CP-violating observables are
SM-like \cite{Buras:2000dm,Buras:2003jf,BBGT}. 
NP effects in this class of models affect the Wilson coefficients at
the scale $O(m_W)$ and collect
NP contributions from scales higher than $m_W$. At the same time, the QCD
 renormalization group evolution down to scales lower than 
$m_W$ is universal for all CMFV models and the same as in the SM.
It can be shown that this class of models is
characterized by strong correlations between observables in $B_d$, $B_s$ and 
$K$ processes \cite{Buras:2003jf}.

The implications of these correlations and of the fact that all
flavour violating interactions are governed by the CKM matrix are striking:
deviations from the SM predictions for most weak-decay observables are
bounded to be at most $50\%$, often much less \cite{Bobeth:2005ck,HaischW07}. 
Moreover, the predictions for
CP asymmetries are generically identical to the SM ones. Consequently, the
distinction between various models of this class on the basis of global
quantities like branching ratios is very challenging.\footnote{%
However, in the case of $B\to X_s\gamma$, 
  $B\to X_s \ell^+\ell^-$ \cite{Buras:2003mk,Haisch:2007vb} and in particular in
 $B\to K^*\ell^+\ell^-$
 \cite{Colangelo:2006vm} sizable effects have been identified in the ACD model
 with one flat universal extra dimension for low compactification scales.}
We will investigate whether more local quantities like the angular observables 
considered in our paper 
could be helpful in this respect. Obviously any measurement of a 
non-standard CP-asymmetry would immediately 
signal new sources of CP violation beyond the CMFV framework.

 The symmetry-based definition of MFV \cite{D'Ambrosio:2002ex} does
 not exclude the appearance of additional non-SM operators. Compared
 to the CMFV framework, large effects can be expected in particular
 from scalar operators. The most popular model in this respect is the 
MFV MSSM which boasts an extensive literature.
In this model the CKM matrix remains the only source of flavour
and CP violation, but, in particular at large $\tan\beta$, additional
scalar and/or pseudoscalar operators,
not present in CMFV models, enter the game. For 
instance in $B_{s,d}\to\mu^+\mu^-$ decays the presence of 
such scalar or pseudoscalar operators which are induced by Higgs
 penguin diagrams can enhance the branching ratio by
one order of magnitude with respect to the SM, models with CMFV 
and the Littlest Higgs model with T-parity. 

The discovery of $B_s\to\mu^+\mu^-$ with a branching ratio of
$O(10^{-8})$ would be a clear signal of non-standard scalar or pseudoscalar 
operators. In Sec.~\ref{sec:6} we shall demonstrate that it is
precisely the angular observables in $B\to K^* (\to K\pi) \mu^+\mu^-$
that allow one to
distinguish whether scalar or pseudoscalar currents would be
responsible for such an enhancement.

\subsection{Flavour-Blind MSSM}

As a modest, but interesting modification of the MSSM with MFV we consider
the so-called Flavour Blind MSSM (FBMSSM) \cite{Baek:1998yn,Bartl:2001wc,Ellis:2007kb,ABP08}. In this
framework, the CKM matrix remains the only source of flavour violation,
while new flavour conserving, but CP-violating  phases are present in the soft
sector. To be more specific, we assume universal soft masses for the different squark flavours and flavour-diagonal trilinear couplings at the electroweak scale, allowing for complex values of the latter parameters.

One would na\"i\/vely expect that CP-violating, but flavour conserving
observables, such as electric dipole moments (EDMs), would be the best 
probes by far for CP violation in the
FBMSSM. However, 
it turns out that  large new CP-violating effects in
flavour physics can still occur at an experimentally visible level. In particular,
there is still a lot of room for CP-violating asymmetries in $B$ decays, like $A_\text{CP}(b\to s\gamma)$ and $S_{\phi K_S}$, while $S_{\psi\phi}$ is well constrained.

The dipole operators play the most relevant role in this context and,
as shown in Ref.~\cite{ABP08}, striking correlations between electric
dipole moments of neutron and electron, the $S_{\phi K_S}$ asymmetry and
$A_\text{CP}(b\to s\gamma)$ are present in this model. In particular this
framework links the explanation of the suppression of $S_{\phi K_S}$ relative
to $S_{\psi K_S}$, measured at the $B$ factories, to a large enhancement 
of $A_\text{CP}(b\to s\gamma)$ and the EDMs of the neutron and the
electron, $d_{n,e}$, over their SM values.

Therefore it is of interest to investigate  whether  the addition of 
flavour conserving, but CP-violating  phases in the soft sector would 
have a visible impact on $B\to K^* (\to K\pi) \mu^+\mu^-$ and consequently whether the FBMSSM could also be tested 
with the help of the observables discussed in the present paper. As we will
see below, very interesting and predictive results are in fact obtained
in this framework.

\subsection{Littlest Higgs Model with T--Parity (LHT)}

Another class of models of interest are those in which the operators remain
as in the SM, but new sources of both flavour and CP violation
beyond the CKM matrix are present. In this class of models the CMFV 
correlations between $B_d$, $B_s$ and $K$ observables are generally violated 
and much larger NP effects than in CMFV models are possible.

A prominent example of this class of models is the Littlest Higgs model with
T-parity \cite{Cheng:2003ju,Cheng:2004yc} in which the interactions between SM quarks and heavy mirror quarks,
mediated by new heavy charged and neutral gauge bosons, involve a new mixing
matrix that differs from the CKM matrix \cite{Hubisz:2005bd} and is 
parameterized by three new mixing angles and three new CP-violating phases \cite{LHTa}.

A number of detailed analyses of FCNC processes in the LHT model has shown that large departures
from SM predictions for FCNC processes are still possible in this model while
satisfying all existing constraints \cite{LHT1,LHT2,LHT3,Goto:2008fj}. 
In particular the CP asymmetry $S_{\psi \phi}$ can be enhanced by an order of magnitude relative to the SM
prediction \cite{LHT1,LHT08} which would be welcome if the data from
the Tevatron \cite{Bsmix} will be confirmed
by more accurate measurements at LHCb. We will investigate whether the LHT 
model can also be tested efficiently by means of the angular observables in
$B\to K^* (\to K\pi) \mu^+\mu^-$. All loop functions with mirror quarks and new heavy 
weak-boson exchanges have been calculated in Refs.~\cite{LHT1,LHT2}. A very recent
paper, Ref.~\cite{Goto:2008fj}, finds additional contributions to $Z$ penguin relative to
Ref.~\cite{LHT2}. We will investigate the importance of these terms in our analysis.

\subsection{General MSSM}

Finally we also consider the MSSM with generic flavour- and CP-violating soft SUSY-breaking terms. In such a framework one is confronted with a large number of free parameters which make it very difficult to perform global analyses.

The flavour-mixing off-diagonal entries in the squark mass matrices,
usually called mass insertions, present in this framework can lead to
complex contributions to the Wilson coefficients of all  operators in
Eqs.~(\ref{eq:O7}) to (\ref{eq:OP}). On the other hand, the mass
insertions are not completely free parameters, but are constrained  by
measurements of many FCNC processes like BR$(B \to X_s \gamma)$, BR$(B
\to X_s \mu^+ \mu^-)$, $\Delta M_s$, $\Delta M_d$, $S_{\psi K_S}$ and
others \cite{GGMS,BGH,CFMS,FOR}.
The remaining parameter space still allows sizeable effects in the
Wilson coefficients governing $B \to K^* \mu^+ \mu^-$. The general
MSSM contributions to these Wilson coefficients have been studied 
for both inclusive decays, $B \to X_s \mu^+ \mu^-$ 
\cite{LMSS99,Gabrielli:2002me}, and exclusive channels, 
$B \to K^{(*)} \mu^+ \mu^-$ \cite{LunghiMatias,Lunghi:1999za}. 
In Sec.~\ref{sec:GMSSM_pheno} we investigate the possible impact of 
these contributions on the observables discussed in the following 
Sec.~\ref{sec:5}, focusing in particular on the question of how to 
distinguish the general MSSM framework from the other models described above.

\section{Observables}\label{sec:5}
\setcounter{equation}{0}

As discussed in Sec.~\ref{sec:3}, the decay $\bar{B}^0 \to
\bar{K}^{*0} (\to K^-\pi^+)\mu^+\mu^-$ is completely described in
terms of twelve angular coefficient functions $I_i^{(a)}$. The
corresponding CP-conjugate mode $B^0 \to K^{*0}(\to K^+\pi^-)
\mu^+\mu^-$ gives access to twelve additional observables, the
CP-conjugate angular coefficient functions $\bar{I}_i^{(a)}$. These
quantities have a clear relation to both experiment and theory:
theoretically they are expressed in terms of transversity amplitudes,
and experimentally they describe the angular distribution. A physical
interpretation of these $I_i^{(a)}$ can be drawn from
Eqs.~(\ref{eq:AC-first}) to (\ref{eq:AC-last}). For example, $I_6^{c}$
depends on scalar operators and $I_7$ to $I_9$ depend on the imaginary
part of the transversity amplitudes, and consequently on their phases,
which come either from QCD effects and enter the QCD factoriation
expressions at $O(\alpha_s)$, see Sec.~\ref{sec:2}, or are
CP-violating SM or NP phases. 

To separate CP-conserving and CP-violating NP effects, we find it more convenient to consider the twelve CP averaged angular coefficients
\begin{equation}
 S^{(a)}_i = \left( I^{(a)}_i + \bar I^{(a)}_i \right) \bigg/ \frac{d(\Gamma+\bar\Gamma)}{dq^2}
\label{eq:Ss}
\end{equation}
as well as the twelve CP asymmetries\footnote{\label{fn:BHP}
Note that our definition of the CP asymmetries differs from Ref.~\cite{Hiller08} by a factor of $\frac{3}{2}$.
}
\begin{equation}
 A^{(a)}_i = \left( I^{(a)}_i - \bar I^{(a)}_i \right) \bigg/ \frac{d(\Gamma+\bar\Gamma)}{dq^2}\,.
\label{eq:As}
\end{equation}
  These are our primary observables that will be used in the
  phenomenological analysis in Sec.~\ref{sec:6}. They offer a clean
  and comprehensive way to analyse the full richness of angular
  distributions in $B\to K^*(\to
  K\pi)\mu^+\mu^-$ decays. We shall show below that all previously
  studied observables, for example the forward-backward asymmetry
  $A_{\rm FB}$, can be easily expressed in terms of our new observables.
$S^{(a)}_i$ and $ A^{(a)}_i$ are normalized to the CP-averaged dilepton mass distribution to reduce both experimental and theoretical uncertainties.
Taking the CP average means that CP-violating effects in the
$S^{(a)}_{i}$ are washed out, resulting in a cleaner
observable. Taking the CP asymmetry, on the other hand, 
means that any non-standard CP violation can be easily identified. 

These CP asymmetries, i.e.\ $A^{(a)}_{i}$, are expected to be small in 
the SM, as previously noted in Ref.~\cite{Hiller08}. This is because 
the only CP-violating phase affecting the decay enters via $\lambda_u$
in Eq.~(\ref{eq:Heff}) and is  doubly Cabibbo-suppressed.
Therefore we are particularly keen to examine these asymmetries in the 
context of CP-violating phases in NP models.

\begin{table}[t]
 \centering
\begin{tabular}{|c|c|c|}
\hline
 & $m_\mu = 0$ & $m_\mu \neq 0$ \\
\hline
SM & 18 & 22 \\
\hline
SM + $\mathcal{O}_S^{(\prime)}$ & 20 & 24 \\
\hline
\end{tabular}
\caption{\small Number of independent observables in $B \to K^{*} (\to K\pi)\mu^+\mu^-$, depending on whether lepton mass effects and/or scalar operators are taken into account.}
\label{tab:num-obs}
\end{table}
It should be stressed that out of these 24 observables, two vanish in the SM, namely $S_6^c$ and $A_6^c$, which are generated only by scalar operators, and four are related in the limit of massless leptons through  $S_1^s = 3 S_2^s$, $S_1^c = -S_2^c$ and  $A_1^s = 3 A_2^s$, $A_1^c = -A_2^c$ (see Sec.~\ref{sec:AC}). Table~\ref{tab:num-obs} summarizes the number of independent observables in these limits.

In addition, even for non-zero lepton mass, only three of the four $S_{1,2}^{s,c}$ are independent, which can be seen as follows. The dilepton mass distribution can be expressed in terms of angular coefficients as
\begin{equation}\label{eq:dGdq2}
 \frac{d\Gamma}{dq^2} = \frac{3}{4} (2 \, I_1^s + I_1^c) - \frac{1}{4} (2 \, I_2^s + I_2^c).
\end{equation}
Therefore, due to the normalization (\ref{eq:Ss}), there is the relation
\begin{equation}\label{eq:S12is1}
 \frac{3}{4} (2 \, S_1^s + S_1^c) - \frac{1}{4} (2 \, S_2^s + S_2^c)=1.
\end{equation}
Consequently, the complete set of 24 independent observables would be 
given by the twelve $A_i^{(a)}$, eleven $S_i^{(a)}$ and the
CP-averaged dilepton mass distribution $d(\Gamma+\bar\Gamma)/dq^2$. 
However, the latter is the only observable for which the normalization 
of the form factors is relevant, so theoretically it is not as clean.

In our opinion, the quantities $S^{(a)}_{i}$ and  $A^{(a)}_{i}$ are
the natural starting point for an experimental analysis. In 
Ref.~\cite{Hurth08}, a detailed investigation was carried out showing 
that a full angular fit was the preferred way to extract observables. 
This would involve fitting Eqs.~(\ref{eq:d4Gamma}) and
(\ref{eq:d4Gammabar}) to data. From such a fit the $I_i^{(a)}$ and 
$\bar{I}_i^{(a)}$  would be found directly, and could be combined 
using Eqs.~(\ref{eq:Ss}) and (\ref{eq:As}) to give the desired quantities.
We suggest that a similar full angular fit could be carried out for
the four-fold spectrum
$d^4(\Gamma\pm\bar{\Gamma)}$, so $S^{(a)}_{i}$ and  $A^{(a)}_{i}$
would be instantly accessible. Note that, due to Eq.~(\ref{eq:IIbar}), the CP-averaged decay distribution $d^4(\Gamma+\bar{\Gamma})$ gives access to $S^{(a)}_{1,2,3,4,7}$ and  $A^{(a)}_{5,6,8,9}$, while the remaining observables can be obtained from $d^4(\Gamma-\bar{\Gamma})$.

Alternatively, $S^{(a)}_{i}$ and  $A^{(a)}_{i}$ can be found by taking
asymmetries and/or integrating $d^4(\Gamma\pm\bar{\Gamma})$ over the
angles $\theta_l$, $\theta_K$ and $\phi$. Details for the extraction
of some of the $A^{(a)}_{i}$ are given in Ref.~\cite{Hiller08}, but we
stress that all our observables can be determined in a similar
manner. To illustrate this point, one case not mentioned in 
Ref.~\cite{Hiller08} is $S_5$, which can be obtained by integrating
over two angles:
\begin{equation}
S_5 = -\frac{4}{3}\left[\int_{\pi/2}^{3\pi/2} - \int_{0}^{\pi/2} - \int_{3\pi/2}^{2\pi}\right]d\phi\left[ \int_0^1 - \int_{-1}^0  \right]d\cos\theta_K \frac{d^3(\Gamma-\bar\Gamma)}{dq^2\, d\cos\theta_Kd\phi} \bigg/ \frac{d(\Gamma+\bar\Gamma)}{dq^2}.
\end{equation}

As stated above, we normalize the $S_i^{(a)}$ and $A_i^{(a)}$ 
to the CP-averaged
dilepton mass distribution in order to reduce the dependence on the form
factors. Our approach described in Secs.~\ref{sec:3.2} and
\ref{sec:NLO} makes use of the full form factors for the dominant
leading-oder 
contribution and the soft form factors for additional suppressed
contributions. Therefore our results are largely independent of the
relation between the soft form factors and the full form
factors. However, to further our understanding of these soft form
factor relations, we investigate them and their $q^2$ dependence in
App.~\ref{app:B}. It is found that relations involving $\xi_\perp$ are
almost independent of $q^2$, whereas those involving $\xi_\parallel$
have a considerable dependence on $q^2$ due to the neglected $1/m_b$
terms. Therefore we stress that the transversity amplitudes
$A_{\perp,\parallel}^{L,R}$ of Sec.~\ref{sec:3.2}, and all angular
observables built from them, should be more or less insensitive to
$1/m_b$ corrections, i.e.\ corrections to QCDF, while $A_0^{L,R}$ and
all corresponding angular variables will be slightly more affected by
such corrections. These findings impact on prior work carried out in this channel, where the transversity amplitudes were given entirely in terms of the soft form factors using QCDF.

 All established observables can be expressed in terms of $S^{(a)}_{i}$ and  $A^{(a)}_{i}$. For example, the CP asymmetry in the dilepton mass distribution is given by (see Eq.~(\ref{eq:S12is1}))
\begin{equation}
 A_\text{CP} = \frac{d(\Gamma-\bar\Gamma)}{dq^2}\bigg/ \frac{d(\Gamma+\bar\Gamma)}{dq^2} = \frac{3}{4} (2 \, A_1^s + A_1^c) - \frac{1}{4} (2 \, A_2^s + A_2^c).
\end{equation}

We prefer to define the normalized forward-backward asymmetry as a
ratio of CP-averaged quantities, to wit
\begin{equation}
 A_\text{FB} = \left[ \int_0^1 - \int_{-1}^0 \right]d\cos\theta_l \frac{d^2(\Gamma-\bar\Gamma)}{dq^2\, d\cos\theta_l} \bigg/ \frac{d(\Gamma+\bar\Gamma)}{dq^2}
 = \frac{3}{8} (2 \, S_6^s + S_6^c).
\label{eq:AFB}
\end{equation}
The CP average is numerically irrelevant in the SM, but makes the connection to experiment more transparent. In addition, this definition is complementary to the forward-backward CP asymmetry \cite{Buchalla:2000sk},
\begin{equation}
 A_\text{FB}^\text{CP} = \left[ \int_0^1 - \int_{-1}^0 \right]d\cos\theta_l \frac{d^2(\Gamma+\bar\Gamma)}{dq^2\, d\cos\theta_l} \bigg/ \frac{d(\Gamma+\bar\Gamma)}{dq^2}
 = \frac{3}{8} (2 \, A_6^s + A_6^c).
\end{equation}

Additional well-established observables are the $K^*$ longitudinal and transverse polarization fractions $F_L$, $F_T$, which are usually defined in terms of transversity amplitudes. We prefer to directly express them in terms of CP-averaged observables and {\em define}
\begin{equation}\label{eq:FLFT}
 F_L = -S_2^c, \qquad F_T = 4 S_2^s.
\end{equation}
The well-known relation $F_T=1-F_L$ is then a consequence of
Eq.~(\ref{eq:S12is1}) in the limit of vanishing lepton mass.

In Refs.~\cite{Kruger:2005ep,Hurth08}, the transverse asymmetries
$A_T^{(i)}$ were introduced. They can be expressed in terms of our
observables as 
\begin{eqnarray}
 A_T^{(2)} & = & \frac{S_3}{2 \, S_2^s}\,,\nonumber\\
 A_T^{(3)} & = & 
\left(\frac{4 \, S_4^2 + S_7^2}{-2 \, S_2^c \, (2 \, S_2^s+S_3)} \right)^{1/2},
\nonumber\\
 A_T^{(4)} & = &
\left( \frac{S_5^2 + 4 \, S_8^2}{4 \, S_4^2 + S_7^2} \right)^{1/2}.
\label{eq:AT2}
\end{eqnarray}

Finally, for some observables it is useful to consider their $q^2$ average. We define
\begin{equation}\label{eq:intS}
\left\langle S_i^{(a)}\right\rangle= \int_{1\,\text{GeV}^2}^{6\,\text{GeV}^2}dq^2\left(
I^{(a)}_i + \bar I^{(a)}_i \right)\bigg/\int_{1\,\text{GeV}^2}^{6\,\text{GeV}^2}dq^2\frac{d(\Gamma+\bar{\Gamma})}{dq^2}\,,
\end{equation}
\begin{equation}\label{eq:intA}
\left\langle A_i^{(a)}\right\rangle= \int_{1\,\text{GeV}^2}^{6\,\text{GeV}^2}dq^2\left(
I^{(a)}_i - \bar I^{(a)}_i \right)\bigg/\int_{1\,\text{GeV}^2}^{6\,\text{GeV}^2}dq^2\frac{d(\Gamma+\bar{\Gamma})}{dq^2}\,.
\end{equation}
The reasons for choosing the interval $1\,{\rm GeV}^2\leq q^2\leq 6\,{\rm
  GeV}^2$ are discussed in Sec.~\ref{2.4}.

We proceed in the next section by studying the predictions for 
$S^{(a)}_{i}$ and  $A^{(a)}_{i}$, keeping in mind the sensitivity to 
hadronic effects. This is carried out first in the SM and later in the  
various NP models described in Sec.~\ref{sec:4}.

%%%%%%%%%%%%% Sec6: Phenomenological Analysis

\section{Phenomenological Analysis}\label{sec:6}
\setcounter{equation}{0}

We are now in a position to perform a phenomenological analysis of
the observables defined in Sec.~\ref{sec:5}, first in the SM, then in
a model-independent manner, and finally for specific NP scenarios.

\subsection{Standard Model}\label{sec:SM-pheno}

\begin{figure}
\centering
\includegraphics[width=0.97\textwidth]{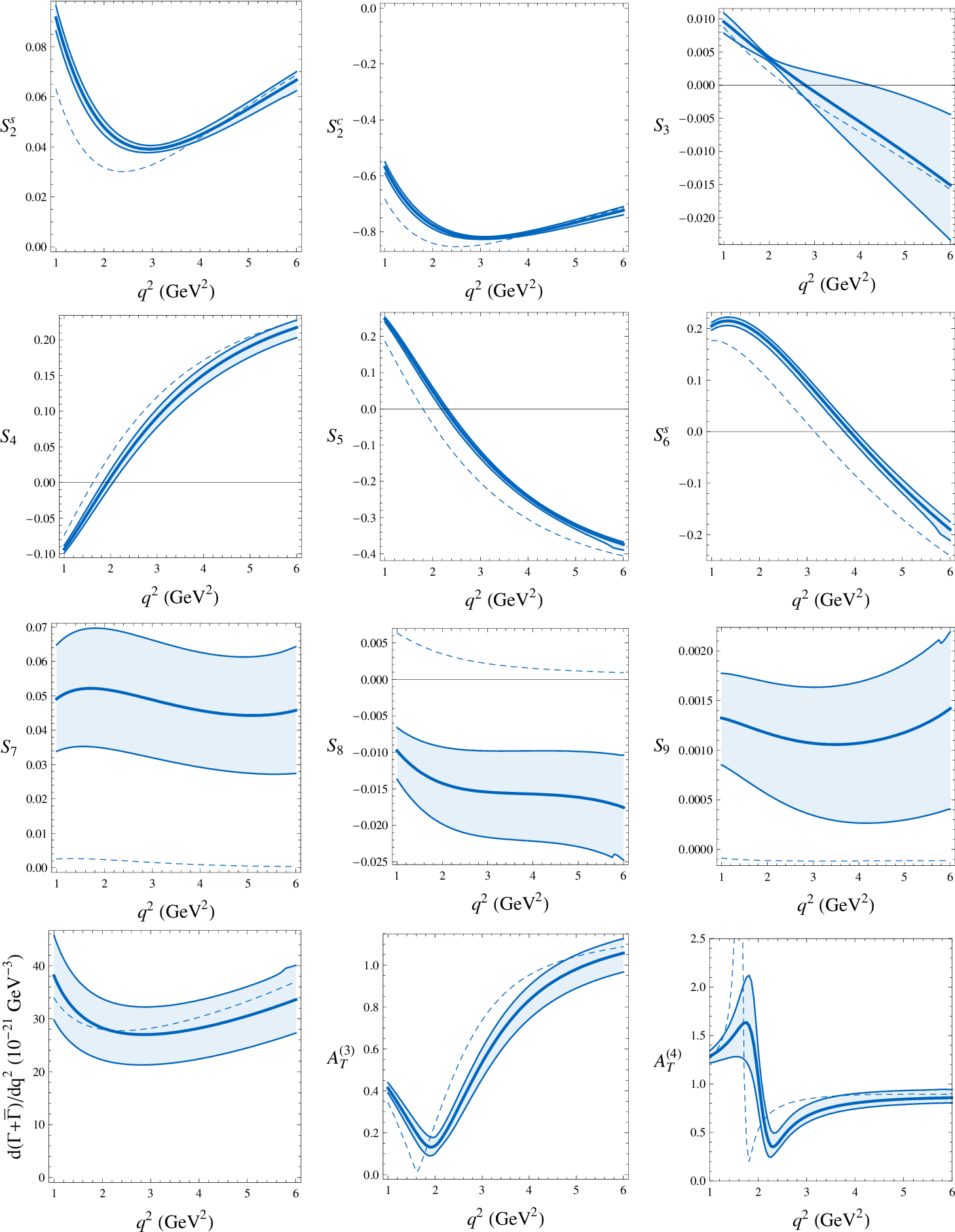}
\caption[]{\small CP-averaged angular coefficients $S_i^{(a)}$, CP-averaged dilepton mass distribution $d(\Gamma+\bar\Gamma)/dq^2$ and transverse asymmetries $A_T^{(3,4)}$ in the SM as a function of $q^2$. The dashed lines are the
  leading-order (LO) contributions, obtained in na\"\i\/ve
  factoriation. The thick solid lines are the full next-to-leading
  order (NLO) predictions from QCD factorization (QCDF), as described
  in Sec.~\ref{2.4}. The blue band defines the total error for
  the NLO result as described in the text.
}\label{fig:Ss-SM}
\end{figure}

Our predictions for the CP-averaged angular coefficients $S_i^{(a)}$ in the SM are shown in Fig.~\ref{fig:Ss-SM}.
 $S_1^{s}$ and $S_1^{c}$ have been omitted since the relations $S_1^s = 3 S_2^s$ and $S_1^c = -S_2^c$ (see Sec.~\ref{sec:AC}) are fulfilled up to lepton-mass effects, which amount to at most 1\%.
$S_{1,2}^{s,c}$ are numerically large as expected.
$S_{4}$, $S_{5}$, $S_{6}^{s}$ are similar in magnitude, but are
 particularly interesting as they each have a zero in $q^2$. All these
 predictions are seen to have small uncertainties, as the
 normalization results in a cancellation of hadronic effects. In
 Tab.~\ref{tab:zeros}, we show our predictions for the positions of
 the zeros of $S_{4}$, $S_{5}$ and $S_{6}^{s}$, denoted by
 $q_0^2(S_i^{(a)})$ from now on.
$S_3$ is numerically small in the SM since it is approximately proportional to the chirality-flipped Wilson coefficient $C_7'$, which is suppressed by a factor $m_s/m_b$.
$S_{7}$, $S_{8}$ and $S_{9}$ are small as well and have a larger error-band as they arise from the imaginary part of the transversity amplitudes.

The error bands have been obtained by adding various uncertainties in quadrature. We estimate the uncertainty due to the form factors by varying the Borel parameter and continuum threshold as discussed in Sec.~\ref{sec:ff}. The renormalization-scale uncertainty is found by varying $\mu$ between $4.0$ and $5.6\;$GeV, where $\mu$ is the scale at which the Wilson coefficients, $\alpha_s$ and the $\overline{\text{MS}}$ masses are evaluated. We also include parametric uncertainties which are estimated by varying the hadronic parameters as indicated in Tab.~\ref{tab:numinput}, the ratio $m_c/m_b$ between $0.25$ and $0.33$, and the CKM angle $\gamma$, which is particularly important for the doubly Cabibbo-suppressed contribution to the CP asymmetries, between $60^\circ$ and $80^\circ$.\footnote{%
The discontinuity in some of the error bands just below
$6\;\text{GeV}^2$ is an unphysical artifact resulting from small charm
quark masses $\sim 1.2\;\text{GeV}$ allowed in the estimation of the
error. This feature was already observed in Ref.~\cite{BFS01}.
}
In addition, we show the leading-order prediction as a dashed line. We
find that the impact of radiative QCDF corrections is moderate for
observables like $S_{2,3,4,5,6}$ that, in the SM, are largely
independent of weak or strong phases, but becomes more prominent for
observables built from imaginary parts, like $S_{7,8,9}$ and $A_i$,
where the main contribution comes from strong phases induced by
$O(\alpha_s)$ corrections in QCDF.
\begin{figure}
\centering
\includegraphics[width=\textwidth]{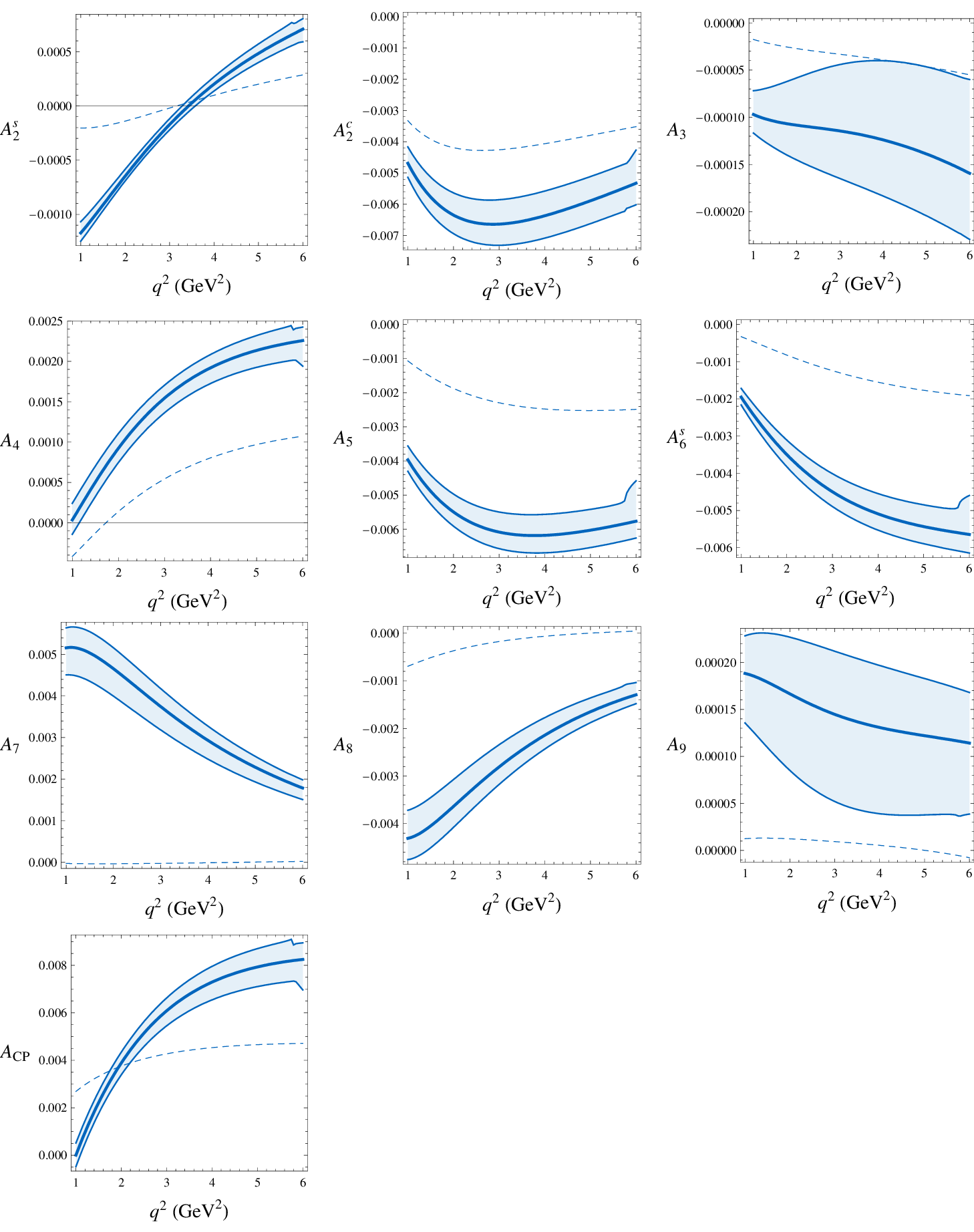}
\caption[]{\small CP asymmetries $A_i^{(a)}$ and $A_\text{CP}$ in the
  SM as a function of $q^2$. The meaning of the curves and
  bands is as in Fig.~\ref{fig:Ss-SM}.}\label{fig:As-SM}
\end{figure}

Some of these $S_i^{(a)}$ can be directly compared to previous results in the literature. $S_2^{s}$ and $S_2^{c}$ correspond to the $K^*$ longitudinal and transverse polarization fractions $F_L$ and $F_T$, see Eq.~(\ref{eq:FLFT}), and $S_6^s$ yields the forward-backward asymmetry $A_\text{FB}$, see Eq.~(\ref{eq:AFB}).
In particular, $q_0^2(S_6^s)$ in Tab.~\ref{tab:zeros} is identical to the zero of the forward-backward asymmetry which has been extensively studied in the literature.
For completeness, in the last row of Fig.~\ref{fig:Ss-SM} we also show
the CP averaged dilepton mass distribution $d(\Gamma+\bar\Gamma)/dq^2$
and the observables $A_T^{(3)}$ and $A_T^{(4)}$ defined in
Ref.~\cite{Hurth08}, see Sec.~\ref{sec:5}. We find that our results
for all these observables compare well to those in the
literature. However, we note that the peak in the plot of $A_T^{(4)}$
is a manifestation of the zero $q_0^2(S_4)$ of $S_{4}$, see
Eq.~(\ref{eq:AT2}). This division by a near-zero quantity induces a
large theoretical uncertainty both in the position of the peak and
its height. We stress that such uncertainties do not arise if the
observables $S_4$ and $S_5$ are considered instead of $A_T^{(3)}$ and 
$A_T^{(4)}$. In fact, as $d\Gamma/dq^2$ is a smooth function in the
range of $q^2$ considered, none of our observables $S_i$ and $A_i$ is
affected by accidental and delicate cancellations in the denominator.

\begin{table}[b]
 \centering
\renewcommand{\arraystretch}{1.4}
\begin{tabular}{|l||c|c|c|}
\hline
Obs. & $S_4$ & $S_5$ & $S_6^s$ \\
\hline
$q^2_0\: [\text{GeV}^2]$ & $1.94^{+0.12}_{-0.10}$  & $2.24^{+0.06}_{-0.08}$  & $3.90^{+0.11}_{-0.12}$ \\
\hline
\end{tabular}
\renewcommand{\arraystretch}{1}
\caption{\small Predictions for the zero positions $q_0^2(S_i^{(a)})$ 
of $S_4$, $S_5$ and $S_6^s$ in the SM.}
\label{tab:zeros}
\end{table}

As explained in Sec.~\ref{sec:5}, the CP asymmetries are close to zero 
in the SM, which is evident from Fig.~\ref{fig:As-SM}, where we show
all the $A_i^{(a)}$ (again except for $A_1^{s,c}$) and the CP
asymmetry in the decay distribution, $A_\text{CP}$. As explained
above, the shift from LO to NLO is substantial. Our results are in 
good agreement with Ref.~\cite{Hiller08}, but do not coincide exactly.  
This can be understood by recalling that we use the full LCSR form
factors and that our  normalization of the soft form factors,
especially $\xi_\parallel$, is different from that used in 
Ref.~\cite{Hiller08}. Also our choice of quark masses, in particular 
$m_c/m_b$ and $m_t$, as well as the scale $\mu$ at which the QCDF 
hard-scattering corrections are evaluated, differs from
\cite{Hiller08}. We stress that, in view of the smallness of the SM
values of $A_i$, these discrepancies become irrelevant once large NP
contributions start to dominate these observables, as we shall see in
the remainder of this section.

In Tab.~\ref{tab:A-int}, 
we list our predictions for the $q^2$-integrated CP-averaged angular coefficients and CP asymmetries as
defined in Eqs. (\ref{eq:intS}) and (\ref{eq:intA}).
$\langle S_2^c \rangle$, $\langle S_6^s \rangle$ and $\langle A_\text{CP} \rangle$ can be directly compared to existing experimental results from BaBar and Belle \cite{BaBar_angular,Wei:2009zv}.

\begin{table}[tb]
 \centering
\renewcommand{\arraystretch}{1.4}
\begin{tabular}{|c|c||c|c|}
\hline
Obs. & $10^{-2} \times \ldots $ &Obs. & $10^{-2} \times \ldots $ \\
\hline\hline
$\langle S_1^s \rangle$ & $16.0^{+0.6}_{-0.6}$ & $\langle S_5 \rangle$ & $-14.2^{+0.8}_{-1.2}$ \\

\hline
$\langle S_1^c \rangle$ & $79.3^{+0.8}_{-0.8}$ & $\langle S_6^s \rangle$ & $3.5^{+0.8}_{-1.1}$ \\

\hline
$\langle S_2^s \rangle$ & $5.3^{+0.2}_{-0.2}$ & $\langle S_7 \rangle$ & $4.8^{+1.7}_{-1.7}$ \\

\hline
$\langle S_2^c \rangle$ & $-76.6^{+0.7}_{-0.7}$ & $\langle S_8 \rangle$ & $-1.5^{+0.6}_{-0.6}$ \\

\hline
$\langle S_3 \rangle$ & $-0.3^{+0.4}_{-0.3}$ & $\langle S_9 \rangle$ & $0.1^{+0.1}_{-0.1}$ \\

\hline
$\langle S_4 \rangle$ & $10.1^{+1.0}_{-1.2}$ & & \\
\hline
\end{tabular}
\qquad
\begin{tabular}{|c|c||c|c|}
\hline
Obs. & $10^{-3} \times \ldots $ &Obs. & $10^{-3} \times \ldots $ \\
\hline\hline
$\langle A_1^s \rangle$ & $-0.2^{+0.2}_{-0.1}$ & $\langle A_5 \rangle$ & $-5.7^{+0.6}_{-0.5}$ \\

\hline
$\langle A_1^c \rangle$ & $6.3^{+0.7}_{-0.8}$ & $\langle A_6^s \rangle$ & $-4.5^{+0.5}_{-0.4}$ \\

\hline
$\langle A_2^s \rangle$ & $-0.1^{+0.1}_{-0.0}$ & $\langle A_7 \rangle$ & $3.4^{+0.4}_{-0.5}$ \\

\hline
$\langle A_2^c \rangle$ & $-6.1^{+0.7}_{-0.6}$ & $\langle A_8 \rangle$ & $-2.6^{+0.4}_{-0.3}$ \\

\hline
$\langle A_3 \rangle$ & $-0.1^{+0.1}_{-0.1}$ & $\langle A_9 \rangle$ & $0.1^{+0.1}_{-0.1}$ \\

\hline
$\langle A_4 \rangle$ & $1.5^{+0.2}_{-0.2}$ & $\langle A_\text{CP} \rangle$ & $5.9^{+0.6}_{-0.6}$ \\
\hline
\end{tabular}
\renewcommand{\arraystretch}{1}
\caption{\small Predictions for the
integrated CP-averaged angular coefficients $\langle S_i^{(a)} \rangle$ (in units of $10^{-2}$)
and the integrated CP asymmetries $\langle A_i^{(a)} \rangle$ (in units of $10^{-3}$)
in the SM. Note the different normalization of the $\langle A_i^{(a)} \rangle$ with respect to Ref.~\cite{Hiller08}, see footnote~\ref{fn:BHP}.
}
\label{tab:A-int}
\end{table}

\subsection{Model-independent Considerations}

Before turning to specific NP scenarios, we investigate the 
model-independent impact of the Wilson coefficients on our observables.
\begin{table}[tb]
\addtolength{\arraycolsep}{3pt}
\renewcommand{\arraystretch}{1.3}
\centering
\begin{tabular}{|l|l|}
\hline
Wilson coefficients & largest effect in \\
\hline\hline
$C_7$, $C_7^\prime$ & $S_1^s$, $S_1^c$, $S_2^s$, $S_2^c$, $S_3$, $S_4$, $S_5$, $S_6^s$,\\
                    & $A_7$, $A_8$, $A_9$,\\
                    & BR$(B \to X_s \gamma)$, BR$(B\to X_s \mu^+ \mu^-)$\\
\hline
$C_9$, $C_9^\prime$, $C_{10}$, $C_{10}^\prime$ & $S_1^s$, $S_1^c$, $S_2^s$, $S_2^c$, $S_3$, $S_4$, $S_5$, $S_6^s$,\\
                                          & $A_7$, $A_8$, $A_9$,\\
                                          & BR$(B\to X_s \mu^+ \mu^-)$\\
\hline
$C_S - C_S^\prime$ & $S_6^c$,\\
                   & BR$(B_s \to \mu^+ \mu^-)$\\
\hline
$C_P - C_P^\prime$ & $S_1^c + S_2^c$,\\
                   & BR$(B_s \to \mu^+ \mu^-)$\\
\hline
\end{tabular}
\caption{\small The Wilson coefficients relevant in $B \to K^* \mu^+ \mu^-$ and the observables they have the largest impact on.}
\label{tab:WC_vs_Obs}
\end{table}

\subsubsection{Impact of Wilson Coefficients on Observables}
\label{sec:impact}

The impact of NP on the angular observables discussed
in our paper is given by the changes of the Wilson coefficients
of the affected operators. One can group these Wilson 
coefficients into three classes:
\begin{itemize}
\item
Dipole coefficients: $C_7$, $C_7^\prime$, $C_8$ and 
$C_8^\prime$. The role of the gluon dipole
operators is subleading in the decay considered.
\item
Semileptonic coefficients: $C_9$, $C_9^\prime$, $C_{10}$ and 
$C_{10}^\prime$.
\item
Scalar coefficients: $C_S-C_S^\prime$ and $C_P-C_P^\prime$.
\end{itemize}
Before entering the discussion of various NP scenarios, it is useful
to study the correlation between the angular coefficients and  
the Wilson coefficients. In Tab.~\ref{tab:WC_vs_Obs} we show which
observables are most affected by a significant change of a given 
coefficient. In Tab.~\ref{tab:Obs_vs_WC} we
show, on the other hand, which  Wilson  coefficients should be altered to produce a large effect in specific observables. 

\begin{table}[tb]
\addtolength{\arraycolsep}{3pt}
\renewcommand{\arraystretch}{1.3}
\centering
\begin{tabular}{|l|l|}
\hline
Observable & mostly affected by \\
\hline\hline
$S_1^s$, $S_1^c$, $S_2^s$, $S_2^c$ & $C_7$, $C_7^\prime$, $C_9$, $C_9^\prime$, 
$C_{10}$, $C_{10}^\prime$\\
\hline
$S_3$ & $C_7^\prime$, $C_9^\prime$, $C_{10}^\prime$\\
\hline
$S_4$ & $C_7$, $C_7^\prime$, $C_{10}$, $C_{10}^\prime$\\
\hline
$S_5$ & $C_7$, $C_7^\prime$, $C_9$, $C_{10}^\prime$\\
\hline
$S_6^s$ & $C_7$, $C_9$\\
\hline
$A_7$ & $C_7$, $C_7^\prime$, $C_{10}$, $C_{10}^\prime$\\
\hline
$A_8$ & $C_7$, $C_7^\prime$, $C_9$, $C_9^\prime$, $C_{10}^\prime$\\
\hline
$A_9$ & $C_7^\prime$, $C_9^\prime$, $C_{10}^\prime$\\
\hline
$S_6^c$ & $C_S - C_S^\prime$\\
\hline
\end{tabular}
\caption{\small The most interesting angular observables in $B \to K^*
  \mu^+ \mu^-$ and the Wilson coefficients they are most sensitive to.}
\label{tab:Obs_vs_WC}
\end{table}

We observe:
\begin{itemize}
\item
$C_7$, $C_7^\prime$, $C_9$, $C_9^\prime$, $C_{10}$ and $C_{10}^\prime$ 
can induce large effects in many observables, or at least in those that do not require the presence of strong phases. To be precise, the $ A_i$ are 
mainly induced by imaginary parts of the Wilson coefficients, while the $S_i$ are 
induced by their real parts.
\item
Only the primed coefficients $C_7^\prime$, $C_9^\prime$ and $C_{10}^\prime$
can significantly affect the observables $S_3$ and $A_9$. As can be seen from Eq.~(\ref{eq:AT2}), $S_3$ corresponds to the transverse asymmetry $A_T^{(2)}$ and the impact of NP physics contributions to $C_7^\prime$ on this observable has been studied for example in Refs.~\cite{Hurth08,Kruger:2005ep,LunghiMatias}.
\item
The scalar operators affect mainly $S_6^c$ and the branching ratio
for $B_s\to \mu^+\mu^-$. This implies interesting correlations
between these two observables as discussed in Sec.~\ref{sec:scalar}.
\end{itemize}

\subsubsection{Model-independent Analysis of \boldmath$S_4$, $S_5$ and $S_6^s$\unboldmath}
\label{sec:zeros_model_independent}

The zero of the forward-backward asymmetry has been the focus of many
experimental and theoretical studies (see for example
Refs.~\cite{LHCbstudy,FM02}) as it is established as being an
observable free from hadronic effects and capable of distinguishing
between NP scenarios. In Sec.~\ref{sec:5} we expressed the CP-averaged
forward-backward asymmetry in terms of $S_6^s$ through
Eq.~(\ref{eq:AFB}), so $S_6^s$ could clearly be studied instead of
$A_\text{FB}$. In addition, from Fig.~\ref{fig:Ss-SM}, we find there
are two more observables with such a zero in $q^2$, $S_4$ and $S_5$. A study of these three observables in a model-independent way could allow us to constrain the NP contributions to the Wilson coefficients.
\begin{figure}[tb]
\centering
\begin{tabular}{cc}
\includegraphics[width=0.47\textwidth]{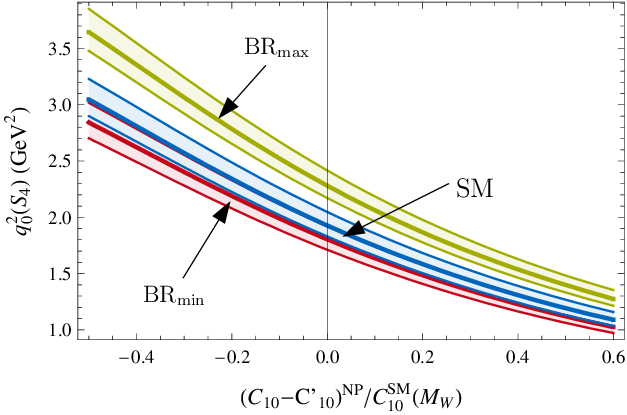} &
\includegraphics[width=0.47\textwidth]{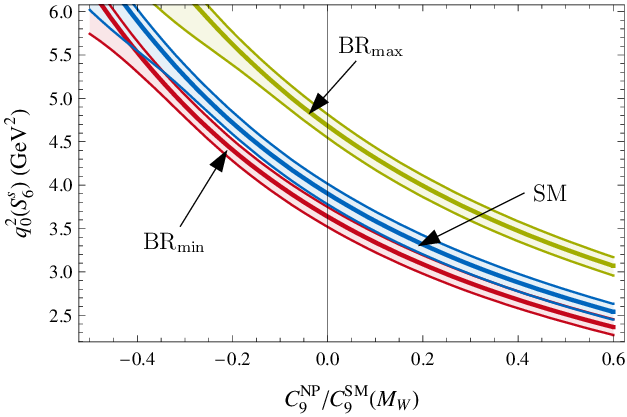} 
\end{tabular}
\vskip-10pt
\caption[]{\small Left: correlation between $q_0^2(S_4)$, the position of
  the zero of $S_4$, and the NP contribution to
  $C_{10}-C_{10}^\prime$. Right: correlation between $q_0^2(S_6^s)$
  and the NP contribution to $C_9$. We use the branching ratio for $B \to X_s\gamma$ to 
constrain the NP contributions to $C_{7}$ and $C_{7}^\prime$.
The green (red) band corresponds to a value of $\text{BR}(B\to X_s\gamma)$ at 
the upper (lower) end of the experimental $2\sigma$ range, the blue band to 
SM values for $C_{7}$, $C_{7}^\prime$.
\label{fig:7}}
\end{figure}
\begin{figure}[tb]
\centering
\includegraphics[width=\textwidth]{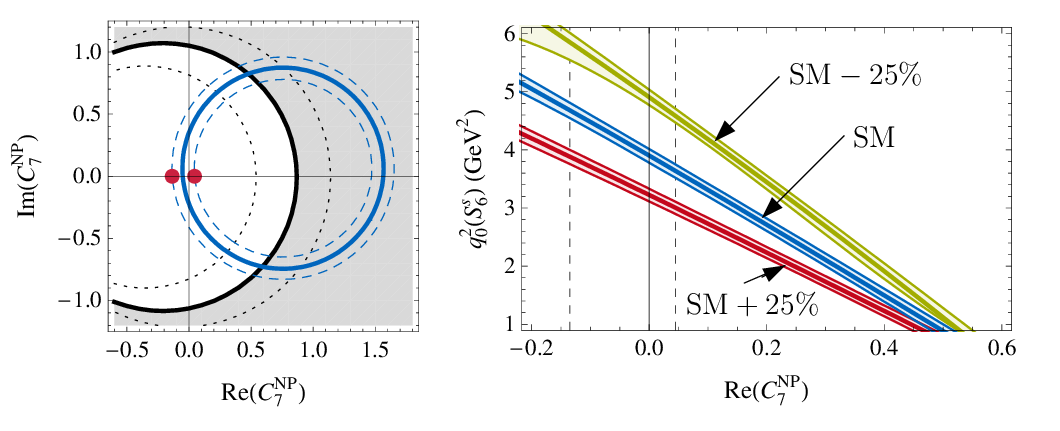}
\caption[]{\small
Left: Experimental constraints on the NP contribution to $C_7$. The blue circles show the constraint from the central and $\pm2\sigma$ values of  $\text{BR}(B\to X_s\gamma)$, assuming $C_7^{\prime\text{NP}}=0$. The black circle corresponds to the $2\sigma$ bound from $\text{BR}(B\to X_s\ell^+\ell^-)$, assuming $C_{10}^{(\prime)\text{NP}}=0$. The solid thick and the dotted lines have been obtained assuming SM and SM$\pm25\%$ values for $C_9$, respectively.
Right: Correlation of the zero in 
 $S_6^s$ with the NP contribution to $\text{Re}(C_7)$. The 
blue, red and green bands indicate SM, SM$+25\%$ and SM$-25\%$ 
values for $C_9$ with the associated theoretical uncertainty.
The vertical dashed lines correspond to the upper and lower bounds on $\text{Re}(C_7)$ in the absence of an imaginary part of $C_7$. (The corresponding points in the left-hand plot are highlighted by red dots.) For an arbitrary imaginary part, the upper bound on $\text{Re}(C_7)$ is removed, and $q_0^2(S_6^s)$ can be at or below $1\,\text{GeV}^2$.
\label{fig:8}}
\end{figure}

From Tab.~\ref{tab:Obs_vs_WC} we see that the zero of $S_4$,
$q_0^2(S_4)$, is largely sensitive to $C_7$, $C_7^\prime$, $C_{10}$
and $C_{10}^\prime$. This dependence arises only through $C_7-C_7^\prime$
and $C_{10}-C_{10}^\prime$. We therefore explore how the position of the zero in $q^2$ is affected by NP modifications to $C_{10}-C_{10}^\prime$ and $C_7$. The current experimental value of the branching ratio of $B\to X_s\gamma$ provides a constraint on $C_7$ and $C_7^\prime$. We find a strong dependence of $q_0^2(S_4)$ on $C_{10}-C_{10}^\prime$, and its measurement would provide very interesting information about these Wilson coefficients. In Fig.~\ref{fig:7}, we show this dependence for real values of $C_7$.

% If the NP introduces an imaginary phase to $C_7$, the bound from $B\to
% X_s\gamma$ is weakened, allowing large effects in the zeros.
% %  which we show in Fig.~\ref{fig:8}.
% In fact, the large values of $\im(C_7)$
% significantly enhance the branching ratio of the decay $B \to X_s
% \gamma$ and in order to be in agreement with the experimental data,
% large positive contributions to $\re(C_7)$ that interfere
% destructively with $C_7^{\rm SM}$ are required\footnote{We note that,
%   for such values of the Wilson coefficients, the branching ratio of
%   the decay $B \to X_s \mu^+ \mu^-$ is largely enhanced to values of
%   around BR$(B \to X_s \mu^+ \mu^-) \simeq 2.5 \times 10^{-6}$, close
%   to the experimental $2\sigma$ upper bound.}. Exactly these large positive contributions to $\re(C_7)$ then unambiguously shift the zeros of $S_4$, $S_5$ and $S_6^s$ towards lower values.

If the NP introduces an imaginary part to $C_7$, the bound from $B\to
X_s\gamma$ is weakened, allowing large effects in the zeros.
In fact, large values of $\im(C_7)$
significantly enhance the branching ratio of the decay $B \to X_s \gamma$
and in order to be in agreement with the experimental data,
large positive contributions to $\re(C_7)$ that interfere
destructively with $C_7^{\rm SM}$ are required.
For such values of the Wilson coefficients, the branching ratio of
  the decay $B \to X_s \mu^+ \mu^-$ is largely enhanced, effectively setting a new upper
 bound on $\re(C_7)$. In the left-hand plot in Fig.~\ref{fig:8}, we show these combined
constraints on $C_7$ in the complex plane.
Exactly the large positive contributions to $\re(C_7)$, which are allowed in the presence of phases in $C_7^\text{NP}$, then unambiguously shift the zeros of $S_4$, $S_5$ and $S_6^s$ towards lower values.
 In the right-hand plot in Fig.~\ref{fig:8}, we show as an example that the allowed range for $q_0^2(S_6^s)$ is greatly enhanced in the case of complex $C_7$.

This analysis can also be applied to $S_6^s$, which depends strongly
on $C_7$ and $C_9$. We examine the dependence of $q^2_0(S_6^s)$ on NP
contributions to $C_{9}$ and $C_7$. This again is restricted by the
experimental value of the branching ratio of $B\to X_s\gamma$. We find
a strong dependence on $C_9$, and for real values of $C_7$ this would
be a clean way to determine infomation about a possible NP
contribution to $C_9$ as seen in Fig \ref{fig:7}. Again, if NP induces
a complex phase of $C_7$, the range in $q_0^2(S_6^s)$ increases dramatically. 

It is a greater challenge to extract information about the Wilson
coefficients from $S_5$ due to its dependence on $C_7$, $C_7^\prime$,
$C_9$ and $C_{10}^\prime$. However, a measurement of $q_0^2(S_5)$
could provide a consistency check with $C_{10}-C_{10}^\prime$ and $C_9$ determined from $S_4$ and $S_6$, provided $C_7$, $C_7^\prime$ are real.
In addition, this might allow one to untangle the effects of $C^{\rm NP}_{10}$ and $C_{10}^{\prime \rm NP}$ in Fig.~\ref{fig:7}.

\subsubsection{Impact of Scalar Currents}\label{sec:scalar}

As mentioned in the introduction, the impact of the scalar and pseudoscalar operators $\mathcal O_{S,P}^{(\prime)}$ on the angular distribution of $B\to K^* (\to K\pi) \mu^+\mu^-$ has been considered before \cite{Kim:2000dq}, and no relevant effects on the observables of interest were found. However, as shown in Sec.~\ref{sec:3.2}, the inclusion of lepton-mass effects\footnote{We stress that we restricted ourselves to muons in our 
numerical analysis.}, which were neglected in previous studies, gives
rise to an additional observable in models with scalar currents, which
can serve as a precision null-test of the SM and, as we will show, in
principle allows one to distinguish between different NP models.

To assess the size of the possible effects generated by these operators, we first consider the allowed ranges for the Wilson coefficients $C_{S,P}^{(\prime)}$.
The most stringent constraint on these coefficients comes from the measurement
of  $B_s\to\mu^+\mu^-$, which is strongly helicity suppressed in the
SM, with a predicted branching ratio of \cite{Buras:2003jf,BBGT}
\begin{equation}\label{eq:Bsmumu-SM}
 \text{BR}(B_s\to\mu^+\mu^-) = (3.37 \pm 0.31) \times 10^{-9}.
\end{equation}
The most recent experimental upper bound still lies, at the 95\%
confidence level, one order of magnitude above the SM \cite{:2007kv}:
\begin{equation}\label{eq:Bsmumu-exp}
 \text{BR}(B_s\to\mu^+\mu^-) < 5.8 \times 10^{-8}\,.
\end{equation}
However, in many models, e.g.\ the MSSM at large $\tan\beta$, this branching ratio can be greatly enhanced.

In a generic NP model, the branching ratio is given by
\begin{equation}
 \text{BR}(B_s\to\mu^+\mu^-) = \tau_{B_s} f_{B_s}^2 m_{B_s}
\frac{\alpha_\text{em}^2 G_F^2 }{16 \pi^3} 
 |V_{tb} V_{ts}^*|^2 \sqrt{ 1 - \frac{4 m_\mu^2}{m_{B_s}^2}} \left[|S|^2 \left( 1 - \frac{4 m_\mu^2}{m_{B_s}^2} \right) + |P|^2\right],
\end{equation}
where
\begin{equation}
S = \frac{m_{B_s}^2}{2} (C_S - C_S') , \qquad
P = \frac{m_{B_s}^2}{2} (C_P - C_P') + m_\mu (C_{10} - C_{10}').
\end{equation}
Considering the experimental bound in Eq.~(\ref{eq:Bsmumu-exp}), these formulae imply the approximate bounds
\begin{equation}\label{eq:CSCP}
 |C_S-C_S'| \lesssim 0.12 \;\text{GeV}^{-1}, \qquad -0.09 \;\text{GeV}^{-1}\lesssim C_P-C_P' \lesssim 0.15\;\text{GeV}^{-1},
\end{equation}
barring large NP contributions to the Wilson coefficients $C_{10}^{(\prime)}$.

Now, inspecting the formulae for the angular coefficients, Eqs.~(\ref{eq:AC-first})--(\ref{eq:AC-last}), one can see that the only terms in which $C_{S}^{(\prime)}$ and $C_{P}^{(\prime)}$ are not suppressed by the lepton mass enter in the angular coefficient $I_1^c$. However, due to the small size of the Wilson coefficients themselves, see (\ref{eq:CSCP}), these terms turn out to be numerically irrelevant in general once the bound from $B_s\to\mu^+\mu^-$ is taken into account.

Since the pseudoscalar operators do not contribute to any other angular coefficient, this implies that they are indeed irrelevant in the phenomenological study of $B\to K^* (\to K\pi) \mu^+\mu^-$. For the scalar operators, however, the situation is different, because of the new angular coefficient $I_6^c$, Eq.~(\ref{eq:I6c}), which is directly proportional to the real part of $(C_S-C_S')$ and thus vanishes in the SM. So, although numerically small, this angular coefficient is an appealing observable because any measurement of a non-zero value would constitute an unambiguous signal of scalar currents at work.

This is in contrast to the process $B_s\to\mu^+\mu^-$, where a large
enhancement of the branching ratio compared to the SM could be caused
by both scalar and pseudoscalar currents. In addition, the measurement of a non-zero $S_6^c$ (the CP-averaged counterpart of $I_6^c$) would allow to determine the sign of $\re(C_S-C_S')$.
In fact, by a combined study of $B_s\to\mu^+\mu^-$ and the observable $S_6^c$, one would be able to constrain the relative sizes of the scalar and pseudoscalar Wilson coefficients, which can serve to distinguish different models of NP. For example, in the MSSM, the ratio of $C_S$ and $C_P$ is
\begin{equation}
 \frac{C_P}{C_S} \approx - \frac{M_{A^0}^2}{M_{H^0}^2} \approx -1
\end{equation}
to a very good accuracy, a relation which could be tested by a measurement of $\text{BR}(B_s\to\mu^+\mu^-)$ and $S_6^c$.

\begin{figure}[tb]
\centering
\includegraphics[width=0.8\textwidth]{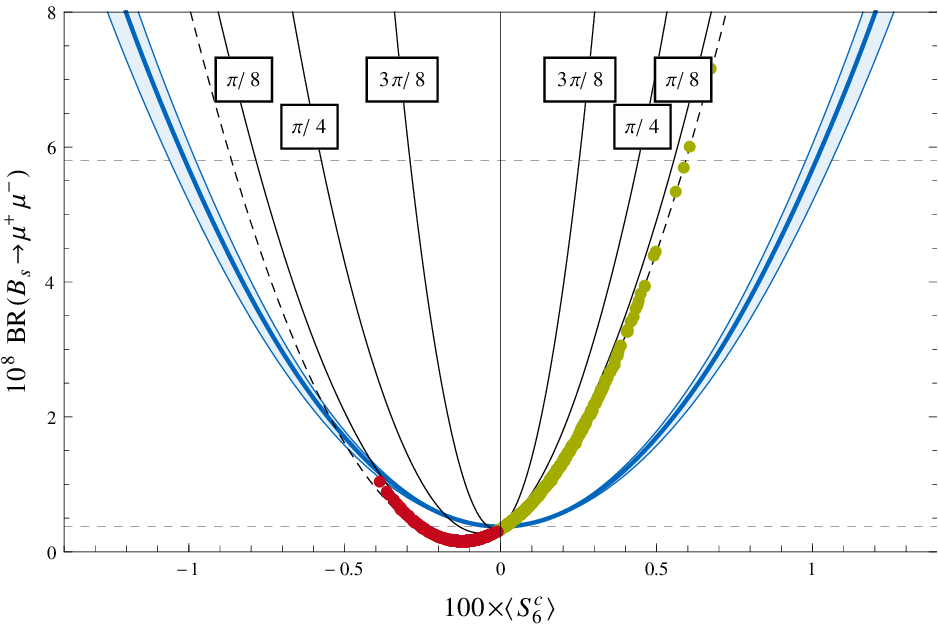}
\caption[]{\small Correlation between the observable $\left\langle
  S_6^c \right\rangle$ and the branching ratio of
  $B_s\to\mu^+\mu^-$. The blue band is obtained by assuming NP
  contributions only to the Wilson coefficient $C_S$, the black curves
  (where error bars are omitted) by assuming $C_P=-C_S$. Different
  values of the phase Arg$(C_S)$ are indicated. The red and green dots
  correspond to points in the CMSSM as described in the text. The
  horizontal dashed lines indicate the SM prediction  for
  $\text{BR}(B_s\to\mu^+\mu^-)$ (\ref{eq:Bsmumu-SM}) and the current experimental upper bound (\ref{eq:Bsmumu-exp}).}
\label{fig:Bsmm-S6c}
\end{figure}

To illustrate this point, we show, in Fig.~\ref{fig:Bsmm-S6c}, the  correlation between $\text{BR}(B_s\to\mu^+\mu^-)$ and $\left\langle S_6^c \right\rangle$ (as defined in Eq.~(\ref{eq:intS})).
The blue band has been obtained by assuming that NP contributions enter only through $C_S$, i.e.\ setting $C_P/C_S=0$, and varying $C_S$ accordingly; the error band takes into account all the sources of error as discussed in Sec.~\ref{sec:SM-pheno}.

Assuming, in contrast, $C_P/C_S=-1$, as would be the case in the MSSM, one obtains the black dashed parabola. As an illustration, the predictions for parameter points in the  constrained MSSM (CMSSM) with large $\tan\beta$ are indicated as red and green dots. These points have been generated by a random scan of the CMSSM parameters in the ranges
\begin{align}
 m_0& \leq 1\,\text{TeV}, &  m_{1/2} & \leq 1\,\text{TeV},\\
 -2m_0 & \leq A_0 \leq 2m_0, & 30 & \leq\tan\beta\leq50,
\end{align}
permitting both signs for the $\mu$-term and discarding points violating existing mass bounds or being incompatible with the measurement of $\text{BR}(B\to X_s\gamma)$. The green dots correspond to $\mu>0$, the red ones to $\mu<0$. It can be seen that the CMSSM points lie on the curve corresponding to $C_S=-C_P$ and, in particular for a positive $\mu$ parameter, could be clearly distinguished from the scenario without pseudoscalar currents, assuming sufficient experimental accuracy.

Since the observable $\left\langle S_6^c \right\rangle$ probes the
real part of $(C_S-C_S')$, the correlation gets modified if one allows
a phase in $C_S$. More precisely, $|\left\langle S_6^c \right\rangle|$
gets reduced for a fixed value of $\text{BR}(B_s\to\mu^+\mu^-)$. This
is illustrated by the black curves corresponding to $C_S=-C_P$, where
both Wilson coefficients are now complex, with the respective phase
$\text{Arg}(C_S)$ indicated by the labels on the curves. This is
precisely what happens in the Flavour Blind MSSM discussed in detail
in Sec.~\ref{sec:FBMSSM-pheno}. In this scenario, the measurement of
the correlation between BR$(B_s\to \mu^+\mu^-)$ and  $\left\langle
S_6^c \right\rangle$ would thus directly probe the phase of the scalar 
Wilson coefficient.

To summarize, while pseudoscalar operators are numerically irrelevant
in the decay $B\to K^* (\to K\pi) \mu^+\mu^-$, a study of the angular
distribution allows one to probe the scalar sector of a theory beyond the SM, in a way that is theoretically clean and complementary to $B_s\to\mu^+\mu^-$.

%%%%%%%%%%%%%%%%%%%%%%%%% Sec6.3: Specific New Physics Scenarios

\subsection{Specific New Physics Scenarios}

With the SM predictions for the CP-averaged angular coefficients 
$S_i^{(a)}$ and the CP asymmetries $A_i^{(a)}$ in hand,
we now investigate how these observables change in 
the NP scenarios discussed in Sec.~\ref{sec:4}.

\subsubsection{Minimal Flavour Violation}\label{sec:MFV-pheno}

In the MFV framework as described in Sec.~\ref{sec:MFV}, no additional CP-violating phases are present and  NP contributions to the Wilson coefficients of the primed operators can be neglected. This implies that all possible effects will arise from real contributions to the Wilson coefficients $C_7$, $C_8$, $C_9$ and $C_{10}$. This implies in turn that the most visible departures from the SM predictions will be in the observables $S_{1,2}^{s,c}$, $S_4$, $S_5$ and $S_6^s$, while the other angular observables and in particular the CP asymmetries will essentially be SM like.\footnote{ In general the concept of MFV does not exclude effects in the scalar Wilson coefficient $C_S$ which affect the observable $S_6^c$. However, as shown in Sec.~\ref{sec:scalar}, these effects can be discussed completely independently and we will not consider them in this section.}

Model-independent studies within the MFV framework show that large NP
contributions to the Wilson coefficients $C_7$, $C_8$, $C_9$ and
$C_{10}$ are still allowed \cite{HIKM08}. In particular, scenarios
in which the sign of these Wilson coefficients is flipped with respect
to the SM cannot yet be excluded.

\begin{figure}[bt]
\centering
\includegraphics[width=\textwidth]{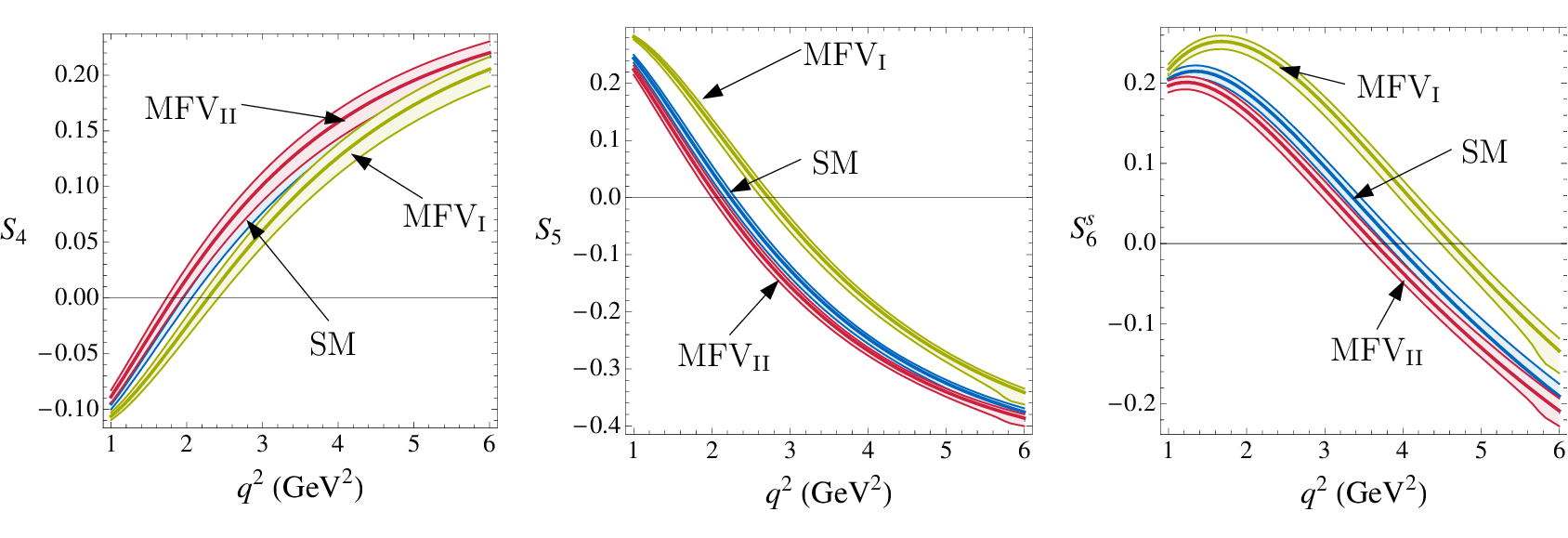}
\vskip-10pt
\caption[]{\small The observables $S_4$, $S_5$ and $S_6^s$ in the SM
  (blue band) and the MFV MSSM scenarios MFV$_{\rm I,II}$ 
described in the text.}\label{fig:MFV}
\centering
\medskip
\includegraphics[width=0.3\textwidth]{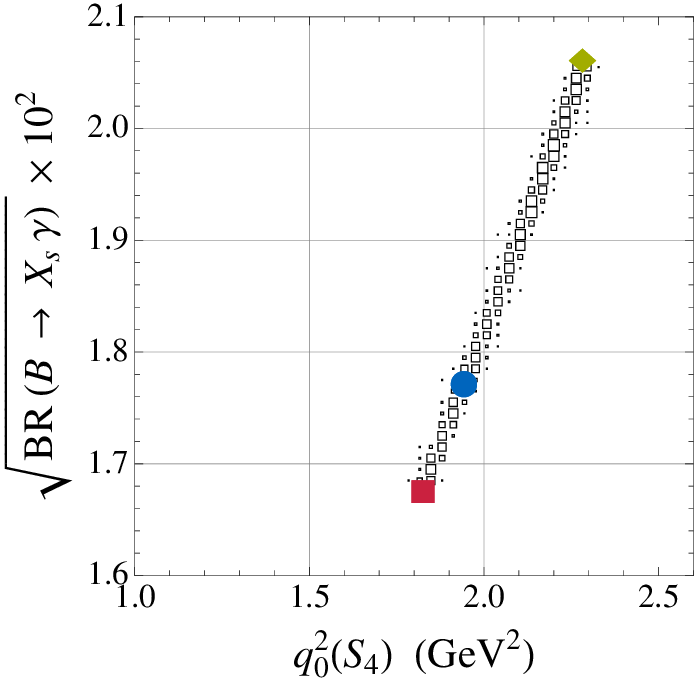}\quad
\includegraphics[width=0.3\textwidth]{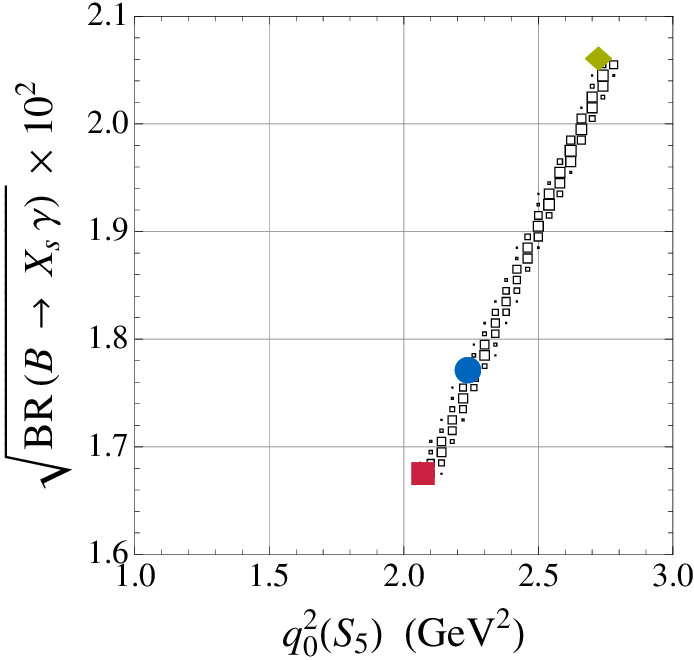}\quad
\includegraphics[width=0.3\textwidth]{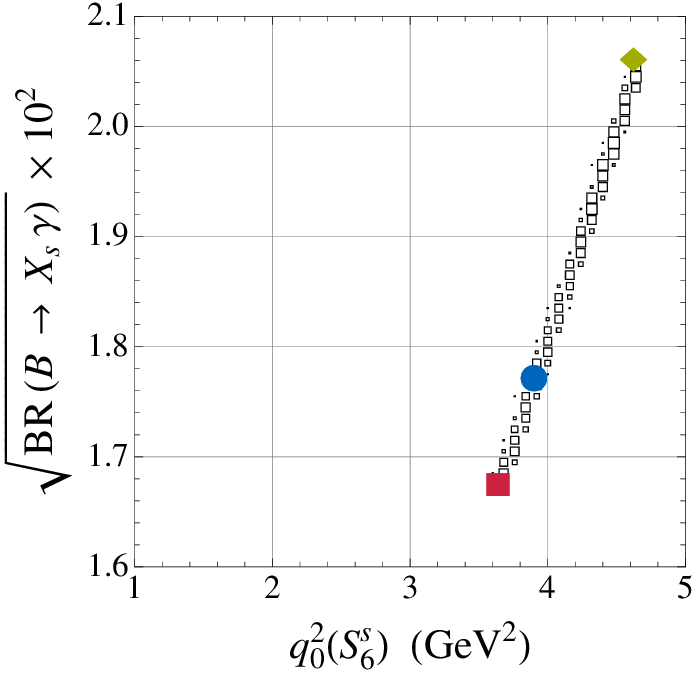}
\vskip-10pt
\caption[]{\small The correlation between the zeros of $S_4$, $S_5$
  and $S_6^s$ and BR$(B \to X_s \gamma)$ in the MFV MSSM. The blue
  circles correspond to the central SM values, while the green
  diamonds represent scenario MFV$_{\rm I}$ and the red squares
  scenario MFV$_{\rm II}$.}\label{fig:MFV_zeros}
\end{figure}

However, in concrete MFV NP models it is usually difficult to generate
large effects in  $C_9$ and $C_{10}$. For
example in the MFV MSSM, NP contributions to $C_9$ and $C_{10}$ are
typically very small \cite{LMSS99,ALGH01}. Therefore, in this model,
the main source of NP effects is $C_7$ whose value can be modified
substantially by chargino-stop loops. For negligible NP contributions
to $C_9$ and $C_{10}$ however, the effects in $C_7$ are strongly
constrained by the data on BR$(B \to X_s \gamma)$ and BR$(B \to X_s
\mu^+ \mu^-)$ and in particular a sign flip in $C_7$ is excluded at
the $3 \sigma$ level \cite{GHM04}. The effects in the $S_i^{(a)}$ are
then quite limited. In Fig.~\ref{fig:MFV}, we show the largest
possible effects in $S_4$, $S_5$ and $S_6^s$:
scenario MFV$_{\rm I}$ (green curves) corresponds to the maximum
allowed negative (i.e.\ constructive) NP contribution to $C_7$
(i.e.\ $C_7^{\rm NP}$) and shifts the zeros of $S_4$, $S_5$ and $S_6^s$
to larger values of $q^2$. Scenario MFV$_{\rm II}$ (red curves), 
on the other hand, corresponds to the largest positive allowed value
of $C_7^{\rm NP}$ and hence shifts the zeros to smaller values.
The separation in $q^2$ between these two curves corresponds to the range shown in Fig.~\ref{fig:7} for $(C_{10}^{\rm NP} - C_{10}^{\prime \rm NP}) = 0$ and $C_9^{\rm NP} = 0$, respectively, where the superscript NP denotes the NP contribution to the Wilson coefficient. The most relevant input parameters corresponding to the two scenarios are collected in Tab.~\ref{tab:parameter_MFVMSSM}.

\begin{table}[th]
\addtolength{\arraycolsep}{3pt}
\renewcommand{\arraystretch}{1.3}
\centering
\begin{tabular}{|l|c|c|c|c|c|c|c|}
\hline
Scenario & $\tan\beta$ & $m_A$ & $m_{\tilde g}$ & $m_{\tilde Q}$ & $m_{\tilde U}$ & $A_{\tilde t}$ & $\mu$ \\
\hline\hline
MFV$_{\rm I}$ & $28$ & $380$ & $530$  & $800$ & $540$ & $-850$ & $860$ \\ \hline
MFV$_{\rm II}$   & $29$ & $530$ & $1000$ & $880$ & $660$ & $ 880$ & $750$ \\ \hline
\end{tabular}
\caption{\small Most relevant parameters of the two MFV MSSM scenarios discussed in the text. $\tan\beta$ is the ratio of the two Higgs VEVs, $m_A$ the mass of the pseudoscalar Higgs, $m_{\tilde g}$ is the gluino mass, $m_{\tilde Q}$ is a universal soft mass for the left handed squark doublets, $m_{\tilde U}$ a universal soft mass for the right handed up squarks, $A_{\tilde t}$ is the stop trilinear coupling and $\mu$ the Higgsino mass parameter. Our conventions for the trilinear coupling are such that the left-right mixing entry in the stop mass matrix is $(m^2)_{LR} = -m_t (A_{\tilde t} + \mu^* \cot\beta)$. All massive parameters are given in GeV.}
\label{tab:parameter_MFVMSSM}
\end{table}

It is well known that in the MFV MSSM, the shift in the zero of the forward-backward asymmetry in $B \to X_s \mu^+ \mu^-$ is highly correlated with a change of the branching ratio of $B \to X_s \gamma$  \cite{Buras:2003mk,BBE04}. In Fig.~\ref{fig:MFV_zeros} we show the corresponding correlation between 
the zeros of $S_4$, $S_5$ and $S_6^s$ and BR$(B \to X_s \gamma)$. 
Any deviation from the lines in the plots would signal
the presence either of NP contributions to Wilson coefficients other
than $C_7$ or of new CP-violating phases that lead to complex values of $C_7$.

\subsubsection{Flavour Blind MSSM}\label{sec:FBMSSM-pheno}

One model with new sources of CP violation is the FBMSSM
discussed in
Refs.~\cite{Baek:1998yn,Bartl:2001wc,Ellis:2007kb,ABP08}. This is a
MSSM where the CKM matrix is the only source of flavour violation, but
additional CP-violating, flavour conserving phases are present in the
soft sector. Within this framework, the majority of non-standard
effects arises though complex NP contributions to the Wilson
coefficient $C_7$. We discuss two scenarios in which the effects are
maximal: scenario FBMSSM$_{\rm I}$ is characterized by large negative
$\im(C_7)$, while scenario FBMSSM$_{\rm II}$  corresponds to a large positive $\im(C_7)$.
The corresponding input parameters are collected in
Tab.~\ref{tab:parameter_FBMSSM}, together with those of a third
scenario, FBMSSM$_{\rm III}$, to be considered later.
\begin{table}
\addtolength{\arraycolsep}{3pt}
\renewcommand{\arraystretch}{1.3}
\centering
\begin{tabular}{|l|c|c|c|c|c|c|c|c|}
\hline
Scenario & $\tan\beta$ & $m_A$ & $m_{\tilde g}$ & $m_{\tilde Q}$ & $m_{\tilde U}$ & $A_{\tilde t}$ & $\mu$ & Arg$(\mu A_{\tilde t})$ \\
\hline\hline
FBMSSM$_{\rm I}$ & $40$ & $400$ & $700$ & $380$ & $700$ & $900$ & $150$ & $-45^\circ$ \\ \hline
FBMSSM$_{\rm II}$  & $40$ & $400$ & $700$ & $380$ & $700$ & $900$ & $150$ & $\phantom{-}50^\circ$ \\ \hline
FBMSSM$_{\rm III}$ & $40$ & $400$ & $700$ & $650$ & $700$ & $900$ & $150$ & $\phantom{-}60^\circ$ \\ \hline
\end{tabular}
\caption{\small Most relevant parameters of the three FBMSSM scenarios discussed in the text. All massive parameters are given in GeV.}
\label{tab:parameter_FBMSSM}
\end{table}
\begin{figure}[tb]
\centering
\includegraphics[width=0.6\textwidth]{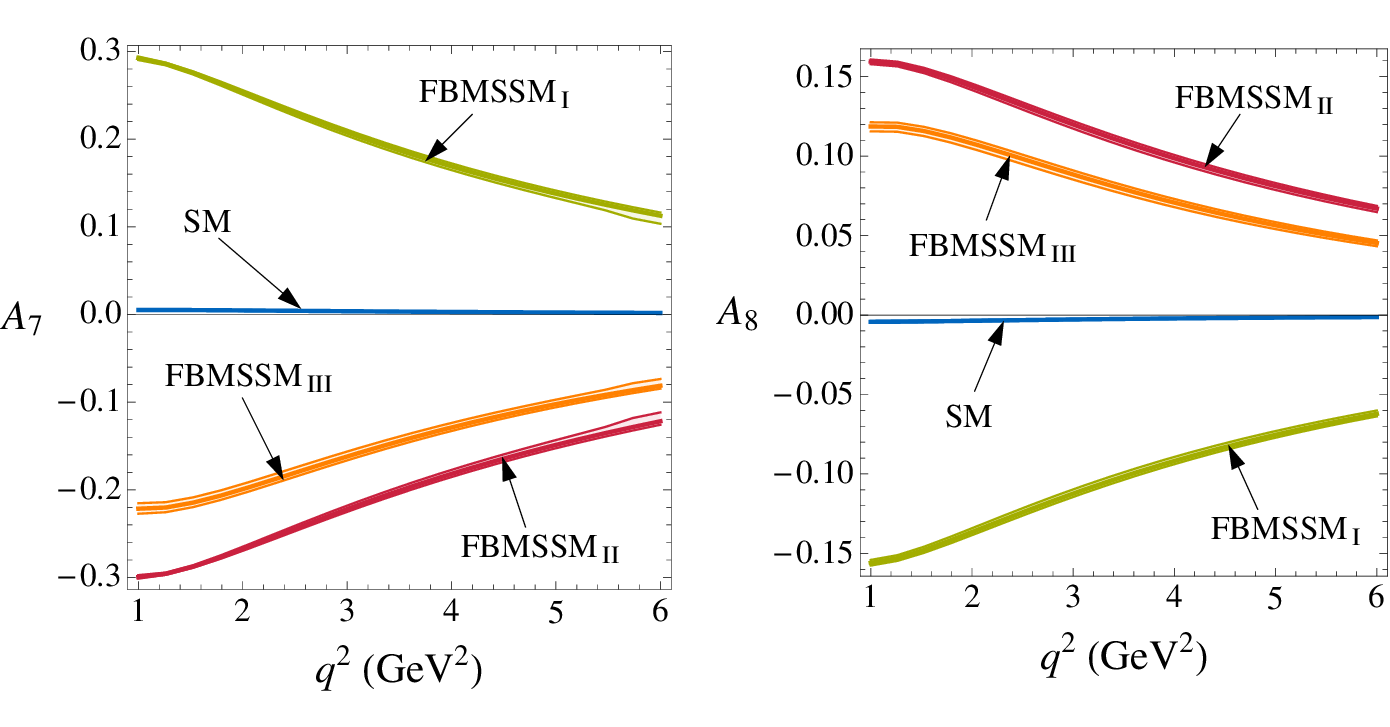}\quad
\raisebox{0.5cm}{\includegraphics[width=0.3\textwidth]{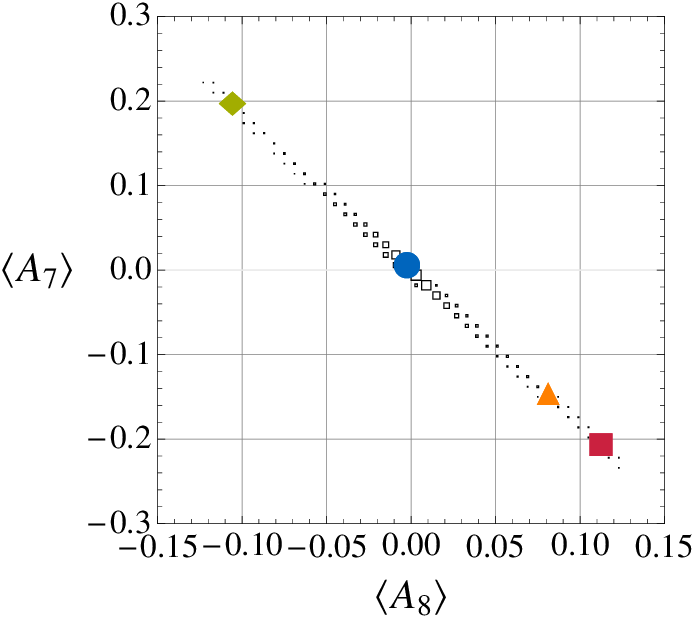}}
\vskip-10pt
\caption[]{\small Left and centre plot: CP asymmetries $A_7$ and $A_8$
  in the SM (blue band) and three FBMSSM scenarios as described in the
  text. Right plot: correlation between the integrated asymmetries
  $\langle A_7 \rangle$ and $\langle A_8 \rangle$ in the
  FBMSSM. Blue circle: SM, green diamond: FBMSSM$_{\rm I}$, red
  square: FBMSSM$_{\rm II}$ , orange triangle: FBMSSM$_{\rm III}$.}
\label{fig:A_ABP}
\centering
\medskip
\includegraphics[width=0.92\textwidth]{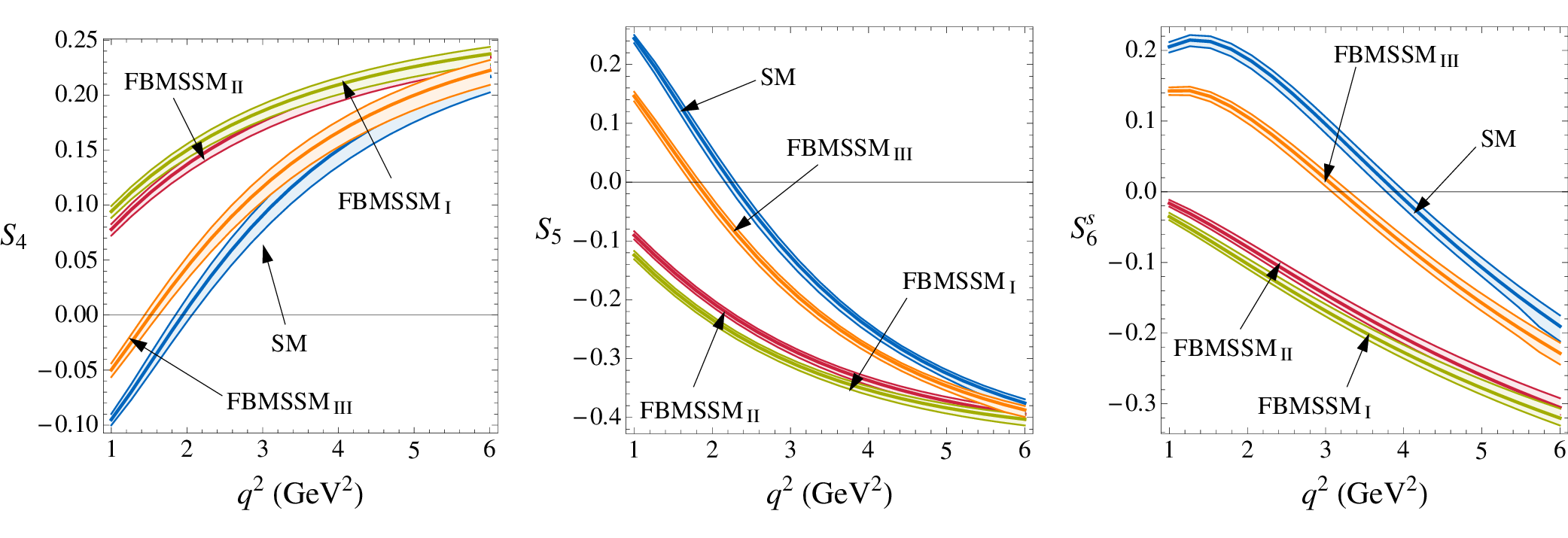}
\vskip-10pt
\caption[]{\small The observables  $S_4$, $S_5$ and $S_6^s$ in the SM
  (blue band) and the three FBMSSM scenarios FBMSSM$_{\rm I,II,III}$.}
\label{fig:S_ABP}
\end{figure}

Concerning the CP asymmetries, we observe that significant departures from the SM predictions can be obtained in $A^s_{1,2}$, $A_5$, $A^s_6$, $A_7$ and $A_8$. The most pronounced effects can be seen in $A_7$ and $A_8$ and these are shown in the left and centre plot of Fig.~\ref{fig:A_ABP}. The effects here are predominantly due to the large imaginary part of $C_7$ and we note that in this case positive values for $A_7$ imply negative ones for $A_8$ and vice versa.
This is also displayed in the right plot of Fig.~\ref{fig:A_ABP}, where we show the almost perfect correlation between the integrated asymmetries $\langle A_7 \rangle$ and $\langle A_8 \rangle$. Any deviation from the line shown in this plot would signal the presence of additional imaginary parts in either $C_7^{\prime}$ or $C_9^{(\prime)}$ and $C_{10}^{(\prime)}$.

In the CP-averaged angular coefficients we find significant departures from the SM in $S_{1,2}^{s,c}$, $S_4$, $S_5$, $S_6^s$ and also in $S_6^c$, while effects in $S_3$, $S_7$, $S_8$ and $S_9$ can hardly be distinguished from the SM. 
Although in the FBMSSM the BR$(B_s \to \mu^+ \mu^-)$ can be close to its experimental upper bound, the effects in $S_6^c$ are smaller than the maximal effects found in the model-independent discussion of Sec.~\ref{sec:scalar}, because the large imaginary part in $C_7$ implies a large phase for the relevant Wilson coefficient $C_S$.
Concerning $S_{1,2}^{s,c}$, we find that while $|S_{1,2}^s|$ is enhanced, $|S_{1,2}^c|$ is suppressed with respect to the SM results.
For $S_4$, $S_5$ and the forward--backward asymmetry $S_6^s$ we find
significant shifts in their zero towards values of $q^2$ lower than
the SM prediction or we even find no zero at all. These effects are
shown in Fig.~\ref{fig:S_ABP} and are much larger than  those possible in the MFV MSSM (see Fig.~\ref{fig:MFV}). The reason for these large shifts are the large values of $\im(C_7)$ in the scenarios  considered, as discussed in Sec.~\ref{sec:zeros_model_independent}.
\begin{figure}[t]
\centering
\includegraphics[width=0.3\textwidth]{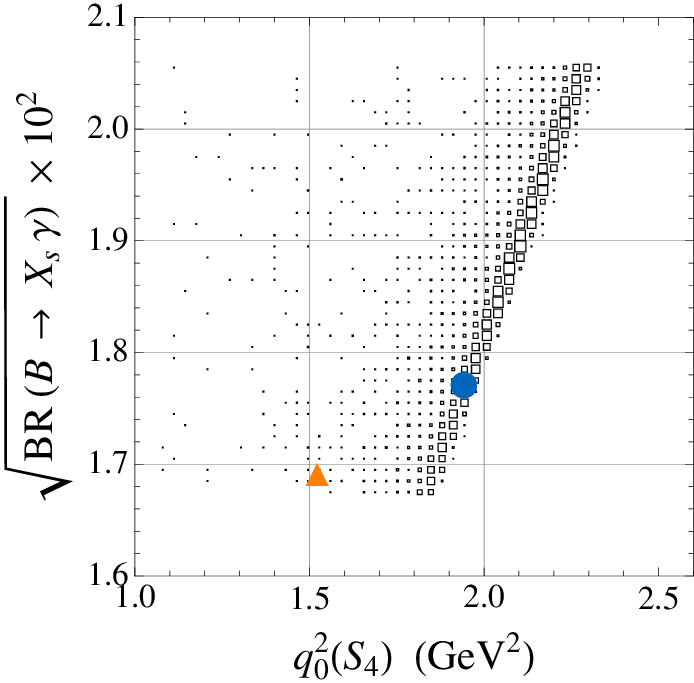}\quad
\includegraphics[width=0.3\textwidth]{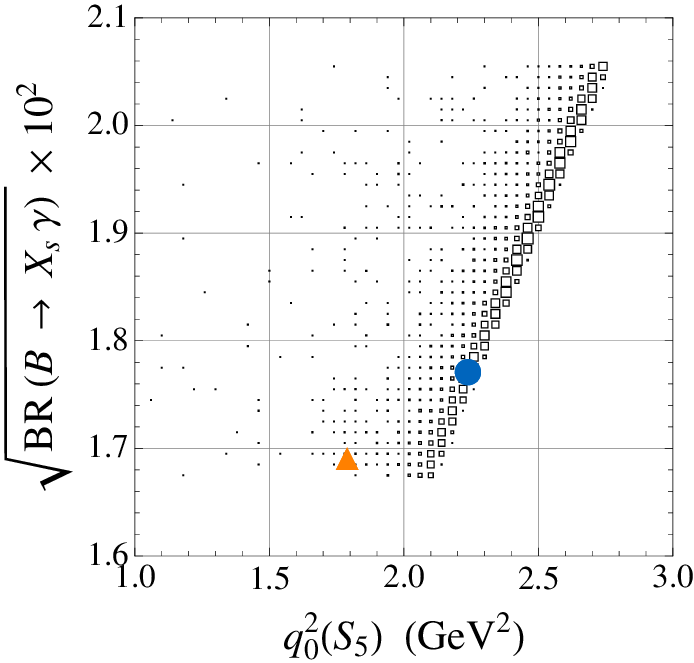}\quad
\includegraphics[width=0.3\textwidth]{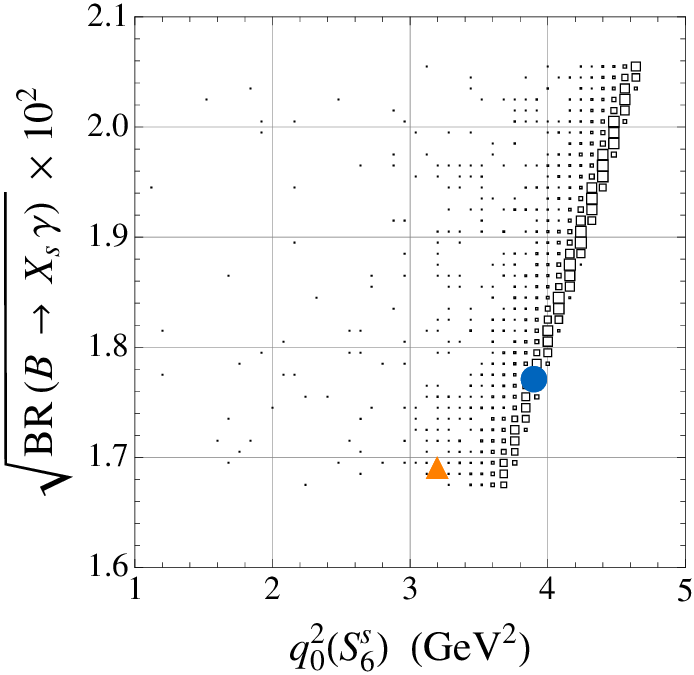}
\includegraphics[width=0.3\textwidth]{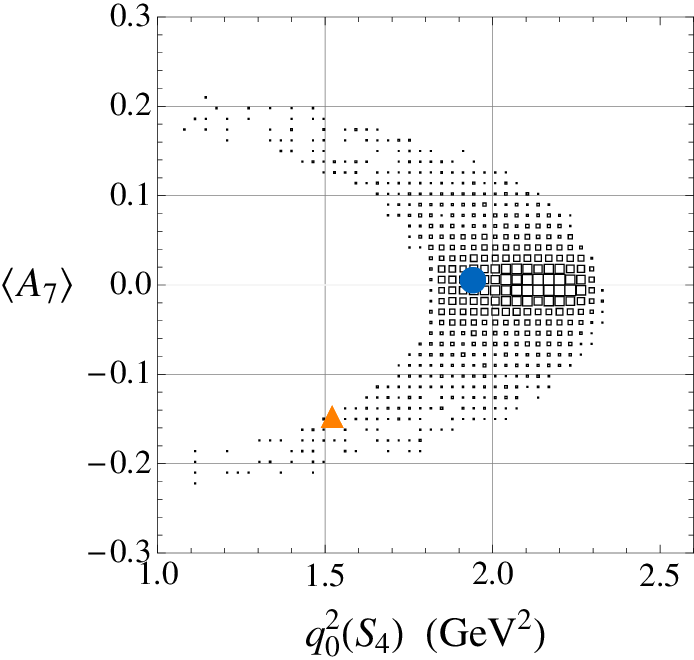}\quad
\includegraphics[width=0.3\textwidth]{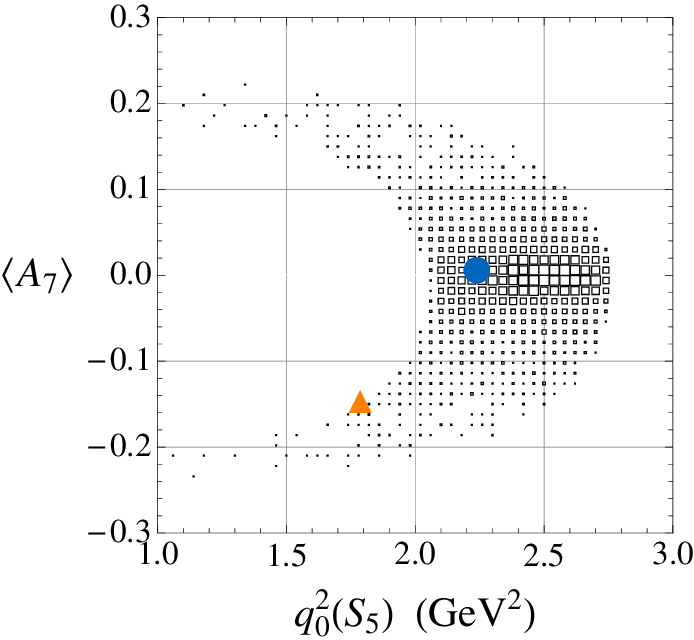}\quad
\includegraphics[width=0.3\textwidth]{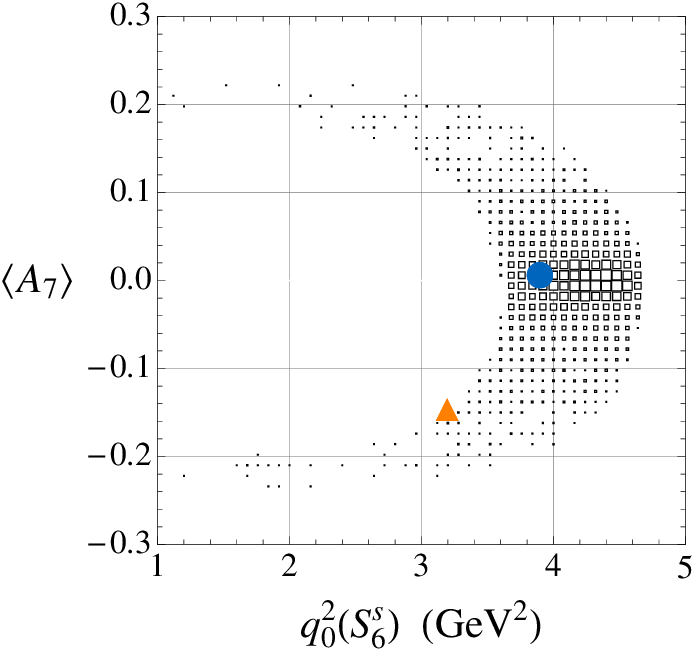}
\caption[]{\small Correlation between the zeros of $S_4$, $S_5$ and $S_6^s$ with the $b \to s \gamma$ branching ratio (upper plots) and with the integrated asymmetry $\langle A_7 \rangle$ (lower plots) in the FBMSSM. The blue circles correspond to the SM predictions. The orange triangles correspond to a FBMSSM scenario that gives $S_{\phi K_S}$ close to the central experimental value $\simeq 0.44$.}
\label{fig:correl_ABP_zeros}
\end{figure}

One finds that the strict correlation between the zeros and BR$(B \to X_s \gamma)$ is lost in the FBMSSM. This is shown in the upper plots of Fig.~\ref{fig:correl_ABP_zeros}. However, as the additional contributions to $b \to s \gamma$ from the imaginary part of $C_7$ can only enhance the branching ratio, one still finds an upper bound on the zeros for a given value of BR$(B \to X_s \gamma)$.
In addition, in the lower plots of Fig.~\ref{fig:correl_ABP_zeros} we show the zeros $q^2_0(S_4)$, $q^2_0(S_5)$ and $q^2_0(S_6^s)$ against the integrated asymmetry $\langle A_7 \rangle$. One observes that large effects in $\langle A_7 \rangle$ are correlated with large shifts in the zeros towards lower values.

In order to identify signs in the CP asymmetries which are favoured in this model  one must include additional observables in the analysis.
\begin{figure}[t]
\centering
\includegraphics[width=0.3\textwidth]{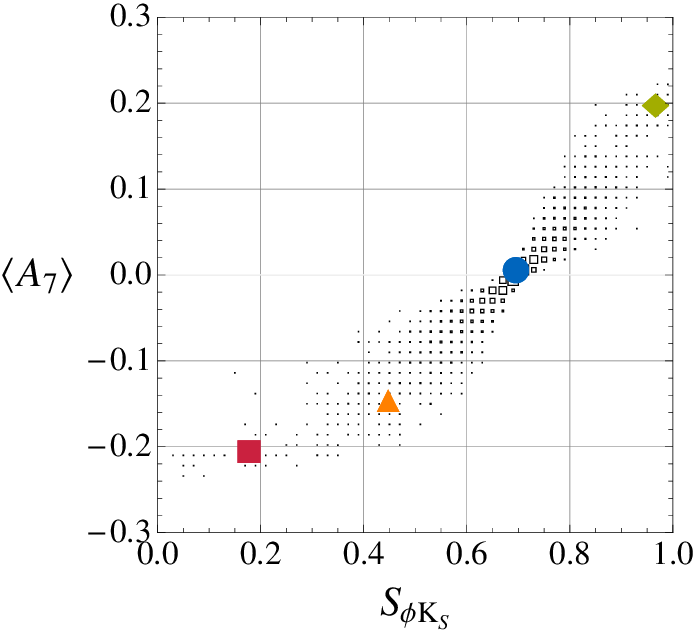}\quad
\includegraphics[width=0.3\textwidth]{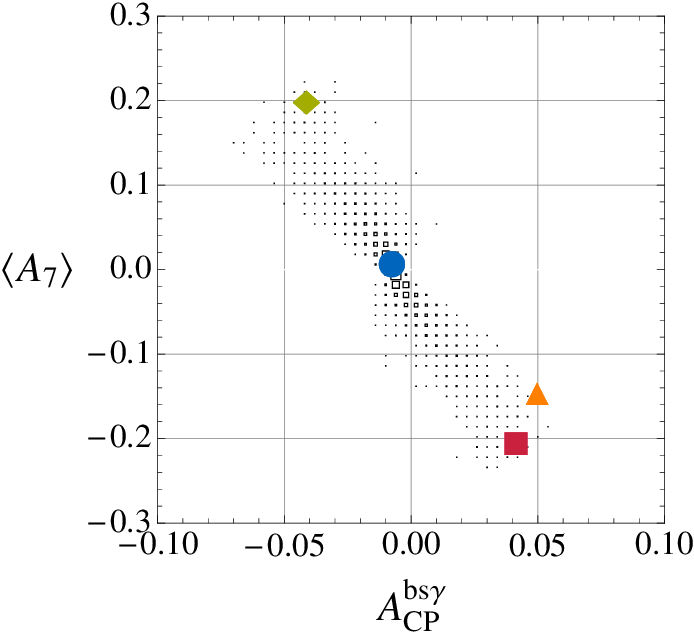}\quad
\includegraphics[width=0.3\textwidth]{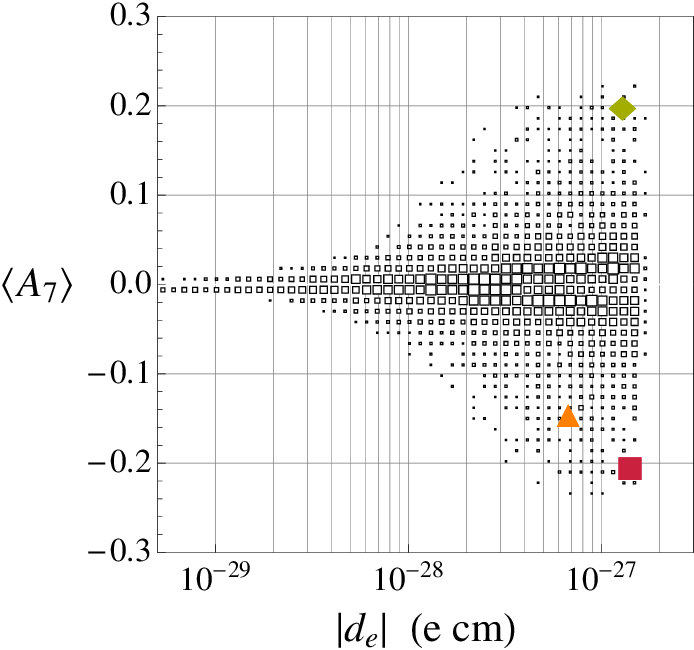}
\caption[]{\small $\langle A_7 \rangle$ vs. $S_{\phi K_S}$ (left plot), 
$\langle A_7 \rangle$ vs. $A_{\rm CP}^{bs\gamma}$ (centre plot) and $\langle 
A_7 \rangle$ vs. $d_e$ (right plot) in the FBMSSM. The blue circles indicate 
the SM values, while the green diamonds, red squares and orange triangles 
correspond to the scenarios FBMSSM$_{\rm I}$, FBMSSM$_{\rm II}$ and 
FBMSSM$_{\rm III}$, respectively.}\label{fig:correl_ABP}
\end{figure}
To this end we also investigate the direct CP asymmetry in the $b \to s \gamma$ decay $A_{\rm CP}(b\to s\gamma)$, the electric dipole moments of the electron and the neutron $d_e$ and $d_n$ and the mixing induced CP asymmetry $S_{\phi K_S}$. We recall that in \cite{ABP08} striking correlations between these observables have been found.
In particular, the desire to explain the anomaly observed in $S_{\phi K_S}$ through the presence of flavour conserving but CP-violating phases implied a positive $A_{\rm CP}(b\to s\gamma)$, by an order of magnitude larger than its SM tiny value and $d_e$, $d_n$ at least as large as $10^{-28}~e$\,cm.

The left plot of Fig.~\ref{fig:correl_ABP} shows the correlation between $\langle A_7 \rangle$ and $S_{\phi K_S}$. We find that a value of $S_{\phi K_S} \simeq 0.44$, as indicated by the present data \cite{HFAG}, implies a  negative value for $\langle A_7 \rangle$ in the range $[-0.2, -0.05]$ and then also a positive value for $\langle A_8 \rangle$ in the range $[0.03, 0.11]$.
In addition to the two scenarios discussed above, we have chosen also
a third scenario, FBMSSM$_{\rm III}$, indicated as orange triangle in the plots of Figs.~\ref{fig:A_ABP}, \ref{fig:correl_ABP_zeros} and \ref{fig:correl_ABP}, that gives $S_{\phi K_S}$ close to the experimental value. This scenario is shown in Figs.~\ref{fig:A_ABP} and \ref{fig:S_ABP} as the orange bands and we find that while one still can get almost maximal effects in $\langle A_7 \rangle$ and $\langle A_8 \rangle$ the effects in $S_4$, $S_5$ and $S_6^s$ are much less pronounced.

In the centre plot of Fig.~\ref{fig:correl_ABP} we report the correlation between $\langle A_7 \rangle$ and $A_{\rm CP}(b\to s\gamma)$. One observes that negative values for $\langle A_7 \rangle$ imply positive values for $A_{\rm CP}(b\to s\gamma)$ that can reach values up to $(5-6)\%$.

Finally, the right plot of Fig.~\ref{fig:correl_ABP} shows the correlation between $\langle A_7 \rangle$ and the EDM of the electron, $d_e$ in the FBMSSM. We find that large values for $\langle A_7 \rangle$ necessarily require large values for the electron EDM close to the current upper bound of $1.6 \times 10^{-27}~e$\,cm \cite{EDM}.

\subsubsection{LHT}

\begin{table}
\begin{center}
\renewcommand{\arraystretch}{1.2}
\begin{tabular}{|l|c|c|c|c|c|c|c|c|c|c|c|} \hline
Scenario &   $f$  & $x_L$ &	$m_H^1$	 & $m_H^2$	 &$m_H^3$ & $\theta^d_{23}$	 &$\theta^d_{13}$	 &$\theta^d_{12}$	 &$\delta^d_{23}$  &$\delta^d_{13}$ & $\delta^d_{12}$ \\ \hline \hline
${\rm LHT}_{\rm I}$  &	1000	 & 0.5   & 565  & 1000  & 770  &  1.60   &  2.50 &   1.35   & 5.70   & 4.20  &  5.80 \\ \hline
${\rm LHT}_{\rm II}$ &	1000	 & 0.5   & 1000  &375   & 425    &1.50   &  1.00  &  4.75  &  4.25  &  0.60   & 2.85 \\ \hline  
\end{tabular}
\end{center}
\caption[]{\small Parameters of the LHT scenarios LHT$_{\rm I,II}$: 
$\theta^d_{ij}$ and $\delta^d_{ij}$ are the parameters of the
CKM-like unitary mixing matrix for the mirror $d$ quarks, $m_H^i$ are the
masses of the mirror quarks, $f$ is the high energy scale and $x_L$ the
mixing parameter of the SM top and the T-even top partner.}\label{tab:LHT}
\end{table}

\begin{figure}
\centering
\includegraphics[width=0.3\textwidth]{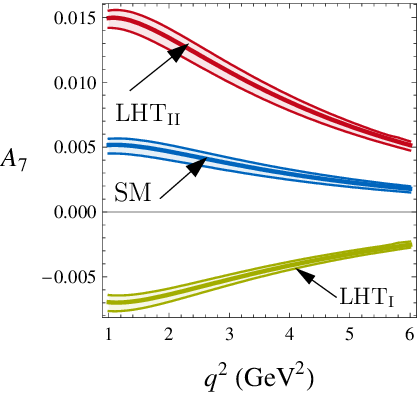}\quad
\includegraphics[width=0.3\textwidth]{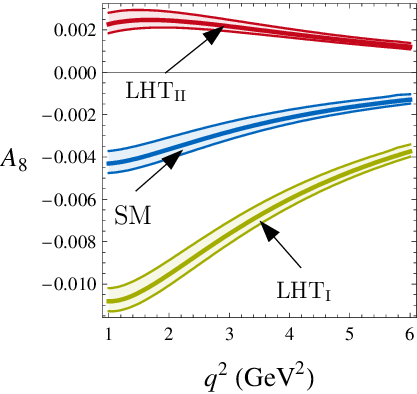}\quad
\raisebox{0.4cm}{\includegraphics[width=0.3\textwidth]{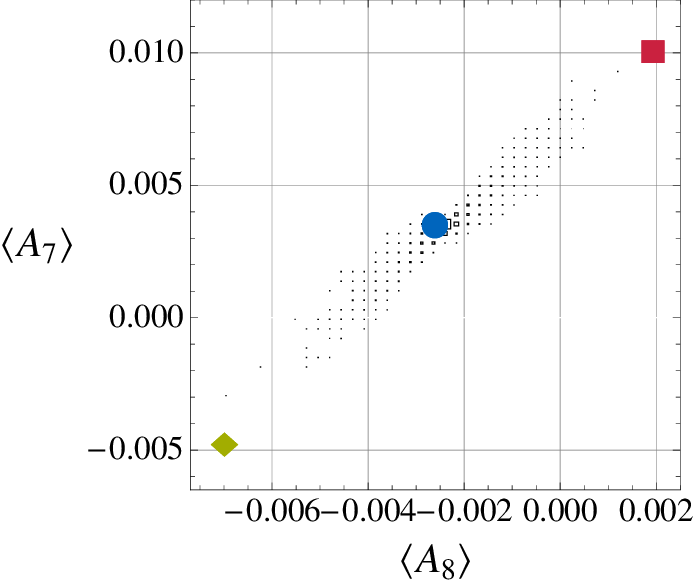}}
\caption[]{\small Left and centre plot: CP asymmetries $A_7$ and $A_8$
  in the SM (blue band) and the LHT scenarios LHT$_{\rm I,II}$. Right
  plot: Correlation between the integrated asymmetries $\langle A_7
  \rangle$ and $\langle A_8 \rangle$ in the LHT. The blue circle
  represents the SM, the green diamond scenario LHT$_{\rm I}$ and
  the red square scenario LHT$_{\rm II}$.}
\label{fig:A78-LHT}
\end{figure}

We analyse the angular observables within the LHT by means of a global parameter scan taking into account all relevant constraints from other flavour observables.
As already anticipated in Ref.~\cite{LHT2}, most NP effects in the observables considered here are found to be small. In particular $S_6^s$, the forward-backward asymmetry, turns out to be very close to the SM. The same applies to all other CP-averaged angular coefficients and most CP asymmetries. 
The largest effects relative to the SM are found in $A_7$ and $A_8$ as
in the SM their absolute values are at most $6 \times 10^{-3}$ and $5
\times 10^{-3}$, respectively. We consider two scenarios, LHT$_{\rm I}$
and LHT$_{\rm II}$, with input parameters as given in
Tab.~\ref{tab:LHT}. In the left and centre plot of
Fig.~\ref{fig:A78-LHT} we show the corresponding asymmetries $A_7$ and
$A_8$ as functions of $q^2$.
The blue curves represent the SM. The green curves labelled LHT$_{\rm
  I}$ correspond to a LHT parameter point that gives the largest
negative NP contribution to $\im(C_9)$ and $\im(C_{10})$, while the LHT$_{\rm
  II}$ curves (red) give the largest positive contribution. Enhancement of both asymmetries by a factor of three is possible for low values of $q^2$ with visible but smaller effects for larger values of $q^2$.

Still, these significant enhancements are one order of magnitude smaller than those found in the FBMSSM.  The reason why much larger effects in $A_7$ and $A_8$ are possible in the latter model is that large NP contributions to the imaginary part of $C_7$ are allowed, comparable in magnitude to the SM contribution. In the LHT model NP contributions to $C_7$ are found to be very small \cite{LHT2}.
As the effects in $A_7$ and $A_8$ are therefore dominantly created by
$\im(C_9)$ and $\im(C_{10})$, the correlation between the integrated
asymmetries $\langle A_7 \rangle$ and $\langle A_8 \rangle$ is
completely different than that found in the FBMSSM (see the right-hand
side plots in Figs.~\ref{fig:A_ABP} and \ref{fig:A78-LHT}).

As a side comment, in our numerical analysis we have used the formulae 
of Ref.~\cite{LHT2} modified by the additional term found in Ref.~\cite{Goto:2008fj}
which remove the UV cutoff dependent terms in $C_9$ and $C_{10}$. 
This modification decreases the two
asymmetries by roughly a factor of 2 to 3. Whether this is the final
result for the LHT model remains to be seen as the structure of the
full heavy-fermion sector in the LHT model is rather involved and a complete 
analysis is still lacking.

\subsubsection{General MSSM}\label{sec:GMSSM_pheno}

Due to the huge number of free parameters in the general MSSM, a comprehensive analysis of this general framework is challenging. As a first step we therefore restrict ourselves to a framework in which NP effects are created dominantly by complex contributions to the Wilson coefficient $C_7^\prime$. Such a situation can easily be achieved in the general MSSM if one introduces flavour violating terms only in the left-right sector of the down squark mass. In particular, a $(\delta_d)_{32}^{LR}$ mass insertion will mostly create contributions to $C_7^\prime$ by means of down squark -- gluino loops, while at the same time leaving the other relevant Wilson coefficients SM like.
\begin{table}
\addtolength{\arraycolsep}{3pt}
\renewcommand{\arraystretch}{1.3}
\centering
\begin{tabular}{|l|c|c|c|c|c|c|c|c|c|c|c|}
\hline
Scenario & $\tan\beta$ & $m_A$ & $m_{\tilde g}$ & $m_{\tilde Q}$ & $m_{\tilde U}$ & $m_{\tilde D}$ & $A_{\tilde u}$ & $A_{\tilde d}$ & $\mu$ & $|(\delta_d)^{LR}_{32}|$ & Arg$(\delta_d)^{LR}_{32}$\\
\hline\hline
$\text{GMSSM}_\text{I}$ & $6$ & $520$ & $500$ & $400$ & $500$ & $380$ & $800$ & $750$ & $470$ & $0.01$ & $-135^\circ $\\ \hline
$\text{GMSSM}_\text{II}$ & $5$ & $740$ & $1000$ & $460$ & $1000$ & $390$ & $1500$ & $440$ & $200$ & $0.03$ & $60^\circ $\\ \hline
\end{tabular}
\caption{\small Most relevant parameters of the two general MSSM scenarios with large $C_7^\prime$ as discussed in the text. $m_{\tilde D}$ a universal soft mass for the right handed down squarks, $A_{\tilde u(\tilde d)}$ universal trilinear couplings for the up (down) squarks and $(\delta_d)^{LR}_{32}$ the left-right mass insertion that generates large effects in $C_7^\prime$. Our conventions for the trilinear coupling are such that the left-right mixing entry in the sbottom mass matrix is $(m^2)_{LR} = -m_b (A_{\tilde d} + \mu^* \tan\beta)$. All massive parameters are given in GeV.}
\label{tab:parameter_GMSSM1}
\end{table}

\begin{figure}
\centering
\includegraphics[width=0.9\textwidth]{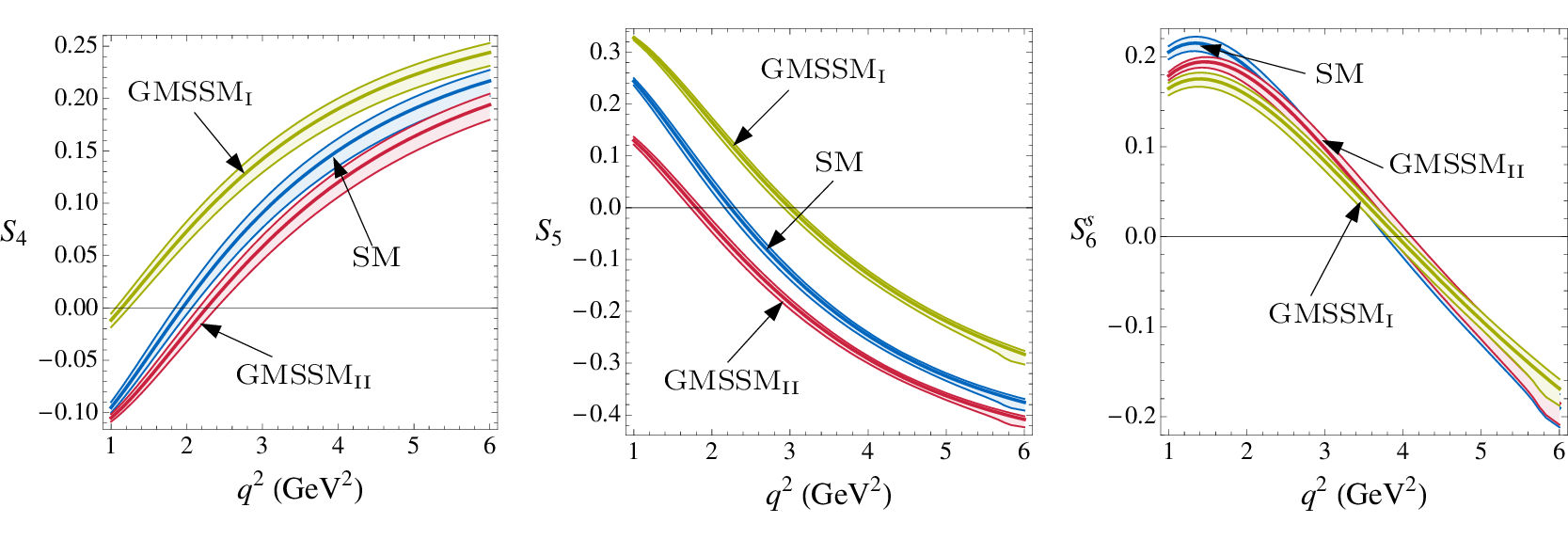}
\caption[]{\small The observables $S_4$, $S_5$ and $S_6^s$ in the SM (blue band) and two GMSSM scenarios with large complex contributions to $C_7^\prime$ as described in the text.}
\label{fig:GMSSMC7p_S4S5S6s}
\centering
\includegraphics[width=0.6\textwidth]{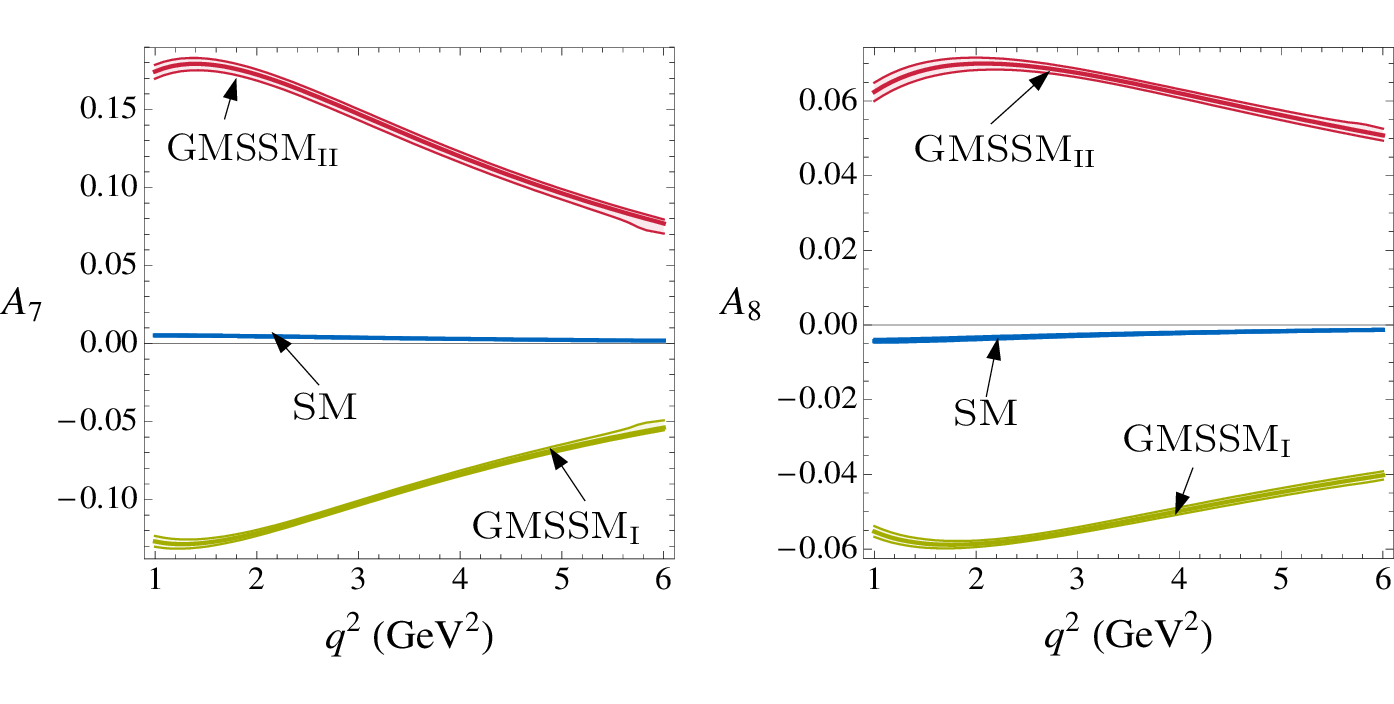}\quad
\raisebox{0.5cm}{\includegraphics[width=0.3\textwidth]{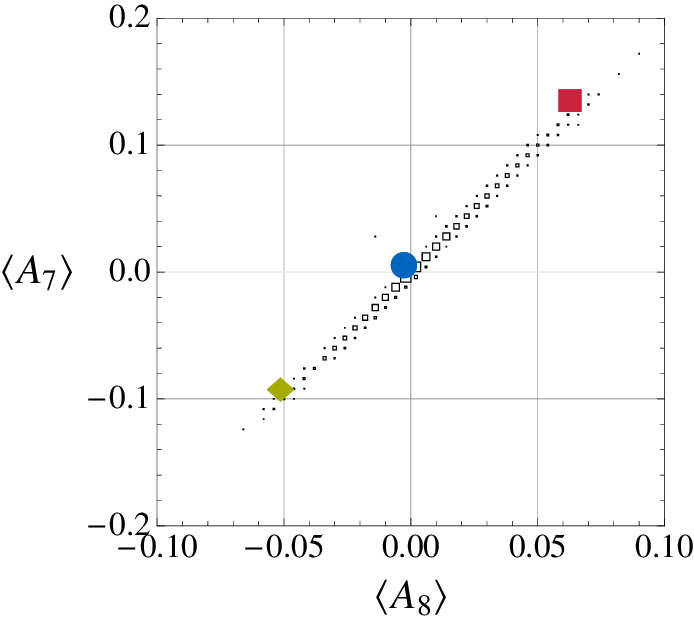}}
\caption[]{\small Left and centre plot: CP asymmetries $A_7$ and $A_8$ in the SM (blue band) and two general MSSM scenarios with large complex contributions to $C_7^\prime$. Right plot: Correlation between the integrated asymmetries $\langle A_7 \rangle$ and $\langle A_8 \rangle$ in the framework of a general MSSM with large complex $C_7^\prime$.  The blue
  circle corresponds to the central SM value, while the green
  diamond represents scenario GMSSM$_{\rm I}$ and the red square
  scenario GMSSM$_{\rm II}$.}
\label{fig:GMSSMC7p_A7A8}
\end{figure}
Fig.~\ref{fig:GMSSMC7p_S4S5S6s} shows possible effects in $S_4$, $S_5$ and $S_6^s$ that arise in this framework due to the real part of $C_7^\prime$, for two example scenarios, $\text{GMSSM}_\text{I}$ and $\text{GMSSM}_\text{II}$.
In Tab.~\ref{tab:parameter_GMSSM1}, we collect the corresponding input parameters.
Compared to the framework of the FBMSSM (see Fig.~\ref{fig:S_ABP}), the shift in the zeros of these observables show a completely different pattern. While the zero of $S_6^s$ remains SM like, a positive shift in $q^2_0(S_4)$ implies a negative shift in $q^2_0(S_5)$ and vice versa.

\begin{figure}
\centering
\includegraphics[width=0.6\textwidth]{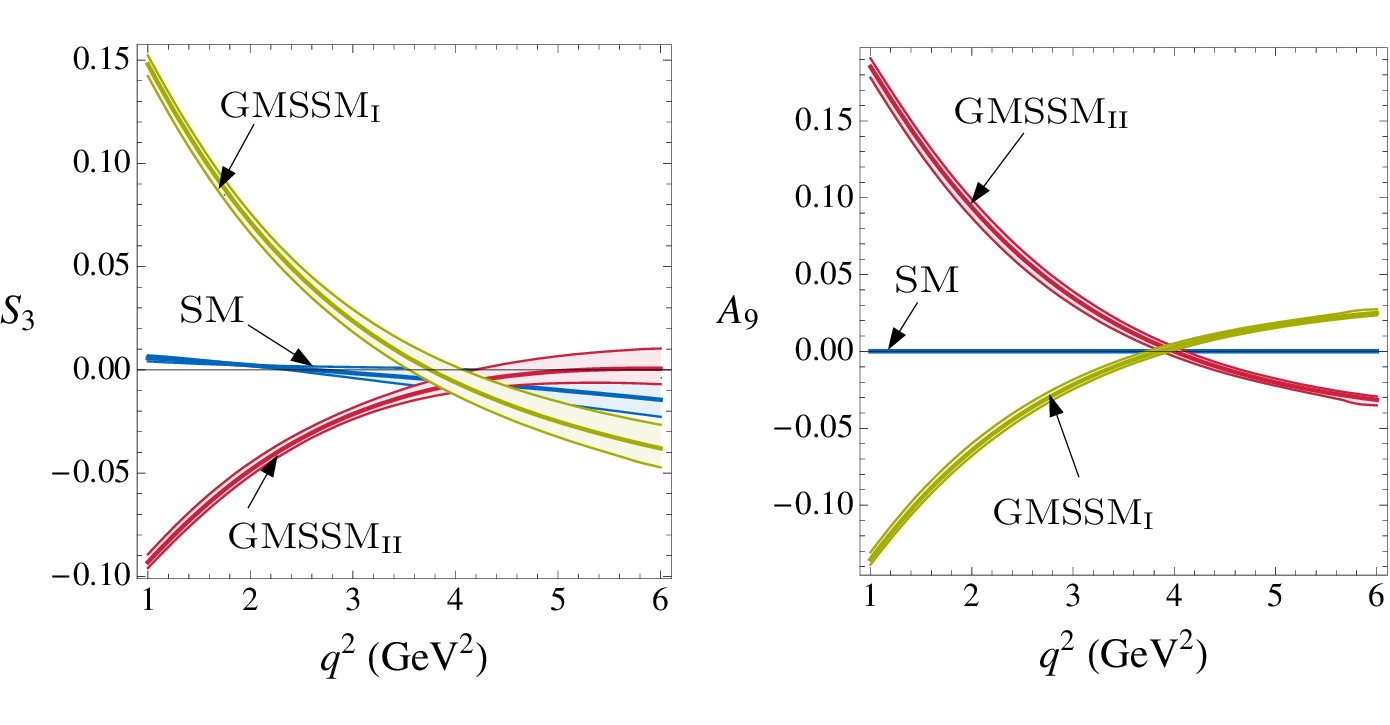}
\vskip-10pt
\caption[]{\small The observables $S_3$ and $A_9$ in the SM (blue
  band) and the two GMSSM scenarios GMSSM$_{\rm I,II}$ 
with large complex contributions to $C_7^\prime$ as described in the text.}
\label{fig:GMSSMC7p_S3A9}
\hskip10pt
\centering
\includegraphics[width=0.9\textwidth]{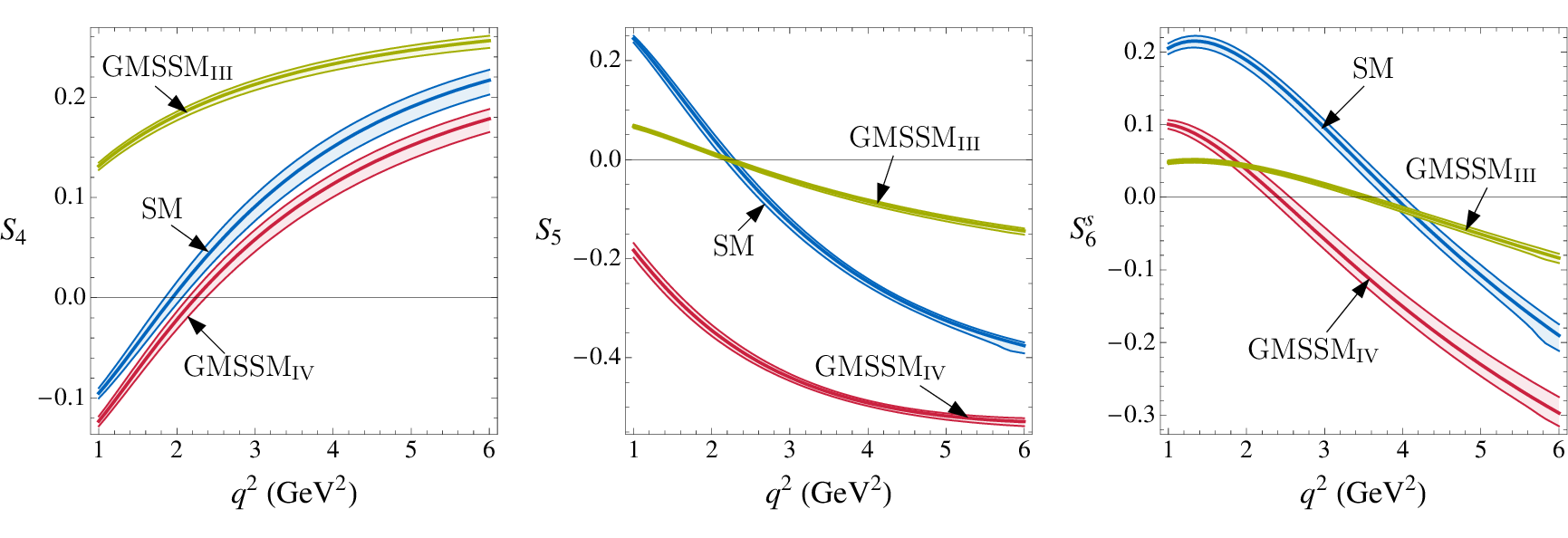}
\includegraphics[width=0.6\textwidth]{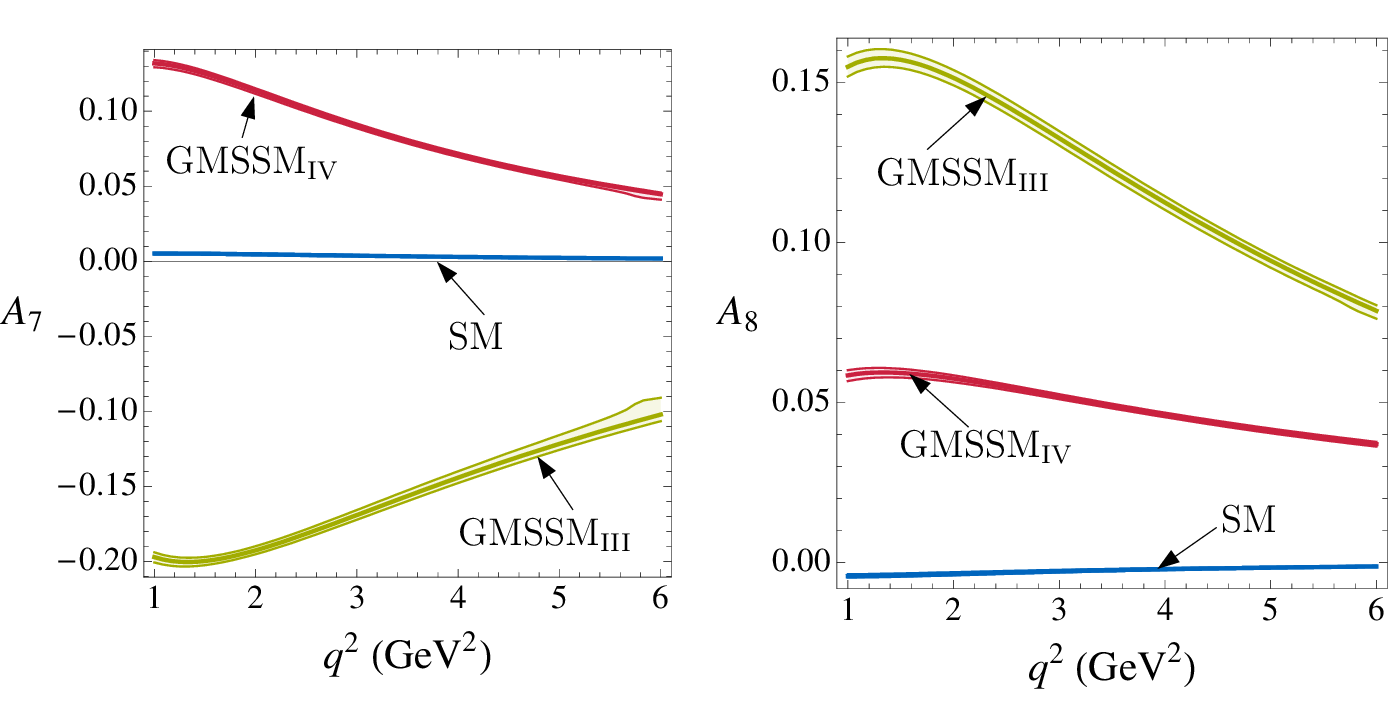}
\includegraphics[width=0.6\textwidth]{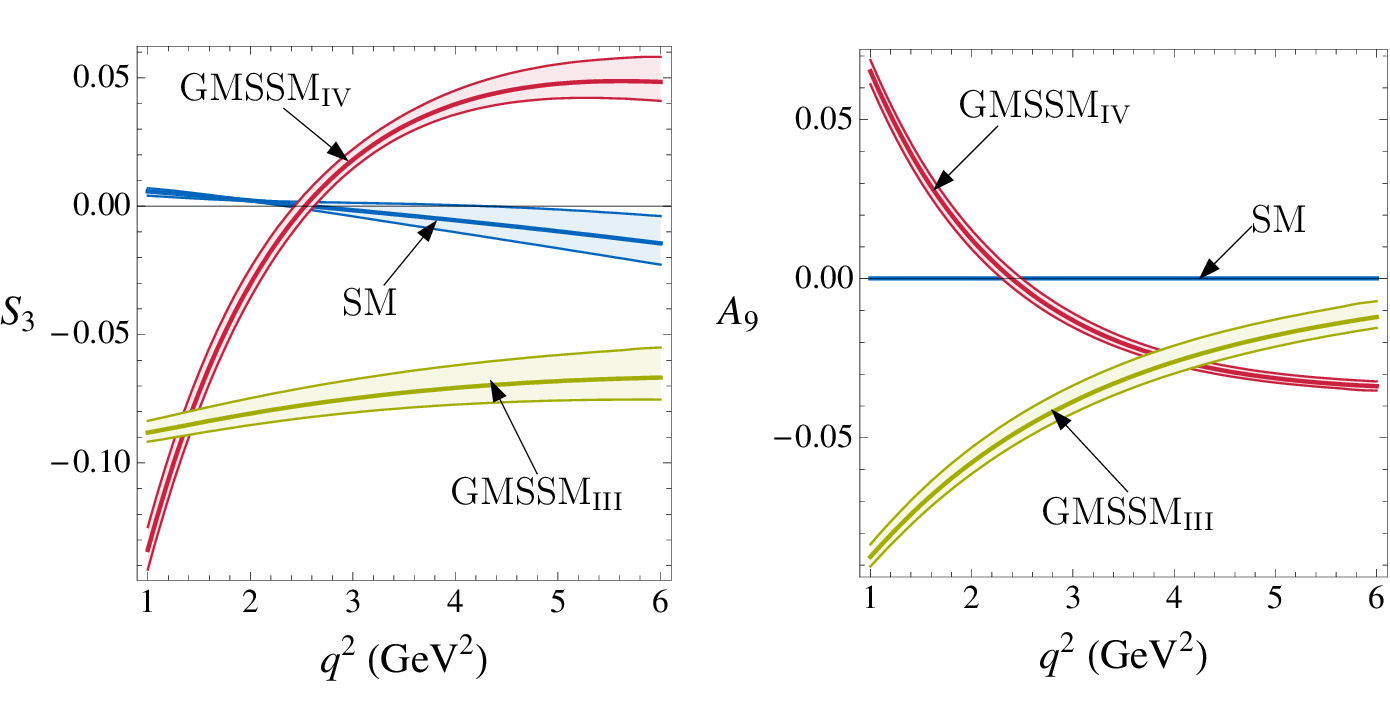}
\vskip-10pt
\caption[]{\small Several observables in the SM (blue band) and two selected GMSSM scenarios that show large non-standard behaviour. See text for details.}
\label{fig:GMSSM}
\end{figure}
Large imaginary parts of $C_7^\prime$ lead to sizeable effects in the asymmetries $A_7$ and $A_8$, but again the pattern of these effects is different to that in the FBMSSM seen in Fig.~\ref{fig:A_ABP}. As shown in Fig.~\ref{fig:GMSSMC7p_A7A8}, a positive (negative) $A_7$ implies also a positive (negative) $A_8$. In particular the correlation plot in the right panel of Fig.~\ref{fig:GMSSMC7p_A7A8} is completely orthogonal to the one in the FBMSSM (see Fig.~\ref{fig:A_ABP}) and thus a clear distinction between these two frameworks is possible.

In addition a large complex $C_7^\prime$ also leads to large non-standard effects in the observables $S_3$ and $A_9$ as shown in Fig.~\ref{fig:GMSSMC7p_S3A9}. In fact, as already mentioned in Sec.~\ref{sec:impact}, effects in $S_3$ and $A_9$ are characteristic for scenarios with large NP contributions to the primed Wilson coefficients. The large effects in $S_3$ are driven by the real part of $C_7^\prime$ and directly correspond to the large effects in the transverse asymmetry $A_T^{(2)}$ that have been analysed in \cite{Kruger:2005ep,LunghiMatias,Hurth08}. Having analysed possible effects in a particular non-minimal flavour violating MSSM framework we finally mention also the case of the general MSSM with generic flavour violating soft terms. Instead of presenting an exhaustive discussion of this framework, we concentrate on two specific scenarios that show effects that go beyond those discussed in the above.

Among the Wilson coefficients that are relevant in the decay $B\to K^*\mu^+\mu^-$ the ones that are most sensitive to NP effects arising from flavour violating down squark masses are $C_7$ and $C_7^\prime$.
In the plots of Fig.~\ref{fig:GMSSM} we show a scenario GMSSM$_{\rm IV}$ that corresponds to large NP contributions to both $C_7$ and $C_7^\prime$. In contrast to the scenario with NP effects dominantly in $C_7^\prime$ discussed above, one observes e.g.\ sizeable effects in the zeros of $S_5$ and $S_6^s$ while the zero in $S_4$ is much less affected. 

One possibility to generate large effects in the Wilson coefficient $C_{10}$ in a supersymmetric framework is through flavour violating entries in the left-right part of the up squark mass \cite{LMSS99,Buchalla:2000sk,ALGH01}. Scenario GMSSM$_{\rm III}$ in Fig.~\ref{fig:GMSSM} corresponds exactly to such a scenario where in addition to large complex NP contributions to $C_7$ and $C_7^\prime$, $C_{10}$ also receives sizeable complex corrections through a $(\delta_u)_{32}^{LR}$ mass insertion. These curves show again a qualitatively different behaviour in various observables. For example large effects in $S_3$ and $A_9$ can be observed, that however do not show a zero in contrast to the red curves discussed above.

%%%%%%%%%%%%%% Sec7: Summary and Conclusions

\section{Summary and Conclusions}\label{sec:7}
\setcounter{equation}{0}

In this paper we have analysed all angular observables in the 
rare decay $B\to K^* (\to K\pi) \mu^+\mu^-$. They can be 
measured at the LHC and later at an upgraded Belle and a 
Super-B facility. These angular observables can be
expressed in terms of CP-conserving and CP-violating
quantities and offer new important tests of the SM and its extentions.
To this end we have improved on previous studies in a number of ways that
have been listed in Sec.~\ref{sec:1}.

Having identified angular observables with small to moderate 
dependence on hadronic quantities and large impact of NP we
have analysed these observables first within the SM and subsequently
within a number of its extentions like 
models with MFV, the flavour-blind MSSM with new flavour conserving, but
CP-violating phases, the LHT model and also
within a general MSSM with generic flavour violating soft terms.
 
The main messages from this study are as follows:
\begin{itemize}
\item 
The most promising and complete set of observables in this channel are our $S_i^{(a)}$ and $A_i^{(a)}$ defined in Sec.~\ref{sec:5}.
\item
Our predictions for the CP-averaged angular coefficients $S_i^{(a)}$ in the SM
are shown in Fig.~\ref{fig:Ss-SM}. Some of these
$S_i^{(a)}$ are found to be large.
\item
On the other hand, as evident from Fig.~\ref{fig:As-SM},
the CP asymmetries $A_i^{(a)}$ are close to zero in the SM.
\item
Our model independent study shows that pseudoscalar operators are numerically 
irrelevant in the decay $B\to K^*\mu^+\mu^-$. On the other hand
a study of the angular distributions allows, in a way which is theoretically
clean and complementary to $B_s\to\mu^+\mu^-$, to probe the scalar
sector of a theory beyond the SM.
\item As one expects, the $A_i^{(a)}$ are SM like in MFV models. On the other hand, 
some of the $S_i^{(a)}$, in particular $S_4$, $S_5$ and $S_6^c$ can show deviations from the SM as seen in Figs.~\ref{fig:MFV} and \ref{fig:MFV_zeros}, where results in the MFV MSSM are shown.
\item
Probably the most interesting results are found in the FBMSSM, in which 
several $S_i^{(a)}$ and  $A_i^{(a)}$ differ significantly, even by orders of magnitude from the SM results,
and there exists a number of striking correlations among the observables
discussed here and also correlations between  $A_7$ (and $A_8$)
and $A_\text{CP}(b\to s\gamma)$ and $S_{\phi K_S}$. All 
these results are shown in Figs.~\ref{fig:A_ABP} to \ref{fig:correl_ABP}.
\item
In the LHT model only the CP asymmetries $A_7$ and $A_8$ differ significantly
from the SM predictions, but these enhancements are smaller than found in
the FBMSSM model. This different pattern of effects could easily distinguish
these two models.
\item
As expected, in a general MSSM, the very large space of parameters does not
allow for clear-cut conclusions. Almost all observables considered
in the present paper can significantly differ from the SM results and
the pattern of deviations can differ from those found in the FBMSSM and
LHT models. This is illustrated in Figs.~\ref{fig:GMSSMC7p_S4S5S6s} to 
\ref{fig:GMSSM}. 
This should allow these three models to be distinguished from
each other.
\end{itemize}

Clearly, it will be very exciting to monitor the upcoming LHC, Belle upgrade and
eventually Super-B factory in this and in the next decade to see whether
the angular observables discussed in our paper will give a hint for 
any of the extensions of the SM.

\section*{Acknowledgements}

The authors would like to thank S.~Recksiegel for providing sets of input 
parameters for the Littlest Higgs model which are compatible with FCNC
constraints, Th.~Feldmann for several useful discussions and G. Hiller and C. Bobeth
for clarifying communication.
P.B. gratefully acknowledges financial support from the Cluster of 
Excellence ``Origin and Structure of the Universe'' at TU Munich. 
A.K.M.B.\  acknowledges 
receipt of a UK STFC studentship and financial support from Lehrstuhl T31, 
TU Munich. This work has been supported in part by the EU network contract 
No.\ MRTN-CT-2006-035482 ({\sc Flavianet}), the Cluster of Excellence 
``Origin and Structure of the Universe'' and the German Bundesministerium 
f\"ur Bildung und Forschung under contract 05HT6WOA.

\appendix

\section*{Appendices}
\renewcommand{\theequation}{\Alph{section}.\arabic{equation}}
\renewcommand{\thetable}{\Alph{table}}
\renewcommand{\thefigure}{\Alph{figure}}
\setcounter{section}{0}
\setcounter{table}{0}
\setcounter{figure}{0}

\section{Kinematics of Four-Body Decays}\label{sec:kinematics}
\setcounter{equation}{0}

In this appendix we collect some relevant results concerning the
kinematics of the four-body decay $X \to Y(\to a b) Z(\to c d)$. For an
excellent discussion see Ref.~\cite{widhalm}.
The translation of the general results to the notation appropriate
for the decay $B \to K^* (\to K\pi) \mu^ + \mu ^ - $ is given in Tab.~\ref{tab:A}. 

The four-body phase space is $4\times4$-dimensional. The on-shell 
conditions of the final states reduce this number to
$4\times3$. Moreover, four-momentum conservation eliminates
further $4$ degrees of freedom. Eventually, exploiting isotropic 
symmetry, one can fix the three Euler angles and ends up with 5 physical 
degrees of freedom. It is customary and convenient to express them by the
following set of variables, introduced first in
Ref.~\cite{CabibboMaksymowicz} for the decay 
$K^ +   \to \pi ^ +  \pi ^ -  e^ +  \nu   $:
\begin{itemize}
	\item $m_{ab}^2$, the effective mass squared of the $ab$
	  system, $m_a+m_b<m_{ab}<m_X-m_c-m_d$;
	\item $m_{cd}^2$, the effective mass squared of the $cd$
	  system, $m_c+m_d<m_{cd}<m_X-m_a-m_b$; note that $m_{ab}+m_{cd}<m_X$;
	\item $\theta_Y$, the angle of the particle $a$ in the
	  c.m. system of the particles $a$ and $b$ with respect to the 
direction of flight of $(a,b)$ in the $X$ rest system, $0<\theta_Y<\pi$;
	\item $\theta_Z$, the angle of the particle $c$ in the
	  c.m. system of the particles $c$ and $d$ with respect to the 
direction of flight of $(c,d)$ in the $X$ rest system, $0<\theta_Z<\pi$;
	\item $\theta_X$, the angle between the plane formed by the
	  decay products $(a,b)$ in the $X$ rest system and the 
corresponding plane of $(c,d)$, $-\pi<\theta_X<\pi$.
\end{itemize}
It is convenient to combine the four-momenta $p_i$, $i=a,b,c,d$,
of the final-state particles
into the following symmetric and antisymmetric momenta:
\begin{eqnarray*}
   {P_{ab} = p_{a}  + p_{b} }\,,  \qquad
   {Q_{ab} = p_{a}  - p_{b} }\,, \\
   {P_{cd} = p_{c}  + p_{d} }\,,  \qquad
   {Q_{cd} = p_{c}  - p_{d} }\,. 
\end{eqnarray*}
The diparticle masses are then given by
\begin{equation*}
P_{ab}^2 = m_{ab}^2\,,\qquad P_{cd}^2 = m_{cd}^2\,.
\end{equation*}
Now we define the angles $\theta_X$, $\theta_Y$ and $\theta_Z$ by
\begin{equation}\label{eq:thetaYZ}
\cos \theta _{Y}  =  - \frac{{\vec Q_{ab} \cdot \vec P_{cd} }}{{\left| 
{\vec Q_{ab} } \right|\left| {\vec P_{cd} } \right|}}\,,~~~ 
\cos \theta _{Z}  =  - \frac{{\vec Q_{cd} \cdot \vec P_{ab} }}{{\left| 
{\vec Q_{cd} } \right|\left| {\vec P_{ab} } \right|}}\,,
\end{equation}
\begin{equation}\label{eq:thetaX}
\sin \theta _{X}  = \frac{{\left( {\vec P_{ab}  \times \vec Q_{ab} }
    \right) 
\times \left( {\vec P_{cd}  \times \vec Q_{cd} } \right)}}{{\left| 
{\vec P_{ab}  \times \vec Q_{ab} } \right|\left| {\vec P_{cd}  \times 
\vec Q_{cd} } \right|}}\,.
\end{equation}
It is important to note that the three-vectors in the above definition 
must be evaluated in the respective rest frames of the particles
$X$, $Y$ and $Z$. 

With the above definitions, it is straightforward to
express all remaining invariant products of the four-vectors  
$P_{ab}$, $P_{cd}$, $Q_{ab}$ and $Q_{cd}$ in terms of the five
variables $\theta_X$, $\theta_Y$, $\theta_Z$, $m_{ab}^2$ and $m_{cd}^2$:
\begin{eqnarray*}
 P_{ab} Q_{ab}  &=& m_a^2  - m_b^2\,,  \\
 P_{ab} P_{cd}  &=& \bar p\,, \\ 
 P_{ab} Q_{cd}  &=&\frac{{m_c^2  - m_d^2 }}{{m_{cd}^2 }}\,\bar p + 
\frac{2}{{m_{cd}^2 }}\,\sigma \sigma _{cd} \cos \theta _{Z}\,, \\
Q_{ab} Q_{ab}   &=& 2(m_a^2  + m_b^2 ) - m_{ab}^2 \,,\\
Q_{ab} Q_{cd}   &=& \frac{1}{{m_{ab}^2 m_{cd}^2 }}\left[ {(m_a^2  -
 m_b^2 ) (m_c^2  - m_d^2 )\bar p}\right. \nonumber\\ 
& &  + 2\sigma \sigma _{ab} (m_c^2  - m_d^2 )\cos \theta _{Y} \nonumber\\
& &  + 2\sigma \sigma _{cd} (m_a^2  - m_b^2 )\cos \theta _{Z} \nonumber\\
& &  + 4\sigma _{ab} \sigma _{cd} \bar p\cos \theta _{Y} \cos \theta
 _{Z}   \nonumber\\ 
& & \left. { + 4\sigma _{ab} \sigma _{cd} m_{ab} m_{cd} \sin \theta
 _{Y} \sin \theta _{Z} \cos \theta _{X} } \right]\,,\\
\varepsilon _{\alpha \beta \gamma \delta } P_{ab}^\alpha  Q_{ab}^\beta  
P_{cd}^\gamma  Q_{cd}^\delta   &=&  - \frac{{4\sigma \sigma _{ab} 
\sigma _{cd} }}{{m_{cd}m_{ab} }}\sin \theta _{Y} \sin \theta _{Z} \sin \theta _{X} \,.
\end{eqnarray*}
The remaining invariants can be obtained from the above by
$(ab)\leftrightarrow (cd)$  and $\theta _{Z} \leftrightarrow \theta _{Y}$.\\
Here we use 
\begin{eqnarray*}
\bar p &=& \frac{1}{2}(m^2_X  - m_{ab}^2  - m_{cd}^2 )\,, \\ 
 \sigma &=& \sqrt {\bar p^2  - m_{ab}^2 m_{cd}^2 } \,,  \\ 
 \bar p_{ab} &=& \frac{1}{2}(m_{ab}^2  - m_a^2  - m_b^2 ) \,, \\ 
 \sigma _{ab} &=&\sqrt {\bar p_{ab} ^2  - m_a^2 m_b^2 }  \
\end{eqnarray*}
and corresponding expressions with $(ab)\rightarrow (cd)$.
Tab.~\ref{tab:A} provides the translation of the general notations
introduced above to the
special case $B \to K^* (\to K\pi) \mu^ + \mu ^ - $.
\begin{table}
\begin{center}
\renewcommand{\arraystretch}{1.2}
\begin{tabular}{|c|c|} \hline
{\rm general} & $B \to K\pi \mu^ + \mu ^ - $\\ \hline\hline
$(ab)$     &  $(\pi K)$\\ \hline
$(cd)$     &   $(\mu^+\mu^-)$\\ \hline
$m_{ab}$   &          $m_{K^*}$  \\ \hline
$m_{cd}$   &         $\sqrt{q^2}$  \\ \hline
$\theta_X$ &  $\phi$      \\ \hline
$\theta_Y$ &  $\theta_{K^*}$      \\ \hline
$\theta_Z$ &  $\theta_\mu$      \\ \hline
$\sigma^2_{ab}$ & $m_{K^*}^4\beta^2/4$ \\ \hline
$\sigma^2_{cd}$ & $q^4\beta^2_{\mu}/4$ \\ \hline
$\sigma^2$ & $\lambda/4$ \\ \hline
$\bar p$ & $(k \cdot q)$\\ \hline
\end{tabular}
\end{center}
\caption[]{\small Translation table between the variables of the
  general four-body decay $X\to Y(\to ab) Z(\to cd)$ discussed in
  App.~\ref{sec:kinematics} and the decay $B\to K^* (\to K\pi) \mu^+\mu^-$. 
Note that $m_{ab}$ is fixed to $m_{K^*}$, which
  leads to an
allowed range of $q^2$ of $\left( {2m_\mu  } \right)^2  < q^2  <
\left( {m_B  - m_{K^ *  } } \right)^2$. $\beta$, $\beta_\mu$ and $\lambda$
are defined in Sec.~\ref{sec:3}.
}\label{tab:A}
\end{table}

\section{Form Factors and QCD Factorization}\label{app:B}
\setcounter{equation}{0}

As we have seen in Sec.~\ref{sec:2.3}, the QCDF formulas for
$B\to K^*\mu^+\mu^-$ are expressed in terms of two soft form factors
$\xi_{\perp,\parallel}$, defined in Eqs.~(\ref{3.17})--(\ref{3.18}),
rather than the seven form factors of
Sec.~\ref{sec:ff}, which implies certain relations between them, valid
in the large energy limit. These relations are the topic of this
appendix and we shall demonstrate that they are indeed fulfilled by
light-cone sum rules (LCSRs) -- 
both analytically and numerically.

 Exploiting Eq.~(\ref{eqF}) for all 7 form factors, one can establish
 a number of relations which are expected to be valid for $K^*$
 energies $E\sim O(m_b)$ \cite{BF00}:
\begin{equation}
  \frac{A_1(q^2)}{V(q^2)} = \frac{2 Em_B}{(m_B+m_{K^*})^2},
\label{B.1}
\end{equation}
\begin{equation}
  \frac{T_1(q^2)}{V(q^2)} = \frac{m_B}{2E} \,
  \frac{T_2(q^2)}{V(q^2)} = \frac{m_B}{m_B+m_{K^*}}   \left(
    1 + \frac{\alpha_s C_F}{4\pi}   \left[ \ln\frac{m_b^2}{\mu^2} -
      L\right] 
      + \frac{\alpha_s C_F}{4\pi} \, \frac{m_B}{4E} \, 
\frac{\Delta F_\perp}{\xi_\perp(q^2)}
   \right),
\label{B.2}
\end{equation}
\begin{eqnarray}
 \frac{(m_B+m_{K^*})/(2E) \, A_1(q^2) - (m_B-m_{K^*})/m_B \,
    A_2(q^2)}{(m_{K^*}/E)\,A_0(q^2)} &=&
\nonumber\\
&&\hspace*{-7cm}
   1 + \frac{\alpha_s C_F}{4\pi}  
    \left[-2+2L\right]
    -\frac{\alpha_s C_F}{4\pi} \, \frac{m_B(m_B-2E)}{(2 E)^2} \, \frac{\Delta
      F_\parallel}{(E/m_{K^*})\,\xi_\parallel(q^2)},\label{B.3}
\end{eqnarray}
\begin{equation}
  \frac{(m_B/2E) \, T_2(q^2) - T_3(q^2)}{(m_{K^*}/E) A_0(q^2)} =
  1 + \frac{\alpha_s C_F}{4\pi}  
     \left[ \ln \frac{m_b^2}{\mu^2} - 2 + 4 L \right]
   - \frac{\alpha_s C_F}{4\pi} \, \left(\frac{m_B}{2E}\right)^2 \frac{\Delta
      F_\parallel}{(E/m_{K^*})\,\xi_\parallel(q^2)},
\label{B.4}
\end{equation}
respectively. The abbreviations on the right hand side are defined as
\begin{eqnarray}
L&=& - \frac{2E}{m_B-2E}\, \ln\,\frac{2E}{m_B}\,,
\nonumber\\
\Delta F_\perp & = & \frac{8\pi^2 f_B f_{K^*}^\perp}{3
  m_B}\,\frac{1}{\lambda_B}\, \int du \,\frac{\phi_\perp(u)}{1-u}\,,
\nonumber\\ 
\Delta F_\parallel & = & \frac{8\pi^2 f_B f_{K^*}^\parallel}{3
  m_B}\,\frac{1}{\lambda_B}\, \int du \,\frac{\phi_\parallel(u)}{1-u}
\end{eqnarray}
in terms of the twist-2 $K^*$ DAs $\phi_{\perp,\parallel}$ and the
first inverse moment of the $B$ meson DA, $1/\lambda_B$. $E$, the
$K^*$'s energy, is
related to $q^2$ by $2 m_B E = m_B^2-q^2$ (terms in $m_{K^*}^2$ are
neglected). Numerical
values of these parameters are given in Tab.~\ref{tab:numinput}.

One requirement for the LCSRs is obviously that they fulfill the
above relations, to the required accuracy, i.e.\ in the large energy
limit, and that the leading contributions to $\xi_{\perp(\parallel)}$ come
from transversal (longitudinal) DAs.  
The above relations imply in particular that both $A_1$ and $A_2$ must
to leading order be given in terms of transversal DAs, while 
these contributions 
must cancel in the combination $(m_B+m_{K^*})/(2E) \, A_1 - 
(m_B-m_{K^*})/m_B \, A_2$. As this is an important point, we write down the 
tree-level contributions to these sum rules, and that for $V$, 
explicitly ($\bar u =
1-u$, $u_0 = (m_b^2-q^2)/(s_0-q^2)$):
\begin{eqnarray}
A_0(q^2) 
& = &
\frac{m_b}{m_B^2 f_B}\,e^{m_B^2/M^2}\left[\frac{f_{K^*}^\parallel m_b}{2}\, 
\int_{u_0}^1 du \, e^{-\frac{m_b^2-\bar u q^2}{uM^2}} \frac{1}{u} 
\left[ g_v(u) + \left\{ 1 + u\,\frac{d}{du}\right\} \left(
  \frac{\Phi(u)}{u} \right) \right] \right.
\nonumber\\
& & \left. + \dots \right],\\
A_1(q^2) 
& = & 
\frac{m_b\,e^{m_B^2/M^2}}{m_B^2 f_B (m_B+m_{K^*})}\int_{u_0}^1 du \,
e^{-\frac{m_b^2-\bar u q^2}{uM^2}} \frac{1}{u}  \left[ f_{K^*}^\perp
  (m_b^2-q^2)\, \frac{\phi_\perp(u)}{2u} + f_{K^*}^\parallel m_b m_{K^*}
  \,g_v(u)\right.
\nonumber\\
& & \left. + \dots\right],\label{3.25}\\
A_2(q^2) 
& = & 
\frac{m_b\,e^{m_B^2/M^2}}{m_B^2 f_B (m_B-m_{K^*})}\int_{u_0}^1 du \,
e^{-\frac{m_b^2-\bar u q^2}{uM^2}} \frac{1}{u}  \left[  f_{K^*}^\perp
  (m_b^2-\bar u q^2)\, \frac{\phi_\perp(u)}{2u} \right.
\nonumber\\
& &  \hspace*{4.5cm}\left. 
-f_{K^*}^\parallel m_b m_{K^*}\,\frac{m_b^2-\bar u
    q^2}{m_b^2-q^2} \left\{ 1 + u \,\frac{d}{du} \right\} 
\frac{\Phi(u)}{u} + \dots \right],\label{3.26}\\
V(q^2) & = & \frac{m_B+m_{K^*}}{2}\, \frac{m_b\,e^{m_B^2/M^2}}{m_B^2
  f_B} \int_{u_0}^1 du \,
e^{-\frac{m_b^2-\bar u q^2}{uM^2}} \frac{1}{u} \left[ f_{K^*}^\perp
  \phi_\perp(u)\right.\nonumber\\
& & \hspace*{5cm} \left. - \frac{u}{2(m_b^2-q^2)} \, f_{K^*}^\parallel 
m_b m_{K^*} \,\frac{d}{du} \,g_a(u) + \dots\right].\label{3.27}
\end{eqnarray}
The dots stand for terms with higher powers of $m_{K^*}$. All DAs
multiplying $f_{K^*}^{(\perp)}$ are longitudinal (transversal). Note
that, as mentioned in Sec.~\ref{sec:ff}, the suppression factor of
higher-twist terms is given by $m_b m_{K^*}/(m_b^2-q^2)$, which
indicates the break-down of the expansion for $q^2\to m_b^2$.
Comparing the above formulas with (\ref{B.3}) one finds that for
$q^2=0$ ($E=m_B/2$) the terms in $\phi_\perp$ cancel exactly in the
combination of $A_1$ and $A_2$ and (\ref{B.3}) is reproduced up to
terms in higher powers of $m_{K^*}$ 
(which are neglected in the large energy limit).
For $q^2\neq 0$, however, there appears to be a non-vanishing
term in $f_{K^*}^\perp$:
\begin{eqnarray}
\lefteqn{(m_B+m_{K^*})/(2E) \, A_1 - (m_B-m_{K^*})/m_B \,
    A_2}\hspace*{2cm}\nonumber\\
& \sim & \frac{m_b\,e^{m_B^2/M^2}}{m_B^2 f_B}
\int_{u_0}^1 du \,
e^{-\frac{m_b^2-\bar u q^2}{uM^2}} f_{K^*}^\perp\,\frac{\phi_\perp(u)}{u} \,
\frac{q^2 ( m_b^2-
    q^2 - u (m_B^2-q^2))}{2 m_B (m_B^2-q^2) u}\,.\label{xyz}
\end{eqnarray}
At this point it is useful to recall that, ideally, if the continuum
model was perfect and all terms in the twist expansion were known, the
LCSR would be independent of $s_0$ and $M^2$. In reality, 
LCSRs are non-trivial functions of both parameters and the form factors
are extracted within an interval of $M^2$ where the dependence on
that parameter is small, i.e.\ the sum rules are
evaluated  near an extremum in $M^2$, if such an
extremum exists. In Ref.~\cite{BZ04PS} we have argued that the central
value of form factors should be evaluated exactly {\em at} the
extremum. This implies that for both an ideal LCSR with completely known
hadronic spectral density $\rho(u)$ and a realistic one with $\rho(u)$
known to a
certain accuracy in the light-cone expansion one requires
\begin{eqnarray}
\lefteqn{\frac{d}{dM^2}\, \int_{u_0}^1 du \
e^{\frac{m_B^2}{M^2}-\frac{m_b^2-\bar u q^2}{uM^2}} \,\rho(u) = 0}\nonumber\\
& \longleftrightarrow &
\int_{u_0}^1 du \, e^{\frac{m_B^2}{M^2}-\frac{m_b^2-\bar u q^2}{uM^2}} 
\frac{u m_B^2 - m_b^2 + \bar u
  q^2}{u}\, \rho(u) = 0\,.
\end{eqnarray}
\begin{figure}
\includegraphics[width=0.44\textwidth]{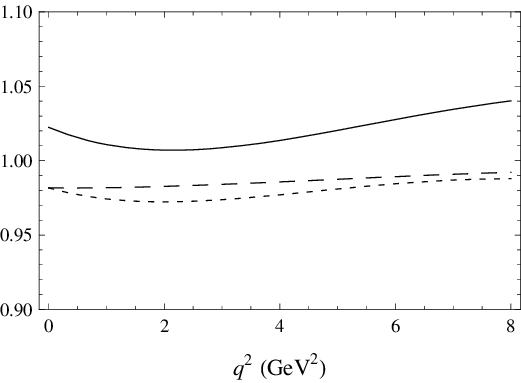}
\qquad
\includegraphics[width=0.44\textwidth]{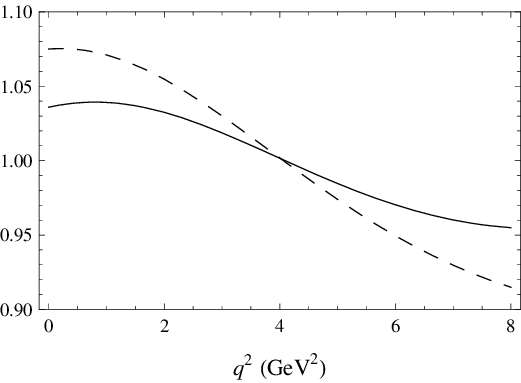}
% \vspace*{-30pt}
\caption[]{\small Form factor ratios (\ref{B.1}) to (\ref{B.4}),
  l.h.s.\ divided by r.h.s., calculated from LCSRs with central values
  of input parameters. In the perfect heavy quark limit, all
  ratios equal 1. Left: form factor ratios based on $\xi_\perp$.
  Solid curve: (\ref{B.1}), long dashes: (\ref{B.2}) with $T_1/V$,
  short dashes: (\ref{B.2}) with $T_2/V$. Right: form factor ratios
  based on $\xi_\parallel$. Solid curve: (\ref{B.3}), long dashes: (\ref{B.4}).
 $\alpha_s$ is evaluated at
the scale $\mu^2=m_B^2-m_b^2 = O(m_b)$.}\label{fig:A}
\end{figure}
As the integral must vanish, one finds that $\rho(u)$ effectively
equals $(m_b^2-q^2)/(m_B^2-q^2)\,\rho(u)/u$
at and near the extremum.
Hence the r.h.s.\ of (\ref{xyz}) {\em vanishes} at the minimum in
$M^2$ {\em if} the l.h.s.\ is treated as LCSR in its own right and not
as sum of two different sum rules (with possibly different extrema
in $M^2$ and different optimal values in $s_0$). This is also the reason
why, in order to obtain values for $A_0(0)$ and $\xi_\parallel$, it is
not useful to add the numbers for the individual form factors $A_1$
and $A_2$ obtained from LCSRs in Ref.~\cite{BZ04}. Instead, the above
combination of $A_1$ and $A_2$ has to be evaluated anew. 
We would like to add that the cancellation of the contribution of
$\phi_\perp$ to $\xi_\parallel$ can also be made manifest by including
the factor $1/E$ into the dispersion relation, using the above
result that factors $u m_B^2 - m_b^2 + \bar u q^2$ return zero under
the integral if it is evaluated at the minimum in $M^2$. One then can
make the replacement
$$\frac{1}{2E} \to \frac{1}{m_B} \,\frac{m_b^2-\bar u q^2}{m_b^2-q^2}$$
upon which the contribution in $\phi_\perp$
cancels explicitly so that
$$\frac{m_B+m_{K^*}}{2E} \, A_1 - \frac{m_B-m_{K^*}}{m_B} \,
    A_2 = \frac{m_{K^*}}{E}\, A_0 + O(\alpha_s) + \dots$$
in the LCSR method. Again, the dots denote terms in higher powers of
    $m_{K^*}$. Similar analyses can be done for the other form factors
    ratios, for instance $A_1/V$, Eq.~(\ref{B.1}). 
Making again use of the fact that an
    additional factor $u$ under the integral equals
    $(m_b^2-q^2)/(m_B^2-q^2)$, one finds from (\ref{3.25}) and
    (\ref{3.27}) that (\ref{B.1}) is fulfilled to twist-2 accuracy,
    but also that the ratio deviates from the r.h.s.\ of (\ref{B.1})
    at twist-3 level, i.e.\ at $O(m_{K^*}/m_b)$.

Turning to the $O(\alpha_s)$ corrections in (\ref{B.1}) to
(\ref{B.4}), we can confirm that to twist-2 accuracy (\ref{B.1})
does not receive any such corrections: they cancel exactly between
numerator and denominator. Reproducing the $O(\alpha_s)$ corrections
in (\ref{B.2}) to (\ref{B.4}) 
is less trivial and requires to explicitly perform
the limit $m_b\to\infty$ of the integral over $u$. We refrain from
doing this analysis explicitly, but refer to Ref.~\cite{mySCET} where
it was shown, for $B\to\pi$ transitions, that LCSRs fulfill the SCET
relations  also at $O(\alpha_s)$.

In Fig.~\ref{fig:A} we plot the form factor ratios as functions of
$q^2$ for central values of the input parameters, separately for
ratios based on $\xi_\perp$ and those based on $\xi_\parallel$. All
ratios include the $\mathcal O(\alpha_s)$ corrections calculated in
QCDF. We find that, overall, the QCDF predictions are fulfilled by the
full QCDF form factors from LCSRs at the level of 10\% or better. 
Nonetheless there is a considerable dependence of the $\xi_\parallel$
ratios on $q^2$. This is due to $1/m_b$ corrections which are
neglected in QCDF, but turn out to be, numerically, larger than the
factorizable $\mathcal O(\alpha_s)$ corrections calculable in QCDF. 
As a consequence, we expect that the transversity amplitudes $A_{
\perp(\parallel)L,R}$ of Sec.~\ref{sec:3.2}, and all angular
observables built from them, should be rather insensitive to $1/m_b$ 
corrections, i.e.\ corrections to QCDF, 
while $A_{0L,R}$, $A_t$, $A_S$ and all corresponding angular variables
will be affected by such corrections to a larger degree.

\end{document}